\documentclass[prb,longbibliography,superscriptaddress,floatfix,overcite]{revtex4-1}
\usepackage{graphicx}
\usepackage{amsmath}
\usepackage{mathtools}
\usepackage{amsfonts}
\usepackage{amssymb}
\usepackage{booktabs}
\usepackage[utf8]{inputenc}

\usepackage{bm}
\begin{document}

\title{Efficient density-fitted explicitly correlated dispersion and exchange dispersion energies}

\author{Monika Kodrycka}
\altaffiliation{Present address: Department of Chemistry, Virginia Tech, Blacksburg, Virginia 24061}
\author{Konrad Patkowski}
\email{patkowsk@auburn.edu}
\affiliation{Department of Chemistry and Biochemistry, Auburn University,
Auburn, AL 36849, United States}

\date{\today}

\begin{abstract}
The leading-order dispersion and exchange-dispersion terms in symmetry-adapted perturbation theory (SAPT),
$E^{(20)}_{\rm disp}$ and $E^{(20)}_{\rm exch-disp}$, suffer from slow convergence to the complete basis
set limit. To alleviate this problem, explicitly correlated variants of these corrections, 
$E^{(20)}_{\rm disp}$-F12 and $E^{(20)}_{\rm exch-disp}$-F12, have been proposed recently. However, the
original formalism (M. Kodrycka {\em et al.}, {\em J. Chem. Theory Comput.} {\bf 2019}, {\it 15}, 5965--5986),
while highly successful in terms of improving convergence, was not competitive to conventional orbital-based
SAPT in terms of computational efficiency due to the need to manipulate several kinds of
two-electron integrals. In this work, we
eliminate this need by decomposing all types of two-electron integrals using robust density fitting.
We demonstrate that the error of the density fitting approximation is negligible when standard auxiliary bases
such as aug-cc-pV$X$Z/MP2FIT are employed. The new implementation allowed us to study all complexes in the A24 
database in basis sets up to aug-cc-pV5Z, and the $E^{(20)}_{\rm disp}$-F12 and $E^{(20)}_{\rm exch-disp}$-F12
values exhibit vastly improved basis set convergence over their conventional counterparts. 
The well-converged 
$E^{(20)}_{\rm disp}$-F12 and $E^{(20)}_{\rm exch-disp}$-F12 numbers can be substituted for conventional
$E^{(20)}_{\rm disp}$ and $E^{(20)}_{\rm exch-disp}$ ones in a calculation of the total SAPT interaction
energy at any level (SAPT0, SAPT2+3, \ldots). 
We show that the addition of F12 terms does not improve
the accuracy of low-level SAPT treatments. However, when the theory errors are minimized in high-level SAPT
approaches such as SAPT2+3(CCD)$\delta$MP2, the reduction of basis set incompleteness errors thanks to the F12
treatment substantially improves the accuracy of small-basis calculations. 
\end{abstract}

\maketitle

\section{Introduction}

Noncovalent intermolecular interactions are ubiquitous, and their quantitative understanding is essential when one investigates diverse physical and chemical phenomena from scattering resonances 
in cold collisions \cite{Klein:17} to crystal structure and polymorphism \cite{Beran:16} to chiral discrimination in biomolecular complexes \cite{Kraka:13}. However, accurate computations of
noncovalent interaction energies are quite nontrivial due to the overwhelming importance of electron correlation. In particular, London dispersion forces, which are the main reason for the
attraction between nonpolar molecules, arise entirely out of electron correlation. Actually, these forces require quite a high-level account of correlation, as the simplest dispersion estimate, contained
in the supermolecular interaction energy computed using the second-order M{\o}ller-Plesset perturbation theory (MP2), often leads to significant overbinding \cite{Hobza:96}. In addition to
the need for a high-level electronic structure treatment (such as the ``gold standard'' \cite{Kodrycka:19} coupled-cluster approach with single, double, and perturbative triple excitations,
CCSD(T) \cite{Raghavachari:89}), capturing the dispersion effects requires large basis sets, as the correlation energy is known for its slow convergence to the complete basis set (CBS) limit.
The basis set convergence of molecular correlation energies can be enhanced by extrapolations \cite{Halkier:98} as well by an explicitly correlated treatment, where an explicit dependence on the
interelectronic distance $r_{12}$ is inserted into the wavefunction {\em Ansatz} \cite{Kutzelnigg:91}. The most practical and successful explicitly correlated variantis use a correlation factor
with nonlinear $r_{12}$ dependence, in particular, the Slater-type expression $e^{-\gamma r_{12}}$, where the parameter $\gamma$ can be varied to control the spatial range of explicit
correlation \cite{Tew:05}. Such an approach is denoted by adding ``-F12'' to the name of the parent electronic structure theory, and many correlated {\em ab initio} approaches have been
``F12'ed'', including MP2 \cite{Tew:05,Werner:07}, CCSD(T) \cite{Shiozaki:08,Kohn:09}, or even multireference methods such as the second-order complete-active-space perturbation theory (CASPT2)
\cite{Shiozaki:10} or multireference configuration interaction (MRCI) \cite{Shiozaki:11}. In the specific case of intermolecular interactions, another way of speeding up the CBS convergence
is an addition of ``midbond'' basis functions centered in between the interacting molecules \cite{Tao:01}. The inclusion of bond functions is particularly helpful in recovering the dispersion
energy \cite{Williams:95}, and the benefits of bond functions can be combined with those of the CBS extrapolation \cite{Jeziorska:08} and of the F12 treatment \cite{Dutta:18}.

The most rigorous definition of dispersion energy is provided by second-order perturbation theory (some higher-order terms can also be classified as pure dispersion, but these terms are typically
small) \cite{Stone:13}. Thus, dispersion can be quantified, separately from other energy contributions such as electrostatics, induction, and exchange, within the framework of symmetry-adapted
perturbation theory (SAPT) \cite{Jeziorski:94,Patkowski:20}. Many variants of SAPT have been proposed in the literature, differing in the approach used to treat the intramolecular electron correlation.
Model SAPT studies for the smallest complexes have been carried out \cite{Korona:97a,Patkowski:01} with the complete, full configuration interaction (FCI) account of intramolecular correlation,
however, this is obviously not possible for realistic systems. Instead, one can start the perturbation expansion from the product of the monomer Hartree-Fock (HF)
determinants. The simplest way to proceed is then to ignore intramolecular correlation completely and carry out a single perturbation expansion in powers of the intermolecular interaction
operator, leading to an approach termed SAPT0. The SAPT0 dispersion energy, closely related to the dispersion part of the supermolecular MP2 interaction energy \cite{Chalasinski:88}, is denoted
as $E^{(20)}_{\rm disp}$, where the consecutive superscripts signify that this correction is of second order in the intermolecular interaction and of zeroth order in the intramolecular correlation.
If quantitative accuracy of the SAPT energy terms is required, one has to go beyond SAPT0 and include, in one way or another, the intramolecular correlation effects. The historically first
method of doing so is the {\em many-body SAPT} \cite{Rybak:91}, in which SAPT becomes a double perturbation theory in powers of both the intermolecular interaction operator $V$ and the 
M{\o}ller-Plesset fluctuation potential $W=W_{\rm A}+W_{\rm B}$, where $W_{\rm X}$ is the difference between the full Hamiltonian for molecule X and its HF approximation. At the highest
developed level of many-body SAPT, the dispersion energy is computed as $E^{(20)}_{\rm disp}+E^{(21)}_{\rm disp}+E^{(22)}_{\rm disp}$, thus, it includes intramolecular correlation through second
order in $W$. Alternatively, an intramolecular correlation correction to SAPT0 dispersion can be evaluated in a nonperturbative manner, using a CC-like partial infinite-order summation of
terms of different orders in $W$ \cite{Williams:95a,Korona:08} or the frequency-dependent density susceptibilities from time-dependent density functional theory, as in the SAPT(DFT) approach
\cite{Hesselmann:05,Misquitta:05a}. The development of improved dispersion expressions in SAPT continues to this day: the most recent accomplishments include the computation of dispersion energies
from the Bethe-Salpeter equation \cite{Holzer:17a} and from the linear response of the complete-active-space self-consistent field (CASSCF) approach \cite{Hapka:19}.

Every electrostatic, induction, and dispersion correction in SAPT is accompanied by an exchange term that arises from the enforcement of the full antisymmetry upon the wave
function of the complex. In particular, second-order dispersion energy has an exchange-dispersion counterpart that cancels a part of the dispersion attraction (typically on the order of 10\%\
at the van der Waals minimum distance). While smaller than dispersion, the exchange dispersion energy converges in relative terms just as poorly as dispersion energy with both the theory level and
the basis set. This is quite unfortunate as all commonly used variants of many-body SAPT approximate the exchange dispersion effects solely by their leading-order
$E^{(20)}_{\rm exch-disp}$ term. The intramolecular correlation contributions to exchange dispersion are approximately included in SAPT(DFT) \cite{Hesselmann:05,Misquitta:05a}, while
benchmark exchange dispersion values for small complexes can be computed at the CCSD level using the approach of Ref.~\onlinecite{Korona:09a}. It should be noted that the exchange dispersion
effects, just like nearly all other SAPT exchange corrections, are usually computed within the so-called {\em single exchange approximation} that neglects intermolecular swaps of more than 
one electron pair when applying the antisymmetry projector (this approach is also called the $S^2$ approximation as it neglects terms beyond the second power in the intermolecular overlap
integrals). While the full nonexpanded $E^{(20)}_{\rm exch-disp}$ expression has been recently derived and implemented \cite{Schaffer:13}, all beyond-$E^{(20)}_{\rm exch-disp}$ exchange
dispersion terms are always computed within the $S^2$ approximation.

As the explicitly correlated F12 approach has been quite successful improving the basis set convergence of supermolecular MP2 and CCSD(T) interaction energies
\cite{Marchetti:08,Marchetti:09,Patkowski:13,Sirianni:17}, it is worth asking if the benefits of the F12 treatment can be transferred to SAPT, in particular, to the slowly converging
dispersion and exchange dispersion energies. The calculation of both effects with the aid of explicitly correlated functions has quite a long history for model four-electron systems,
and the highly accurate CBS values of $E^{(20)}_{\rm disp}$, $E^{(21)}_{\rm disp}$, and $E^{(20)}_{\rm exch-disp}$ for the helium dimer have been established 
\cite{Szalewicz:79,Korona:97,Jeziorska:07} using Gaussian-type geminal (GTG) explicitly correlated bases \cite{Mitroy:13}. The first step towards a system-independent F12 extension of
SAPT was recently taken by Przybytek \cite{Przybytek:18}. He proposed an algorithm to improve the CBS convergence of $E^{(20)}_{\rm disp}$ using an {\em Ansatz} similar to
MP2-F12, where ordinary dispersion amplitudes are augmented by a set of F12 amplitudes approximately taking into account the excitations out of the molecular-orbital (MO) space. The 
F12 amplitudes in Przybytek's approach were obtained by a minimization of a suitable dispersion-only Hylleraas functional. Thus, the best variational approximation to the CBS value of
$E^{(20)}_{\rm disp}$ at a given basis set level was obtained, and the resulting CBS convergence improvement relative to conventional $E^{(20)}_{\rm disp}$ was truly impressive,
comparable to increasing the orbital basis set (of the augmented correlation-consistent aug-cc-pV$X$Z variety \cite{Kendall:92}) by two cardinal numbers. However, the approach 
of Ref.~\onlinecite{Przybytek:18} is not very practical as the complete amplitude optimization scales like the eighth power of the system size and sometimes suffers from numerical
instabilities. Therefore, our collaborators and we have recently proposed \cite{Kodrycka:19a} several approximate {\em Ans\"{a}tze} for $E^{(20)}_{\rm disp}$-F12 including the
optimized diagonal {\em Ansatz} (ODA, defined precisely in the next section) that scales like $N^5$ just like conventional $E^{(20)}_{\rm disp}$, is numerically stable, and produces
F12 dispersion energies nearly as accurate as the full F12 amplitude optimization. In the same work \cite{Kodrycka:19a}, we have employed the F12 dispersion amplitudes (computed
using any {\em Ansatz}) in the calculation of an explicitly correlated exchange dispersion energy, $E^{(20)}_{\rm exch-disp}$-F12 (within the $S^2$ approximation), substantially
improving the convergence of conventional $E^{(20)}_{\rm exch-disp}$ to its CBS limit.

The F12 extensions to $E^{(20)}_{\rm disp}$ and $E^{(20)}_{\rm exch-disp}$ proposed in Refs.~\onlinecite{Przybytek:18} and \onlinecite{Kodrycka:19a} significantly improve the
basis set convergence of these SAPT0 corrections. However, in the original implementations, they are not competitive with conventional $E^{(20)}_{\rm disp}$ and 
$E^{(20)}_{\rm exch-disp}$ calculations in larger basis sets. The reason for the suboptimal performance of these algorithms, besides the fact that only modest effort was put
into optimizing the initial proof-of-concept implementations, is the necessity to evaluate and transform multiple types of two-electron integrals over functions from the atomic-orbital (AO) basis set and
the complementary auxiliary basis set (CABS) approximately spanning the orthogonal complement to the AO space \cite{Valeev:04}. The union of the AO and CABS bases is meant to provide a reasonably
complete representation of the entire one-electron space so that all three- and four-electron integrals can be expressed by two-electron
integrals using resolution of identity (RI). 
A further increase in the efficiency of the F12 approaches has been possible thanks to expanding the two-electron integrals and
other four-index quantities in three-index auxiliary tensors using density fitting (DF) \cite{Whitten:73,Dunlap:79,Manby:03,May:04a}. Density-fitted electronic structure
theories benefit from their significantly lowered CPU time and storage requirements as the computation and transformation of two-electron integrals is avoided (even though
the formal scaling of the underlying approach is rarely reduced). It should be noted that while the terms RI and DF are often used interchangeably in other contexts, they
have separate and well-defined meanings within the F12 formalism. Both RI and DF have been the essential factors in turning MP2-F12 and approximate 
CCSD(T)-F12 \cite{Adler:07,Knizia:09,Hattig:10} into robust practically useful approaches that combine improved basis set convergence with only a modest increase in the
computational cost relative to the parent MP2 and CCSD(T) methods. On the contrary, our initial implementation of $E^{(20)}_{\rm disp}$-F12 and $E^{(20)}_{\rm exch-disp}$-F12
\cite{Kodrycka:19a} uses RI but not DF; therefore, while the CBS convergence of these SAPT terms is vastly improved, the computational complexity is substantially higher
than for the parent $E^{(20)}_{\rm disp}$ and $E^{(20)}_{\rm exch-disp}$ corrections due to the need to compute and transform several types of two-electron integrals. Moreover,
these integrals involve not just AO basis indices, but up to two indices running over the RI basis that is typically several times larger than the AO set. As a result, 
the $E^{(20)}_{\rm disp}$-F12 and $E^{(20)}_{\rm exch-disp}$-F12 implementations cannot be competitive with their conventional SAPT0 counterparts until all four-index
quantities present are decomposed by density fitting. This DF decomposition of the $E^{(20)}_{\rm disp}$-F12 and $E^{(20)}_{\rm exch-disp}$-F12 expressions, leading to
an efficient and practical computer implementation, is the topic of this manuscript.

Before we proceed with the development of the density fitted SAPT-F12 expressions, it is important to note, perhaps counterintuitively, that a stand-alone CBS convergence
improvement of the dispersion and exchange-dispersion terms is likely to make the total SAPT0 interaction energies worse, not better. Just like supermolecular MP2 tends
to overestimate the binding, especially in aromatic complexes, the MP2-level $E^{(20)}_{\rm disp}$ approximation tends to overestimate the magnitude of the true dispersion
energy. As a result, it is usually preferable to perform SAPT0 calculations far away from CBS, so that the overbinding of the SAPT0 dispersion partially cancels with
the underbinding due to an incomplete basis set. The ``calendar'' basis set jun-cc-pVDZ \cite{Papajak:11} has been observed to provide a particularly consistent error
cancellation and is the recommended choice for SAPT0 calculations \cite{Parker:14}. The replacement of small-basis $E^{(20)}_{\rm disp}$ and $E^{(20)}_{\rm exch-disp}$
values by their F12 counterparts (which are much closer to the CBS limit) disturbs the error cancellation and thus is likely to diminish the accuracy of SAPT0 interaction
energies. Instead, the true utility of the $E^{(20)}_{\rm disp}$-F12 and $E^{(20)}_{\rm exch-disp}$-F12 corrections is in higher-level SAPT calculations, where the
intramolecular correlation effects (in the form of $E^{(21)}_{\rm disp}+E^{(22)}_{\rm disp}$ or, even better, the coupled-cluster-level dispersion energy \cite{Korona:08})
prevent the overestimation of dispersion. While an F12 {\em Ansatz} for any intramolecular correlation terms in dispersion energy has not been proposed yet (except for 
$E^{(21)}_{\rm disp}$ in the specific case of interactions between two-electron systems \cite{Szalewicz:79,Korona:97,Jeziorska:07}), one can combine an F12-enhanced CBS
estimate of the leading-order $E^{(20)}_{\rm disp}$ term with an intramolecular correlation correction computed conventionally in a moderate basis set.
Such a ``composite'' treatment, in the spirit of the popular MP2/CBS+$\delta$CCSD(T) approach to supermolecular interaction energies (with a moderate-basis 
$\delta$CCSD(T)$=$CCSD(T)$-$MP2 correction added on top of the MP2 CBS limit \cite{Riley:10,Marshall:11}) or the many successful composite approaches to thermochemistry
\cite{Tajti:04,Karton:06}, is completely rigorous (no double counting occurs) and is likely to improve benchmark high-level SAPT dispersion energies and the
resulting total SAPT interaction energies. For the latter, intramolecular correlation corrections to dispersion are included in the SAPT2+, SAPT2+(3),
and SAPT2+3 levels of theory (as defined in Ref.~\onlinecite{Parker:14}), and these high-order SAPT variants are the primary target for improvement by our
F12 approach to $E^{(20)}_{\rm disp}$ and $E^{(20)}_{\rm exch-disp}$.

In this work, we derive and implement efficient expressions for the $E^{(20)}_{\rm disp}$-F12(ODA) and $E^{(20)}_{\rm exch-disp}$-F12(ODA) explicitly correlated
SAPT0 corrections employing \textit{robust} density fitting \cite{Manby:03, May:04a} in all two-electron quantities. The initial implementation of the new expressions
(just like for the non-DF formulas of Ref.~\onlinecite{Kodrycka:19a}) is facilitated by the {\sc Psi4NumPy} framework \cite{Smith:18} that combines the low-level
functionality (integrals, HF vectors, \ldots) of the {\sc Psi4} electronic structure program \cite{Parrish:17,Smith:20} with the tensor manipulation and linear algebra
capabilities of the {\sc NumPy} library. In the near future, the new DF-$E^{(20)}_{\rm disp}$-F12(ODA) and DF-$E^{(20)}_{\rm exch-disp}$-F12(ODA) functionality
will be implemented in the development version of the {\sc Psi4} code itself. The new DF formalism enables us to extend benchmark studies of the explicitly correlated
SAPT dispersion and exchange dispersion to substantially larger systems and basis sets, and we first exploit this capability by extending the SAPT-F12
calculations \cite{Przybytek:18,Kodrycka:19a} for the A24 database of noncovalent complexes \cite{Rezac:13a} to larger basis sets, including bases with midbond
functions. Then, we investigate the CBS convergence of $E^{(20)}_{\rm disp}$ and $E^{(20)}_{\rm exch-disp}$, and the effect of its F12 enhancement, on the total 
SAPT2+, SAPT2+(3), and SAPT2+3 interaction energies, on the same noncovalent databases as the SAPT benchmarking study of Ref.~\onlinecite{Parker:14}. Such an 
elimination of the leading basis set incompleteness error in high-level SAPT data is expected to provide a clearer picture of the directions for the
further improvement of the SAPT methodology, both for the dispersion energy and for other terms.

\section{Theory}

In Sections \ref{sec:Disp-F12} and \ref{sec:Exch-disp-F12} we summarize the main equations which define the non-DF $E^{(20)}_{\rm disp}$-F12(ODA) and $E^{(20)}_{\rm exch-disp}$-F12(ODA) energies, respectively. In Section \ref{sec:Fitting} the DF approximation is applied to obtain the corresponding expressions for DF-SAPT-F12. 

Throughout this paper, the index convention from Ref.~\onlinecite{Kodrycka:19a} is adopted. Thus, separate sets of indices are defined for monomers A and B, that is  $(i,k,m),a,r,x,\alpha$ and  $(j,l,n),b,s,y,\beta$, respectively. 
The range of each orbital index is as follows: indices $i,j,k,l,m,n$ run over occupied orbitals; $a,b$, virtual (unoccupied) orbitals; $r,s$, all molecular orbitals (MOs), both occupied and virtual; $x$,$y$, the complementary auxiliary (CA) functions approximately spanning the orthogonal complement of the MO space; $\alpha,\beta$, the formally complete orthonormal set, which is constructed as the union of the MO and CA subspaces; and $A,B$, the DF basis set. 
We will assume that a full dimer basis set is employed for both the orbital and CABS bases: as a result, the orbitals $r$ and $s$ span the same space and so do $x$ and $y$.
A summation over each repeated index is implied in all expressions. The index notation is  summarized in Table~\ref{orb_spaces} for easy reference.

\begin{table}[!ht]
\begin{center}
\caption{Orbital spaces used in this work.}
\begin{tabular}{ c| c| c } 
 \hline
 \hline
 Orbital space & Monomer A & Monomer B \\
 \hline
 Occupied orbitals& $i$, $k$, $m$ & $j$, $l$, $n$ \\ 
 Virtual orbitals & $a$          & $b$ \\ 
 Any molecular orbitals & $r$    & $s$ \\  
 Complementary auxiliary orbitals & $x$ & $y$ \\
 Complete orthonormal or RI basis & $\alpha$ & $\beta$ \\
 DF basis & \multicolumn{2}{c}{$A$, $B$} \\
 \hline
 \hline
\end{tabular}
\label{orb_spaces}
\end{center}
\end{table}

\subsection{$E^{(20)}_{\rm disp}$-F12(ODA)}\label{sec:Disp-F12}

The starting point for the development of the explicitly correlated dispersion energy is a ``dispersion-only" Hylleraas functional
\begin{eqnarray}\label{hyller0}
J_{\rm disp}[\chi]&=&\langle\chi|{\cal Q}_{\rm A}{\cal Q}_{\rm B}
(F^{\rm A}+F^{\rm B}-E^0_{\rm A}-E^0_{\rm B}){\cal Q}_{\rm A}{\cal Q}_{\rm B}
|\chi\rangle
\nonumber \\ &&
+\langle\chi|{\cal Q}_{\rm A}{\cal Q}_{\rm B}(V-\langle V\rangle)|
\phi^{\rm A}_{\rm HF}\phi^{\rm B}_{\rm HF}\rangle+
\langle\phi^{\rm A}_{\rm HF}\phi^{\rm B}_{\rm HF}|(V-\langle V\rangle)
{\cal Q}_{\rm A}{\cal Q}_{\rm B}|\chi\rangle
\nonumber \\ &&
\end{eqnarray}
where $F^{X}$ is the Fock operator, $E^{0}_X$ is the sum of the occupied orbital energies $\epsilon^{\rm X}_{i(j)}$,  $\langle V\rangle\equiv\langle \phi^{\rm A}_{\rm HF}\phi^{\rm B}_{\rm HF}|
V|\phi^{\rm A}_{\rm HF}\phi^{\rm B}_{\rm HF}\rangle$ is the first-order
electrostatic correction $E^{(10)}_{\rm elst}$, and ${\cal Q}_{\rm X}$  is the operator projecting out the ground state for a given monomer $X \in \{A,B\}$.
It has been shown \cite{Przybytek:18, Kodrycka:19a} that $E^{(20)}_{\rm disp}$-F12 is obtained variationally by minimizing the above functional using a trial function 
\begin{equation}\label{trialf12}
\chi=T^{ij}_{ab}|\Phi^{ab}_{ij}\rangle+T^{ij}_{kl}{\cal F}^{kl}_{\alpha\beta}
|\Phi^{\alpha\beta}_{ij}\rangle
\end{equation}
where $|\Phi^{\alpha\beta}_{ij}\rangle=\hat{E}_{\alpha i}\hat{E}_{\beta j}
|\phi^{\rm A}_{\rm HF}\phi^{\rm B}_{\rm HF}\rangle$ is a doubly excited
(once on A, once on B) determinant. The target of F12 methods is to include the full space of doubly excited configurations, which is achieved via a set of amplitudes $T^{ij}_{kl}$ and a
suitable internal contraction
\begin{equation}
{\cal F}^{kl}_{\alpha\beta}=\langle kl|\hat{F}_{12}\hat{Q}_{12}|\alpha\beta
\rangle
\end{equation}
where $\hat{F}_{12}$ is a correlation factor and $\hat{Q}_{12}$ is the strong-orthogonality projector. A correlation factor plays a central role in the F12 methods since it introduces the explicit dependence of the function on the interelectronic distance $r_{12}$. Here, the standard exponential expression is assumed \cite{Hattig:12}
\begin{equation}
\hat{F}_{12}\equiv F(r_{12}) = \exp(-\gamma r_{12})
\end{equation}
where $\gamma$ stands for a length-scale parameter. 
The operator $\hat{Q}_{12}$ is chosen as {\em Ansatz} 3 of Ref. \onlinecite{Werner:07},
\begin{equation}
\hat{Q}_{12}=\left(1_1-\sum_i |i\rangle\langle i|_1\right)
\left(1_2-\sum_j |j\rangle\langle j|_2\right)
\left(1_{12}-\sum_{ab} |ab\rangle\langle ab|_{12}\right)
\end{equation}
with the subscripts indicating which electron coordinates are affected
by a given part of the projector. 

The final expression for the Hylleraas dispersion functional, valid for arbitrary amplitudes
$T_{ij}^{ab}$ and $T_{ij}^{kl}$, reads
\begin{align}\label{hyllf123}
J_{\rm disp}[\chi]=&\phantom{+}
4T_{ij}^{ab}T^{ij}_{ab}(\epsilon^{\rm A}_a+\epsilon^{\rm B}_b-\epsilon^{\rm A}_i-\epsilon^{\rm B}_j)
+8T^{ij}_{ab}K^{ab}_{ij}
\nonumber \\ &
+4T_{ij}^{kl}T^{ij}_{mn}B_{kl,mn}
-4\left(\epsilon^{\rm A}_{i}+\epsilon^{\rm B}_{j}\right)T_{ij}^{kl}T^{ij}_{mn}X_{kl,mn}
+8T_{ij}^{kl}V^{ij}_{kl}
\nonumber \\ &
+8T_{ij}^{ab}T^{ij}_{kl}C^{kl}_{ab}
\end{align}
where a handful of intermediates  were introduced analogously to the MP2-F12(3C) approach \cite{Werner:07}
\begin{eqnarray}
K^{ab}_{ij} &=&\langle ab | r^{-1}_{12} | ij \rangle \label{Kabij} \\
V^{ij}_{kl} &=& \langle ij | r^{-1}_{12} \widehat{Q}_{12} \widehat{F}_{12}| kl \rangle \label{Vijkl}\\
B_{kl,mn} &=& \langle kl | \widehat{F}_{12} \widehat{Q}_{12} (\widehat{f}_{A1} + \widehat{f}_{B2}) \widehat{Q}_{12}\widehat{F}_{12} | mn \rangle \label{Bijkl}\\
X_{kl,mn} &=& \langle kl | \widehat{F}_{12} \widehat{Q}_{12}\widehat{F}_{12} |mn \rangle \label{Xijkl}\\
C^{kl}_{ab} &=& \langle kl | \widehat{F}_{12} \widehat{Q}_{12}(\widehat{f}_{A1} + \widehat{f}_{B2})|ab \rangle \label{Cklab}
\end{eqnarray} 
Note that in the above equations, the Fock operators $\widehat{f}_1$ and $\widehat{f}_2$ for the first and second electron, respectively, pertain to different molecules. 
The presence of the projector operator  $\hat{Q}_{12}$ in Eqs.~(\ref{Vijkl})--(\ref{Cklab}) leads to three- and four-electron integrals; thus, RI is employed to factorize these terms into products of two-electron integrals.
In addition, we assumed the generalized Brillouin condition (GBC) and the so-called approximation 3C \cite{Werner:07} to evaluate the $C$ and $B$ matrix elements, containing products of the correlation factor $\hat{F}_{12}$ and the Fock operators.  

The conventional $(T_{ij}^{ab})$ and explicitly correlated $(T_{ij}^{kl})$ dispersion amplitudes are determined by minimizing Eq.~(\ref{hyllf123}). 
Since the full minimization comes at a cost of  $N^{8}$, too expensive to be practical, some approximations were proposed and tested in Ref. \onlinecite{Kodrycka:19a}.
In this work, one of these approximations, the Optimized Diagonal {\em Ansatz} (ODA) \cite{Kodrycka:19a} is applied, which assumes that the $T_{ij}^{kl}$ amplitudes are diagonal, that is, 
\begin{equation}
T_{ij}^{kl}= T^{ij}_{ij }\delta_{ik}\delta_{jl}.
\end{equation} 
Within the ODA approximation, the dispersion amplitudes are computed from the following equations \cite{Kodrycka:19a} (with the summations written explicitly this time)
\begin{equation}\label{oda11}
T_{ij}^{ab}
=\frac{K^{ab}_{ij}+T^{ij}_{ij}C^{ij}_{ab}}{\epsilon^{\rm A}_i+\epsilon^{\rm B}_j-\epsilon^{\rm A}_a-\epsilon^{\rm B}_b}
\end{equation}
\begin{equation}\label{oda2}
T_{ij}^{ij}\left[B_{ij,ij}
-\left(\epsilon^{\rm A}_{i}+\epsilon^{\rm B}_{j}\right)X_{ij,ij}
+\sum_{ab}
\frac{C^{ij}_{ab}C^{ij}_{ab}}{\epsilon^{\rm A}_i+\epsilon^{\rm B}_j-\epsilon^{\rm A}_a-\epsilon^{\rm B}_b}\right]
=
-V^{ij}_{ij}
-\sum_{ab}
\frac{K^{ab}_{ij}C^{ij}_{ab}}{\epsilon^{\rm A}_i+\epsilon^{\rm B}_j-\epsilon^{\rm A}_a-\epsilon^{\rm B}_b}
\end{equation}
The ODA formalism in $E^{(20)}_{\rm disp}$-F12 is closely related to the original approach to explicitly
correlated MP2 \cite{Kutzelnigg:91,Tew:06} which also retained only diagonal F12 amplitudes. 
Such a selection is not invariant with 
respect to a unitary transformation of occupied orbitals. Thus, our formalism is restricted to canonical 
HF orbitals on each monomer; however, the same restriction holds for conventional $E^{(20)}_{\rm disp}$ 
and $E^{(20)}_{\rm exch-disp}$ as well, so this is not a practical problem. Since the occupied orbitals 
in SAPT pertain to one monomer at a time, our ODA approach most closely relates to supermolecular MP2-F12 
with an orbital-variant Ansatz using localized orbitals, which demonstrated good performance, a lack of 
geminal basis set superposition error, and a correct long-range behavior in the early study of weak MP2-F12 
interaction energies \cite{Tew:06}. In supermolecular F12 calculations, orbital invariance is commonly 
assured by means of a diagonal Ansatz with the explicitly correlated amplitudes fixed by cusp 
conditions \cite{Tenno:04}. We have previously tested a similar fixed-amplitude Ansatz (with an 
optimized common amplitude) in $E^{(20)}_{\rm disp}$-F12 and $E^{(20)}_{\rm exch-disp}$-F12
\cite{Kodrycka:19a}; however, the resulting accuracy, while clearly superior to conventional $E^{(20)}_{\rm disp}$ and $E^{(20)}_{\rm exch-disp}$, did not match the accuracy of ODA. As both Ans\"{a}tze exhibit virtually the same computational cost, only the more accurate ODA variant was used in the current work.

\subsection{$E^{(20)}_{\rm exch-disp}$-F12(ODA)}\label{sec:Exch-disp-F12}

The second-order exchange dispersion correction in the F12 formalism decomposes into
two terms. The first one is evaluated using the conventional exchange-dispersion formula \cite{Rybak:91} involving the amplitudes $T^{ij}_{ab}$ (which, in this case, are obtained from the ODA expression (\ref{oda11})).
The second term is an explicitly correlated correction involving the $T^{ij}_{kl}$ amplitudes 
\cite{Kodrycka:19a}:
\begin{align}
\delta E^{(20)}_{\rm exch-disp}\mbox{-F12} = & 2T^{ij}_{kl} \left[
 F^{kl}_{xb} K^{xb'}_{i'j}S^{b}_{i}S^{i'}_{b'}
-2F^{kl}_{xb} K^{xb'}_{ij}S^{b}_{i'}S^{i'}_{b'}
+F^{kl}_{ay} K^{a'y}_{ij'}S^{a}_{j}S^{j'}_{a'}-2F^{kl}_{ay} K^{a'y}_{ij}S^{a}_{j'}S^{j'}_{a'}
\right.
\nonumber \\ &
-2F^{kl}_{xb} (\omega_B)^{x}_{i}S^{b}_{i'}S^{i'}_{j}
+F^{kl}_{xb} (\omega_B)^{x}_{i'}S^{b}_{i}S^{i'}_{j}-F^{kl}_{ay} (\omega_B)^{y}_{i}S^{a}_{j}
\nonumber \\ &
-2F^{kl}_{ay} (\omega_A)^{y}_{j}S^{a}_{j'}S^{j'}_{i}
+F^{kl}_{ay} (\omega_A)^{y}_{j'}S^{a}_{j}S^{j'}_{i}-F^{kl}_{xb} (\omega_A)^{x}_{j}S^{b}_{i}
\nonumber \\ &
-(K^F)^{kl}_{i'j'} S^{j'}_{i}S^{i'}_{j}
+2(K^F)^{kl}_{ij'} S^{j'}_{i'}S^{i'}_{j}
+2(K^F)^{kl}_{i'j} S^{j'}_{i}S^{i'}_{j'}
-(K^F)^{lk}_{ij}
\nonumber \\ &
+F^{lk}_{i'j'}K^{i'j'}_{ij}
+F^{lk}_{aj'}K^{aj'}_{ij}
+F^{lk}_{i'b}K^{i'b}_{ij}
+F^{lk}_{xj'}K^{xj'}_{ij}
+F^{lk}_{i'y}K^{i'y}_{ij}
\nonumber \\ & 
+F^{kl}_{xn} K^{ax}_{ij}S^{n}_{a}
+F^{kl}_{my} K^{yb}_{ij}S^{m}_{b}
+F^{kl}_{rs} K^{ab}_{ij}S^{s}_{a}S^{r}_{b}
\nonumber \\ &
+F^{kl}_{xn} K^{xn}_{i'j'}S^{j'}_{i}S^{i'}_{j}
+F^{kl}_{my} K^{my}_{i'j'}S^{j'}_{i}S^{i'}_{j}
+F^{kl}_{rs} K^{rs}_{i'j'}S^{j'}_{i}S^{i'}_{j}
\nonumber \\ &
-2 F^{kl}_{xn} K^{xn}_{ij'}S^{j'}_{i'}S^{i'}_{j}
-2 F^{kl}_{my} K^{my}_{ij'}S^{j'}_{i'}S^{i'}_{j}
-2 F^{kl}_{rs} K^{rs}_{ij'}S^{j'}_{i'}S^{i'}_{j}
\nonumber \\ &
\left.
-2 F^{kl}_{xn} K^{xn}_{i'j}S^{j'}_{i}S^{i'}_{j'} 
-2 F^{kl}_{my} K^{my}_{i'j}S^{j'}_{i}S^{i'}_{j'} 
-2 F^{kl}_{rs} K^{rs}_{i'j}S^{j'}_{i}S^{i'}_{j'} 
\right]
\label{finalform}
\end{align}
where $F^{kl}_{\alpha\beta}=\langle kl|\hat{F}_{12}|\alpha\beta\rangle$ and 
$(K^F)^{kl}_{\alpha\beta}=\langle kl|\hat{F}_{12}r_{12}^{-1}|\alpha\beta\rangle$
are two common matrix elements involving the correlation factor, 
$S^{i}_{j}=\langle i|j\rangle$ is the overlap integral and $(\omega_B)^{a}_{i}$ is the matrix element of the electrostatic
potential of monomer B, that is,
\begin{equation}\label{omegaB}
(\omega_B)^{a}_{i}=\langle a|v_B|i\rangle + 2K^{aj}_{ij},
\end{equation}
and $v_B$ is the nuclear potential of molecule B (the
$(\omega_A)^{b}_{j}$ matrix elements are defined similarly). All integral types in Eq.~(\ref{finalform}) already appear in the calculation of intermediates for the $ E^{(20)}_{\rm disp}$-F12 energy, however, the integrals containing $\hat{F}_{12}$ and $\hat{F}_{12}r_{12}^{-1}$ can now be of the exchange type: $F^{\bm{BA}}_{\bm{AB}}$. 
Moreover, the $T_{ij}^{ab}$ and $T_{ij}^{kl}$ amplitudes are obtained from the preceding $E^{(20)}_{\rm disp}$-F12 calculations (Eqs.~(\ref{oda11}) and ~(\ref{oda2})); thus, in the ODA approach employed in this work, the $T_{ij}^{kl}$ tensor is diagonal.

\subsection{Robust density-fitted formulas for $E^{(20)}_{\rm disp}$-F12 and $E^{(20)}_{\rm exch-disp}$-F12}\label{sec:Fitting}

We now employ density fitting (DF) \cite{Manby:03,May:04a} to decompose the 4-index integrals into contractions of 3- and 2-index quantities, thereby reducing the computational cost as well as storage requirements. The main idea of DF is replacing the one-particle orbital product densities $|pq) $ by approximated densities $|\widetilde{pq})$. The latter are expanded in a set of auxiliary functions (denoted as $A$) as follows
\begin{equation}
|pq)\approx |\widetilde{pq})= D_{pq}^{A} \vert A)
\end{equation}
The expansion coefficients $D_{pq}^{A}$ can be obtained by minimizing the difference between the actual and fitted product densities  \cite{Whitten:73, Dunlap:79}
\begin{equation}\label{residual}
\Delta^{w}_{pq} = (pq - \widetilde{pq}|\widehat{w}_{12}|pq - \widetilde{pq}),
\end{equation}
 with a suitable positive definite $\widehat{w}_{12}$ metric. 
It has been demonstrated \cite{Manby:03} that the Coulomb operator $r^{-1}_{12}$ is the most convenient metric for fitting not only two-electron integrals, but also other integrals of the F12 theory, hence it will be used herein. Once the Coulomb metric is applied, the fitting coefficients are given by
\begin{equation}\label{eq:Dpqa}
D_{pq}^A  = [{\rm \textbf{J}^{-1}}]_{AB} (B| r^{-1}_{12}|pq)
\end{equation}
where $[{\rm \textbf{J}^{-1}}]$ is the inverse of the two-center electron repulsion integral (ERI) matrix
\begin{equation}
[\textbf{J}]_{AB}=\int A({\mathbf r}_1) \frac{1}{r_{12}} B({\mathbf r}_2)\, d{\mathbf r}_1 d{\mathbf r}_2
= (A| r^{-1}_{12}|B)
\end{equation}
 and $(B| r^{-1}_{12}|pq)$ are the three-center ERIs.
A crucial goal of density fitting is avoiding errors in integrals that are linear in the density errors. 
This is achieved by the so-called \textit{robust} fit proposed by Dunlap \cite{Dunlap:79,Manby:03}:
\begin{equation}\label{robust}
(pq|\widehat{v}_{12}|rs)_{robust} \approx (\widetilde{pq}|\widehat{v}_{12}|rs)+(pq|\widehat{v}_{12}|\widetilde{rs})-(\widetilde{pq}|\widehat{v}_{12}|\widetilde{rs})
\end{equation}
where $\widehat{v}_{12}$ is any particular kernel operator for an integral.
Equation~(\ref{robust}) allows us to obtain a fitting error in the integrals that is quadratic with respect to the fitting error in the densities, regardless of whether $\widehat{v}_{12}$ and $\widehat{w}_{12}$ 
are the same or different \cite{Manby:03,May:04a}.
When applying the DF algorithm to ordinary ERIs, they become approximated as
\begin{equation}
(pq|r^{-1}_{12}|rs)\approx(\widetilde{pq}|r^{-1}_{12}|\widetilde{rs}) = J_{pq}^{A}D_{rs}^{A}
\end{equation}
with the $D$ intermediate defined in Eq.~(\ref{eq:Dpqa}), and the $J$ intermediate computed as
\begin{equation}
J^{A}_{pq} = (A|r^{-1}_{12}|pq)  
\end{equation}
In the DF-SAPT-F12 theory, in addition to ERIs, we need to fit four types of  explicitly correlated two-electron integrals with the operators $\widehat{F}_{12}$, $\widehat{F}^{2}_{12}$, $\widehat{F}_{12}r^{-1}_{12}$, and the double commutator between the kinetic energy ($\widehat{t_1}+\widehat{t_2}$) and the correlation factor, $[[\widehat{F}_{12},\widehat{t_1}+\widehat{t_2}],\widehat{F}_{12}]$ \cite{Hoefener:08}. To do so, the explicitly robust formulas (Eq.~(\ref{robust})) combined with the Coulomb metric are employed. 
This leads to several additional intermediates that we need to form before we evaluate $E^{(20)}_{\rm disp}$-F12(ODA) (Eqs.~(\ref{Kabij})--(\ref{Cklab})) as well as the $\delta E^{(20)}_{\rm exch-disp}$-F12(ODA) correction to the second-order exchange dispersion energy (Eq.~(\ref{finalform})) using the DF approximation:

\begin{eqnarray}
J_{AB} &=& (A|r^{-1}_{12}|B) \\
F_{AB} &=& (A|\widehat{F}_{12}|B) \\
K^{F}_{AB} &=& (A|\widehat{F}_{12}r^{-1}_{12}|B) \\
F^{2}_{AB} &=& (A|\widehat{F}^{2}_{12}|B) \\
U^{F}_{AB} &=& (A|[[\widehat{F}_{12},\widehat{t_1}+\widehat{t_2}],\widehat{F}_{12}]|B)\\
F^{A}_{ij} &=& (A|\widehat{F}_{12}|ij)  \\
K^{F}_{A,ij} &=& (A|\widehat{F}_{12}r^{-1}_{12}|ij) \\
F^{2}_{A,ij} &=& (A|\widehat{F}^{2}_{12}|ij) \\
U^{F}_{A,ij} &=& (A|[[\widehat{F}_{12},\widehat{t_1}+\widehat{t_2}],\widehat{F}_{12}]|ij) 
\end{eqnarray}
These definitions allow us to rewrite the $V$, $X$, $B$ and $C$ matrix elements, Eqs.~(\ref{Vijkl})--(\ref{Cklab}), explicitly. 
The resulting formulas contain many instances of the $F^{\alpha\beta}_{\alpha'\beta'}$ matrix element for different types of indices $\alpha,\alpha',\beta,\beta'$. To keep the expressions reasonably compact, we will list this intermediate in its non-DF, four-index form. In the actual evaluation of these expressions, the robust density-fitted form of $F^{\alpha\beta}_{\alpha'\beta'}$ is always used:
\begin{equation}\label{df_of_F}
F^{\alpha\beta}_{\alpha'\beta'}=
D^{A}_{\alpha\alpha'}F^{A}_{\beta\beta'}+F^{A}_{\alpha\alpha'}D^{A}_{\beta\beta'}
-D^{A}_{\alpha\alpha'}F_{AB}D^{B}_{\beta\beta'}
\end{equation} 
Note that, for optimal scaling and computational efficiency, this evaluation requires treating the
three terms resulting from Eq.~(\ref{df_of_F}) separately, and optimizing the order of contractions between the individual two- and three-index tensors, both those forming a part of $F^{\alpha\beta}_{\alpha'\beta'}$ and those
arising from other matrices. The resulting expressions are
\begin{equation}\label{dfVijkl}
V^{ij}_{kl} = D^{A}_{ik}K^{F}_{A,jl}+K^{F}_{A,ik}D^{A}_{jl} - D^{A}_{ik}K^{F}_{AB}D^{B}_{jl} 
 - J^{A}_{ir}D^{A}_{js}F^{kl}_{rs} 
 - J^{A}_{ix}D^{A}_{jn}F^{kl}_{xn} 
 - J^{A}_{im}D^{A}_{jy}F^{kl}_{my}
\end{equation}
\begin{equation}\label{dfXklmn}
X^{kl}_{mn} = D^{A}_{km}F^{2}_{A,ln}+F^{2}_{A,km}D^{A}_{ln}-D^{A}_{km}F^{2}_{AB}D^{B}_{ln}
 - F^{kl}_{rs}F^{rs}_{mn} 
 - F^{kl}_{xj}F^{xj}_{mn} 
 - F^{kl}_{iy}F^{iy}_{mn} 
\end{equation}
\begin{equation}\label{dfCklab}
C^{kl}_{ab} = f_{ax}^{\bm{A}}F^{kl}_{xb}+ F^{kl}_{ay}f_{yb}^{\bm{B}} 
\end{equation}
\begin{equation}\label{BtoA}
B_{kl,mn}=\frac{1}{2}(A_{kl,mn}+A_{mn,kl})
\end{equation}

\begin{eqnarray}\label{dfBklmn}
A_{kl,mn} &=& D^{A}_{km}U^{F}_{A,ln} +U^{F}_{A,km}D^{A}_{ln}-D^{A}_{km}U^{F}_{AB}D^{B}_{ln}
 \nonumber \\
&+& (D^{A}_{xm}(f_{xk}^{\bm{A}}+k_{xk}^{\bm{A}}) + D^{A}_{rm}(f_{rk}^{\bm{A}}+k_{rk}^{\bm{A}}))F^{2}_{A,ln} 
 +  (F^{2}_{A,xm}(f_{xk}^{\bm{A}}+k_{xk}^{\bm{A}}) + F^{2}_{A,rm}(f_{rk}^{\bm{A}}+k_{rk}^{\bm{A}}))D^{A}_{ln} \nonumber \\ 
&-& (D^{A}_{xm}(f_{xk}^{\bm{A}}+k_{xk}^{\bm{A}}) + D^{A}_{rm}(f_{rk}^{\bm{A}}+k_{rk}^{\bm{A}})) F^{2}_{AB}D^{B}_{ln} \nonumber \\
&+& D^{A}_{km}(F^{2}_{A,yn}(f_{yl}^{\bm{B}}+k_{yl}^{\bm{B}})+F^{2}_{A,sn}(f_{sl}^{\bm{B}}+k_{sl}^{\bm{B}})) 
 +  F^{2}_{A,km}(D^{A}_{yn}(f_{yl}^{\bm{B}}+k_{yl}^{\bm{B}}) + D^{A}_{sn}(f_{sl}^{\bm{B}}+k_{sl}^{\bm{B}})) \nonumber \\
&-& D^{A}_{km}F^{2}_{AB}(D^{B}_{yn}(f_{yl}^{\bm{B}}+f_{yl}^{\bm{B}}) + D^{B}_{sn}(f_{sl}^{\bm{B}}+k_{sl}^{\bm{B}})) \nonumber \\
&-& (D^{A}_{kr}k_{rx}^{\bm{A}}+D^{A}_{kx^{'}}k_{x^{'}x}^{\bm{A}})F^{A}_{lb}F^{mn}_{xb} 
 -  (F^{A}_{kr}k_{rx}^{\bm{A}}+F^{A}_{kx^{'}}k_{x^{'}x}^{\bm{A}}) D^{A}_{lb}F^{mn}_{xb} 
 +  (D^{A}_{kr}k_{rx}^{\bm{A}}+D^{A}_{kx^{'}}k_{x^{'}x}^{\bm{A}})F_{AB}D^{B}_{lb}F^{mn}_{xb} \nonumber \\
&-& D^{A}_{kx}(F^{A}_{ls}k_{sb}^{\bm{B}}+F^{A}_{ly}k_{yb}^{\bm{B}})F^{mn}_{xb} 
 -  F^{A}_{kx}(D^{A}_{ls}k_{sb}^{\bm{B}}+D^{A}_{ly}k_{yb}^{\bm{B}})F^{mn}_{xb} 
 +  D^{A}_{kx}F_{AB} (D^{B}_{ls}k_{sb}^{\bm{B}}+D^{B}_{ly}k_{yb}^{\bm{B}})F^{mn}_{xb} \nonumber \\
&-& (D^{A}_{kr}k_{ra}^{\bm{A}}+D^{A}_{kx}k_{xa}^{\bm{A}})F^{A}_{ly}F^{mn}_{ay} 
 -    (F^{A}_{kr}k_{ra}^{\bm{A}}+F^{A}_{kx}k_{xa}^{\bm{A}})D^{A}_{ly}F^{mn}_{ay} 
 +    (D^{A}_{kr}k_{ra}^{\bm{A}}+D^{A}_{kx}k_{xa}^{\bm{A}})F_{AB}D^{B}_{ly}F^{mn}_{ay} \nonumber \\
&-& D^{A}_{ka}(F^{A}_{ls}k_{sy}^{\bm{B}}+F^{A}_{ly^{'}}k_{y^{'}y}^{\bm{B}})F^{mn}_{ay} 
 -    F^{A}_{ka}(D^{A}_{ls}k_{sy}^{\bm{B}}+D^{A}_{ly^{'}}k_{y^{'}y}^{\bm{B}})F^{mn}_{ay} 
 +    D^{A}_{ka}F_{AB}(D^{B}_{ls}k_{sy}^{\bm{B}}+D^{B}_{ly^{'}}k_{y^{'}y}^{\bm{B}})F^{mn}_{ay} \nonumber \\
&-& (D^{A}_{kr}k_{rx}^{\bm{A}}+D^{A}_{kx^{'}}k_{x^{'}x}^{\bm{A}})F^{A}_{ly}F^{mn}_{xy} 
 -  (F^{A}_{kr}k_{rx}^{\bm{A}}+F^{A}_{kx^{'}}k_{x^{'}x}^{\bm{A}})D^{A}_{ly}F^{mn}_{xy}  
 +  (D^{A}_{kr}k_{rx}^{\bm{A}}+D^{A}_{kx^{'}}k_{x^{'}x}^{\bm{A}})F_{AB}D^{B}_{ly}F^{mn}_{xy} \nonumber \\     
&-& D^{A}_{kx}(F^{A}_{ls}k_{sy}^{\bm{B}}+F^{A}_{ly^{'}}k_{y^{'}y}^{\bm{B}})F^{mn}_{xy} 
 -  F^{A}_{kx}(D^{A}_{ls}k_{sy}^{\bm{B}}+D^{A}_{ly^{'}}k_{y^{'}y}^{\bm{B}})F^{mn}_{xy} 
 +  D^{A}_{kx}F_{AB}(D^{B}_{ls}k_{sy}^{\bm{B}}+D^{B}_{ly^{'}}k_{y^{'}y}^{\bm{B}})F^{mn}_{xy} \nonumber \\
&-& ( D^{A}_{kr^{'}}(f_{r^{'}r}^{\bm{A}}+k_{r^{'}r}^{\bm{A}})+D^{A}_{kx}(f_{xr}^{\bm{A}}+k_{xr}^{\bm{A}})) F^{A}_{ls}F^{mn}_{rs} 
 -  ( F^{A}_{kr^{'}}(f_{r^{'}r}^{\bm{A}}+k_{r^{'}r}^{\bm{A}})+F^{A}_{kx}(f_{xr}^{\bm{A}}+k_{xr}^{\bm{A}})) D^{A}_{ls}F^{mn}_{rs} \nonumber \\
&+& ( D^{A}_{kr^{'}}(f_{r^{'}r}^{\bm{A}}+k_{r^{'}r}^{\bm{A}})+D^{A}_{kx}(f_{xr}^{\bm{A}}+k_{xr}^{\bm{A}})) F_{AB}D^{B}_{ls}F^{mn}_{rs} \nonumber \\
&-& ( D^{A}_{kr}(f_{rx}^{\bm{A}}+k_{rx}^{\bm{A}})+D^{A}_{kx^{'}}(f_{x^{'}x}^{\bm{A}}+k_{x^{'}x}^{\bm{A}})) F^{A}_{lj}F^{mn}_{xj} 
 -  (F^{A}_{kr}(f_{rx}^{\bm{A}}+k_{rx}^{\bm{A}})+F^{A}_{kx^{'}}(f_{x^{'}x}^{\bm{A}}+k_{x^{'}x}^{\bm{A}})) D^{A}_{lj}F^{mn}_{xj} \nonumber \\
&+& (D^{A}_{kr}(f_{rx}^{\bm{A}}+k_{rx}^{\bm{A}})+D^{A}_{kx^{'}}(f_{x^{'}x}^{\bm{A}}+k_{x^{'}x}^{\bm{A}})) F_{AB}D^{B}_{lj}F^{mn}_{xj} \nonumber \\
&-& (D^{A}_{kr}(f_{ri}^{\bm{A}}+k_{ri}^{\bm{A}})+D^{A}_{kx}(f_{xi}^{\bm{A}}+k_{xi}^{\bm{A}})) F^{A}_{ly}F^{mn}_{iy} 
 -  (F^{A}_{kr}(f_{ri}^{\bm{A}}+k_{ri}^{\bm{A}})+F^{A}_{kx}(f_{xi}^{\bm{A}}+k_{xi}^{\bm{A}}))D^{A}_{ly} F^{mn}_{iy} \nonumber \\
&+& (D^{A}_{kr}(f_{ri}^{\bm{A}}+k_{ri}^{\bm{A}})+D^{A}_{kx}(f_{xi}^{\bm{A}}+k_{xi}^{\bm{A}})) F_{AB}D^{B}_{ly}F^{mn}_{iy} \nonumber \\
&-& D^{A}_{kr}(F^{A}_{ls^{'}}(f_{s{'}s}^{\bm{B}}+k_{s{'}s}^{\bm{B}}) + F^{A}_{ly}(f_{ys}^{\bm{B}}+k_{ys}^{\bm{B}})) F^{mn}_{rs} 
 -  F^{A}_{kr}(D^{A}_{ls^{'}}(f_{s{'}s}^{\bm{B}}+k_{s{'}s}^{\bm{B}}) + D^{A}_{ly}(f_{ys}^{\bm{B}}+k_{ys}^{\bm{B}})) F^{mn}_{rs} \nonumber \\
&+& D^{A}_{kr}F_{AB}(D^{B}_{ls^{'}}(f_{s{'}s}^{\bm{B}}+k_{s{'}s}^{\bm{B}}) + D^{B}_{ly}(f_{ys}^{\bm{B}}+k_{ys}^{\bm{B}})) F^{mn}_{rs} \nonumber \\
&-& D^{A}_{kx}(F^{A}_{ls}(f_{sj}^{\bm{B}}+k_{sj}^{\bm{B}})+F^{A}_{ly}(f_{yj}^{\bm{B}}+k_{yj}^{\bm{B}})) F^{mn}_{xj} 
 -  F^{A}_{kx}(D^{A}_{ls}(f_{sj}^{\bm{B}}+k_{sj}^{\bm{B}})+D^{A}_{ly}(f_{yj}^{\bm{B}}+k_{yj}^{\bm{B}})) F^{mn}_{xj} \nonumber \\
&+& D^{A}_{kx}F_{AB}(D^{B}_{ls}(f_{sj}^{\bm{B}}+k_{sj}^{\bm{B}})+D^{B}_{ly}(f_{yj}^{\bm{B}}+k_{yj}^{\bm{B}})) F^{mn}_{xj}\nonumber \\
&-& D^{A}_{ki}(F^{A}_{ls}(f_{sy}^{\bm{B}}+k_{sy}^{\bm{B}})+F^{A}_{ly^{'}}(f_{y^{'}y}^{\bm{B}}+k_{y^{'}y}^{\bm{B}})) F^{mn}_{iy} 
 -  F^{A}_{ki}(D^{A}_{ls}(f_{sy}^{\bm{B}}+k_{sy}^{\bm{B}})+D^{A}_{ly^{'}}(f_{y^{'}y}^{\bm{B}}+k_{y^{'}y}^{\bm{B}})) F^{mn}_{iy} \nonumber \\
&+& D^{A}_{ki}F_{AB}(D^{B}_{ls}(f_{sy}^{\bm{B}}+k_{sy}^{\bm{B}})+D^{B}_{ly^{'}}(f_{y^{'}y}^{\bm{B}}+k_{y^{'}y}^{\bm{B}})) F^{mn}_{iy} \nonumber \\
&-& F^{kl}_{ab}f_{ax}^{\bm{A}}F^{xb}_{mn} - F^{kl}_{ab}F^{ay}_{mn}f_{yb}^{\bm{B}}
\end{eqnarray}
The $f^{\rm \textbf{X}}_{kx}$ and $k^{\rm \textbf{X}}_{kx}$ matrices in Eqs.~(\ref{dfCklab}) and (\ref{dfBklmn}) denote the matrix elements of the Fock and exchange operators for monomer \textbf{X}, respectively. 
Note that the explicit symmetrization of Eq.~(\ref{BtoA}) is not required in the ODA case as only the 
diagonal elements $B_{kl,kl}=A_{kl,kl}$ are needed.

Analogously, the density-fitted explicitly correlated correction to the second-order exchange-dispersion energy, $\delta E^{(20)}_{\rm exch-disp}$-F12 (Eq.~(\ref{finalform})), is computed as

\begin{eqnarray}\label{delta_dfexchf12}
\delta E^{(20)}_{\rm exch-disp}\textit{-}\rm F12 &=& 2T^{ij}_{kl}\left[ F^{kl}_{xb}
J^{A}_{xi^{'}}D^{A}_{b^{'}j}S^{b}_{i}S^{i^{'}}_{b^{'}} \right.
 -  2F^{kl}_{xb}
J^{A}_{xi}D^{A}_{b^{'}j}S^{b}_{i^{'}}S^{i^{'}}_{b^{'}} \nonumber \\
&+& F^{kl}_{ay}
J^{A}_{a^{'}i}D^{A}_{yj^{'}}S^{a}_{j}S^{j^{'}}_{a^{'}}
 - 2F^{kl}_{ay}
J^{A}_{a^{'}i}D^{A}_{yj}S^{a}_{j^{'}}S^{j^{'}}_{a^{'}}\nonumber \\
&-& 2F^{kl}_{xb}
((V_{B})^{x}_{i} + 2J^{A}_{xi}D^{A}_{nn}) S^{b}_{i^{'}}S^{i^{'}}_{j} 
 +  F^{kl}_{xb}
((V_{B})^{x}_{i^{'}} + 2J^{A}_{xi^{'}}D^{A}_{nn}) S^{b}_{i}S^{i^{'}}_{j} \nonumber \\
&-&F^{kl}_{ay}
((V_{B})^{y}_{i} + 2J^{A}_{yi}D^{A}_{nn}) S^{a}_{j} 
 - 2F^{kl}_{ay}
((V_{A})^{y}_{j} + 2J^{A}_{mm}D^{A}_{yj}) S^{a}_{j^{'}}S^{j^{'}}_{i} \nonumber \\
&+&F^{kl}_{ay}
((V_{A})^{y}_{j^{'}} + 2J^{A}_{mm}D^{A}_{yj^{'}}) S^{a}_{j}S^{j^{'}}_{i} 
 - F^{kl}_{xb}
((V_{A})^{x}_{j} + 2J^{A}_{mm}D^{A}_{xj}) S^{b}_{i} \nonumber \\
&-& (D^{A}_{ki^{'}}K^{F}_{A,lj^{'}}+K^{F}_{A,ki^{'}}D^{A}_{lj^{'}}
-D^{A}_{ki^{'}}K^{F}_{AB}D^{B}_{lj^{'}})S^{j^{'}}_{i}S^{i^{'}}_{j}\nonumber \\
&+& 2(D^{A}_{ki}K^{F}_{A,lj^{'}}+K^{F}_{A,ki}D^{A}_{lj^{'}}
-D^{A}_{ki}K^{F}_{AB}D^{B}_{lj^{'}})S^{j^{'}}_{i^{'}}S^{i^{'}}_{j}\nonumber \\
&+& 2(D^{A}_{ki^{'}}K^{F}_{A,lj}+K^{F}_{A,ki^{'}}D^{A}_{lj}
-D^{A}_{ki^{'}}K^{F}_{AB}D^{B}_{lj})S^{j^{'}}_{i}S^{i^{'}}_{j^{'}}\nonumber \\
&-& (D^{A}_{li}K^{F}_{A,kj} +K^{F}_{A,li}D^{A}_{kj}-D^{A}_{li}K^{F}_{AB}D^{B}_{kj})\nonumber \\
&+& F^{lk}_{i'j'}
J^{A}_{i^{'}i}D^{A}_{j^{'}j}
 +  F^{lk}_{aj'}
J^{A}_{ai}D^{A}_{j^{'}j}
 +  F^{lk}_{i'b}
J^{A}_{i^{'}i}D^{A}_{bj}
 +  F^{lk}_{xj'}
J^{A}_{xi}D^{A}_{j^{'}j}\nonumber \\
&+& F^{lk}_{i'y}
J^{A}_{i^{'}i}D^{A}_{yj}
 +  F^{kl}_{xn}
J^{A}_{ai}D^{A}_{xj}S^{n}_{a}
 +  F^{kl}_{my}
J^{A}_{yi}D^{A}_{bj}S^{m}_{b}\nonumber \\
&+& F^{kl}_{rs}
J^{A}_{ai}D^{A}_{bj}S^{s}_{a}S^{r}_{b}
 +  F^{kl}_{xn}
J^{A}_{xi^{'}}D^{A}_{nj^{'}}S^{j^{'}}_{i}S^{i^{'}}_{j}
 +  F^{kl}_{my}
J^{A}_{mi^{'}}D^{A}_{yj^{'}}S^{j^{'}}_{i}S^{i^{'}}_{j}\nonumber \\
&+& F^{kl}_{rs}
J^{A}_{ri^{'}}D^{A}_{sj^{'}}S^{j^{'}}_{i}S^{i^{'}}_{j}
 -  2F^{kl}_{xn}
J^{A}_{xi}D^{A}_{nj^{'}}S^{j^{'}}_{i^{'}}S^{i^{'}}_{j}
 -  2F^{kl}_{my}
J^{A}_{mi}D^{A}_{yj^{'}}S^{j^{'}}_{i^{'}}S^{i^{'}}_{j}\nonumber \\
&-& 2F^{kl}_{rs}
J^{A}_{ri}D^{A}_{sj^{'}}S^{j^{'}}_{i^{'}}S^{i^{'}}_{j}
 -  2F^{kl}_{xn}
J^{A}_{xi^{'}}D^{A}_{nj}S^{j^{'}}_{i}S^{i^{'}}_{j^{'}}
 -  2F^{kl}_{my}
J^{A}_{mi^{'}}D^{A}_{yj}S^{j^{'}}_{i}S^{i^{'}}_{j^{'}}\nonumber \\
&-& \left. 2F^{kl}_{rs}
J^{A}_{ri^{'}}D^{A}_{sj}S^{j^{'}}_{i}S^{i^{'}}_{j^{'}}\right]
\end{eqnarray}

One may notice that Eqs.~(\ref{dfVijkl})--(\ref{delta_dfexchf12}) have many common intermediates, which are computed only once and reused. The computational cost of the $V$, $X$, $B$ and $C$ intermediates as well as of DF-$\delta E^{(20)}_{\rm exch-disp}$-F12 scales as $N^{5}$.

\section{Computational details} \label{sec:details}

The expressions for DF-$E^{(20)}_{\rm disp}$-F12(ODA) and DF-$E^{(20)}_{\rm exch-disp}$-F12(ODA) were implemented and tested within the {\sc Psi4NumPy} framework \cite{Smith:18}.  A production-level density-fitted SAPT0-F12 code is an ongoing project and will be available in the {\sc Psi4} program \cite{Parrish:17,Smith:20}.

It is a common practice to carry out the F12 calculations with three auxiliary basis sets in addition to the AO basis: the CABS set added to the AO one for the RI approximations, the DF basis in the calculation of correlated pair functions, and the JKFIT basis for the DF of Coulomb and exchange operators \cite{Knizia:09}.  In this work, the orbital basis sets were the augmented correlation consistent
aug-cc-pV$X$Z $\equiv $  a$X$Z sets of Dunning and coworkers \cite{Dunning:89, Kendall:92}, with $X$ ranging from D to 6.  
 The CABS bases were chosen as the a$X$Z-RIFIT sets, a.k.a. a$X$Z-MP2FIT \cite{Weigend:02,Hattig:05}.
The DF and JKFIT basis sets required for density fitting were also chosen as a$X$Z-MP2FIT \cite{Weigend:02,Hattig:05}.  We also examined the accuracy of the DF-SAPT-F12 calculations with the a$X$Z-JKFIT \cite{Weigend:02b} sets utilized for fitting the Fock matrices, that is, with a$X$Z-MP2FIT as the DF set and a$X$Z-JKFIT as the JKFIT set. 
The cardinal number $X$ of the auxiliary bases was equal or larger than the corresponding $X$ for the orbital set.
As we will write triplets such as aDZ/aDZ-RIFIT/aDZ-MP2FIT for the orbital/CABS/DF basis set combination, for clarity we will employ the notation a$X$Z-MP2FIT and a$X$Z-JKFIT rather than the more common a$X$Z/MP2FIT and a$X$Z/JKFIT.
The F12 integrals were computed employing the exponential correlation factor $\exp(-\gamma r_{12})$ with the length-scale parameter $\gamma$ set to 1.0 $a_0^{-1}$. This correlation factor was fitted to a sum of 6 Gaussian terms \cite{Tew:05}. 

We observed that the density-fitting approximation introduces numerical instabilities in pair correlation energies for core orbitals, where the coefficient multiplying the amplitude $T^{ij}_{ij}$ in Eq.~(\ref{oda2}) is very small, making an accurate determination of this amplitude challenging. While these instabilities were never an issue in the non-DF calculations with the ODA {\em Ansatz} in Ref.~\onlinecite{Kodrycka:19a}, we were not able to obtain reliable core-orbital $T^{ij}_{ij}$ amplitudes with density fitting.
Therefore, all calculations in this work were performed within the frozen core (FC) scheme. If all-electron (AE) dispersion and exchange dispersion energies are required, one can combine the FC values of $E^{(20)}_{\rm disp}$-F12 and $E^{(20)}_{\rm exch-disp}$-F12 with the conventional (non-F12) contributions of pair correlation energies involving one or two core orbitals. 
Alternatively, the F12 pair correlation energies for core orbitals could be computed using the fixed-amplitude Ansatz from Ref.~\onlinecite{Kodrycka:19a}, which is more approximate than ODA but avoids the instability of Eq.~(\ref{oda2}). 
While we have never observed such instabilities in frozen-core DF-$E^{(20)}_{\rm disp}$-F12(ODA) calculations anywhere close to the van der Waals minimum separation, they do show up for some pair correlation energies at large distances so that the fixed-amplitude Ansatz might be preferable in this case. Fortunately, conventional $E^{(20)}_{\rm disp}$ is already quite accurate at these distances, and $E^{(20)}_{\rm exch-disp}$ is essentially zero.
The FC approximation is the main reason for small discrepancies between the results presented below and the corresponding AE data of Ref.~\onlinecite{Kodrycka:19a}. The secondary reason for the discrepancies in $E^{(20)}_{\rm exch-disp}$-F12 are the small errors in our original implementation of this correction \cite{Kodrycka:19a} that led to minor differences (up to about 0.005 kcal/mol) with respect to the correct results -- see the Erratum to Ref.~\onlinecite{Kodrycka:19a} for more details.

The first class of systems tested are five small complexes in the same geometries as in our previous study~\cite{Kodrycka:19a}: He--He, Ne--Ne, Ar--Ar, H$_2$O--H$_2$O, and CH$_4$--CH$_4$. Subsequently, we performed a comprehensive analysis of DF-SAPT-F12 on the entire A24 benchmark database \cite{Rezac:13a} with basis sets up to a5Z, comparing the results of the conventional and explicitly correlated dispersion and exchange dispersion energies. 
The SAPT0 calculations were carried out by means of the {\sc Psi4} \cite{Hohenstein:10,Parrish:17}  quantum chemistry package with basis sets up to a6Z, with and without midbond functions, and utilized density fitting with the a$X$Z-MP2FIT auxiliary basis. 
The reference non-DF $E^{(20)}_{\rm disp}$-F12(ODA) and $E^{(20)}_{\rm exch-disp}$-F12(ODA) energies were computed using the FC scheme with all technical details presented in Ref.~\onlinecite{Kodrycka:19a}. 

\begin{figure}[!h]
	\centering
	\caption{Convergence of the frozen-core non-DF and DF $E_{\rm disp}^{(20)}$-F12 results as a function of basis set for the H$_2$O--H$_2$O complex from the A24 database. 
The hydrogenic functions from the same a$X$Z orbital basis set or the constant (3s3p2d2f) set of functions are chosen for midbond functions. 
The reference value marked by a black dashed line was obtained at the $E_{\rm disp}^{(20)}$/(a6Z+(a6Z),a7Z+(a7Z)) extrapolated level.
 The yellow dashed lines indicate the benchmark uncertainty. }
	\label{figure:h2odim_DFE20disp}
	\includegraphics[height=0.42\textheight]{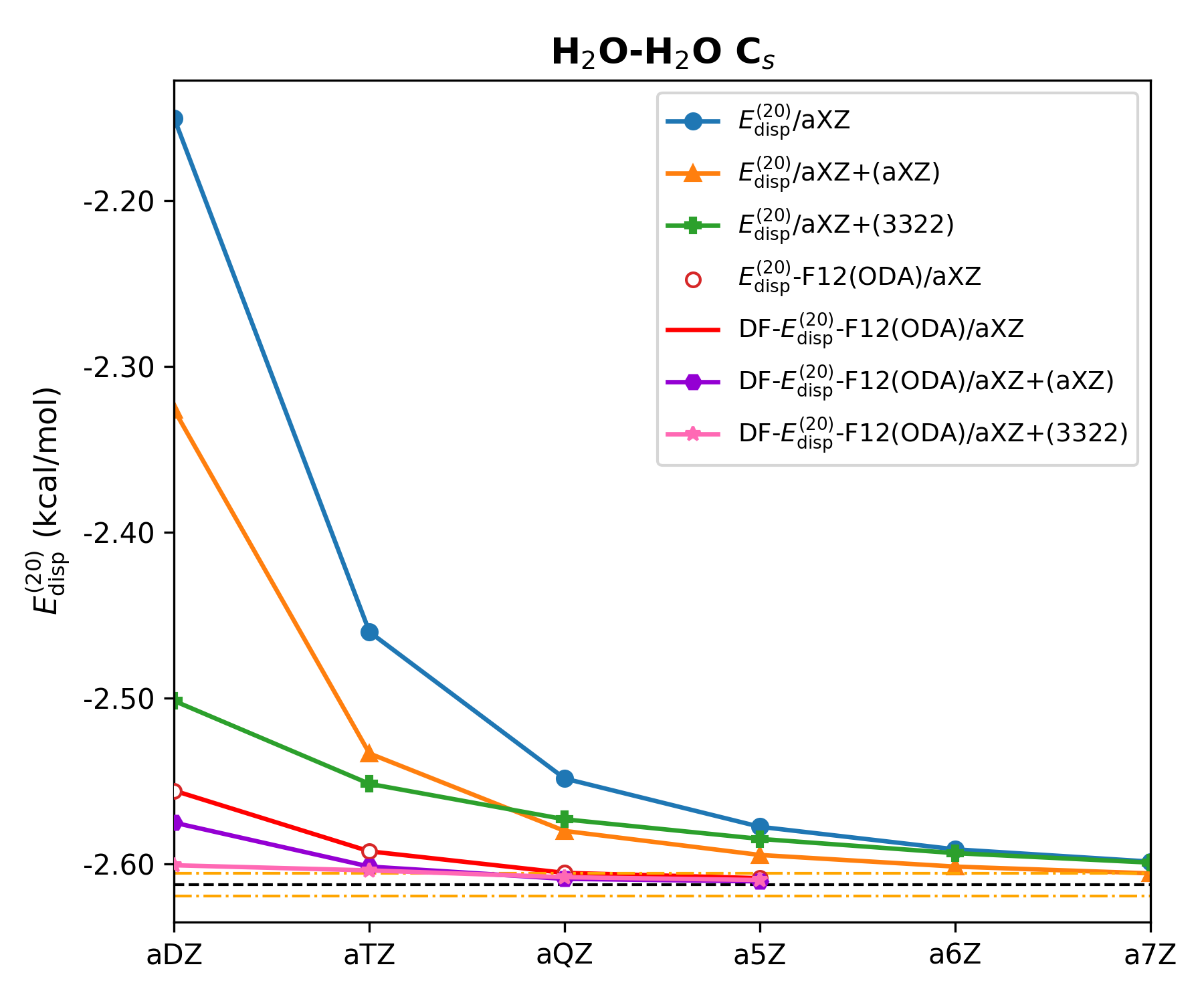}
\end{figure}

\begin{figure}[!h]
	\centering
	\caption{Convergence of the frozen-core non-DF and DF $E_{\rm exch-disp}^{(20)}$-F12 results as a function of basis set for the H$_2$O--H$_2$O complex from the A24 database. 
The hydrogenic functions from the same a$X$Z orbital basis set or the constant (3s3p2d2f) set of functions are chosen for midbond functions. 
The reference value marked by a black dashed line was obtained at the $E_{\rm exch-disp}^{(20)}$/(a6Z+(a6Z),a7Z+(a7Z)) extrapolated level.  The yellow dashed lines indicate the benchmark uncertainty. }
	\label{figure:h2odim_DFsum}
	\includegraphics[height=0.42\textheight]{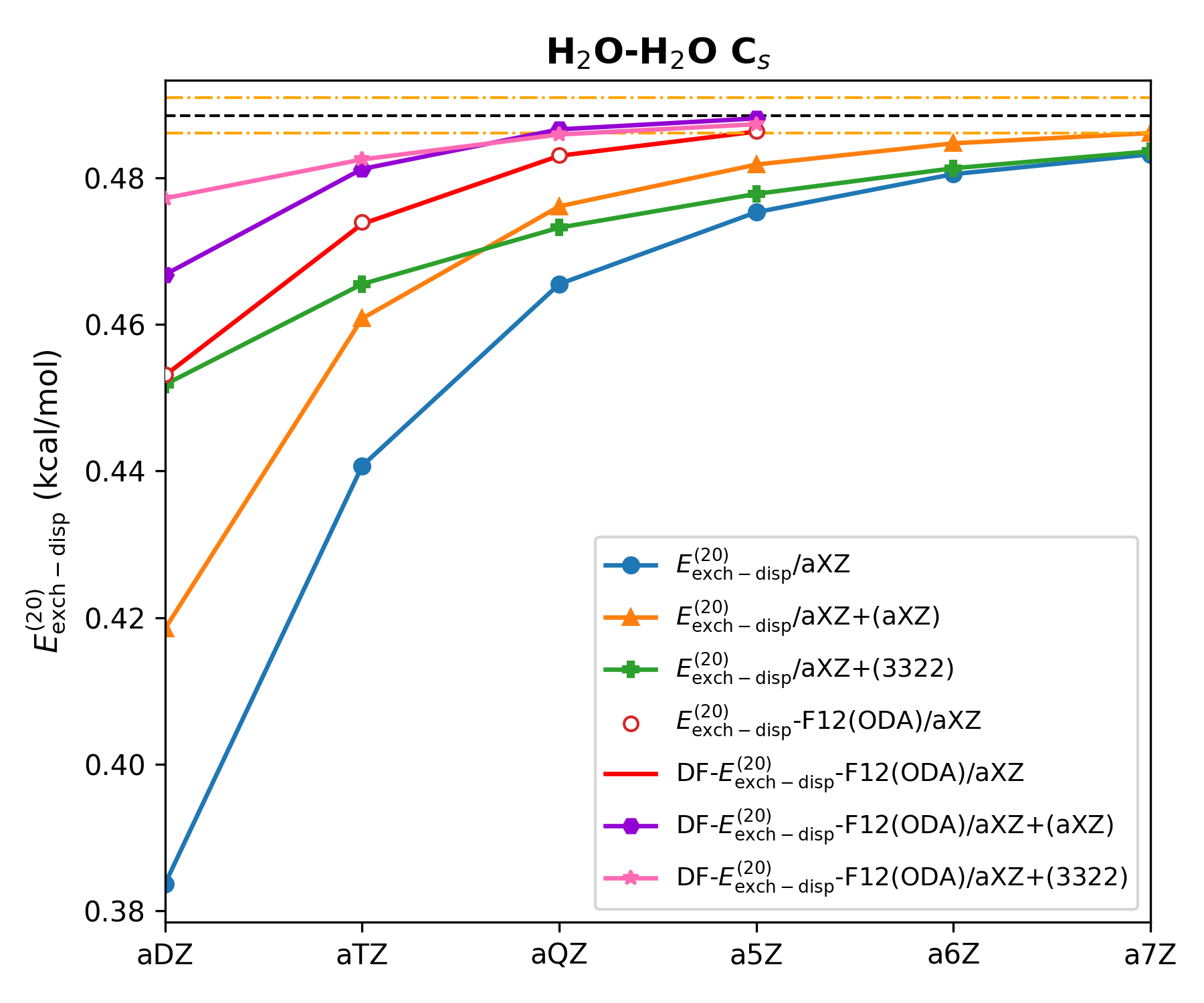}
\end{figure}

In this work, we increased the benchmark quality relative to Ref.~\onlinecite{Kodrycka:19a}. The CBS limits of $E^{(20)}_{\rm disp}$ and $E^{(20)}_{\rm exch-disp}$ were now obtained by the standard $X^{-3}$ extrapolation technique \cite{Halkier:98,Helgaker:97}. Specifically, the conventional $E^{(20)}_{\rm disp}$ and $E^{(20)}_{\rm exch-disp}$ SAPT corrections were extrapolated from the a5Z atom-centered orbital basis augmented by a hydrogenic a5Z set of midbond functions, and  the a6Z atom-centered orbital basis augmented by a hydrogenic a6Z set of midbond functions (such an extrapolation will be denoted as (a5Z+(a5Z),a6Z+(a6Z)).
The uncertainty of the benchmark result was assumed as the absolute difference between the reference extrapolated value and the one computed in the larger of the two bases, that is, a6Z+(a6Z).
In addition, to test the quality of the benchmark, we selected the water dimer and obtained an improved reference value by performing even larger calculations (also including midbond functions) using the aug-cc-pV7Z and aug-mcc-pV7Z \cite{Mielke:02} bases for the oxygen and hydrogen atoms, respectively. The reference value for this system employed in this large-basis convergence test (Figs. \ref{figure:h2odim_DFE20disp} and \ref{figure:h2odim_DFsum}) was estimated by extrapolating the a6Z+(a6Z) and a7Z+(a7Z) results.  However, whenever the water dimer is investigated as a part of the entire A24 dataset, we will continue using the (a5Z+(a5Z),a6Z+(a6Z)) benchmark for consistency: the improved septuple-zeta reference will only be employed when explicitly stated.

In addition to using midbond functions for benchmark calculations, their
benefits for the convergence rate of DF-$E^{(20)}_{\rm disp}$-F12 and DF-$E^{(20)}_{\rm exch-disp}$-F12 were investigated for the five test case dimers and  for all systems from the A24 database. In this work, we considered the constant (3s3p2d2f) set of midbond functions as well as a set of hydrogenic a$X$Z midbond functions with the cardinal number $X$ that varies in accordance with the atomic basis set. 
Herein, the shorthand notation of a$X$Z+(3322) is utilized for the a$X$Z atom-centered basis set augmented with the (3s3p2d2f) bond functions, while the variable-midbond calculations are denoted by a$X$Z+(a$X$Z).
The midbond function exponents for the (3s3p2d2f) set were (0.9, 0.3, 0.1) for sp and (0.6, 0.2) for df. The corresponding auxiliary basis, used in both the CABS and DF contexts, contains (1.8, 1.2, 0.6, 0.4, 0.2) exponents for spd functions, (1.5, 0.9, 0.5, 0.3) exponents for f, and (1.5, 0.9, 0.3) exponents for g \cite{Podeszwa:06a}.
The location of the midbond center was chosen as \cite{Akin-Ojo:03}
\begin{equation}
{\mathbf r}_{bond} = \frac{\sum_{a \in A} \sum_{b \in B} w_{ab} \frac{{\mathbf r}_a + {\mathbf r}_b}{2} }{\sum_{a \in A} \sum_{b \in B} w_{ab} } \quad w_{ab}=\vert {\mathbf r}_{a} -{\mathbf r}_{b} \vert ^{-6}
\end{equation}
where the summations run over all atoms $a$ (located at ${\mathbf r}_{a}$) in molecule $A$ and all atoms $b$ (located at ${\mathbf r}_{b}$) in molecule $B$. Such an approach of placing midbonds has been recommended  to avoid issues when one monomer is much longer than the other and the midpoint between the molecular centers of mass is still within one of the monomers \cite{Patkowski:17}.

After examining the CBS convergence of the individual $E^{(20)}_{\rm disp}$ and $E^{(20)}_{\rm exch-disp}$ terms, in the last part of this manuscript, we illustrate how the inclusion of $E^{(20)}_{\rm disp}$-F12 and $E^{(20)}_{\rm exch-disp}$-F12 in place of standard $E^{(20)}_{\rm disp}$ and $E^{(20)}_{\rm exch-disp}$ affects the accuracy of total SAPT interaction energies. 
For this purpose, we build on the study of Ref.~\onlinecite{Parker:14}, which examined the errors of interaction energies computed using various levels of SAPT with respect to reference CCSD(T)/CBS data. Specifically, Ref.~\onlinecite{Parker:14} investigated, in addition to SAPT0, the higher-level SAPT2, SAPT2+, SAPT2+(3), and SAPT2+3 variants including, to a various degree, the effects of intramolecular electron correlation. 
In addition, these higher-level SAPT treatments can be combined with a coupled cluster doubles (CCD) account of dispersion \cite{Williams:95a,Parrish:13} in place of the conventional one based on double perturbation theory, and/or with a ``$\delta$MP2" correction accounting for, in particular, some higher-order couplings between induction and dispersion. 
The accuracy of all resulting variants of SAPT has been investigated in Ref.~\onlinecite{Parker:14} using a benchmark dataset which is the union of the S22 \cite{Jurecka:06,Marshall:11}, HBC6 \cite{Thanthiriwatte:11}, NBC10 \cite{Sherrill:09}, and HSG \cite{Faver:11} databases. 
One should note that the datasets containing potential energy curves (that is, HBC6 and NBC10) were truncated in Ref.~\onlinecite{Parker:14} to at most five points per system (the van der Waals minimum, two shorter separations, and two longer separations), and we impose exactly the same truncation.
Moreover, the exchange-dispersion energies in Ref.~\onlinecite{Parker:14} are scaled to approximately account
for the effects missing in the single exchange approximation:
\begin{equation}\label{eq:s2scale}
E^{(20)}_{\rm exch-disp}{\rm (scaled)}=E^{(20)}_{\rm exch-disp}(S^2)\cdot\left(
\frac{E^{(10)}_{\rm exch}{\rm (full)}}{E^{(10)}_{\rm exch}(S^2)}\right)^{\alpha}
\end{equation}
with $\alpha=1.0$, and we will employ the same scaling for consistency.
In this work, we consider an F12 version of each of the SAPT variants from Ref.~\onlinecite{Parker:14}, obtained from the parent method by replacing the sum $E^{(20)}_{\rm disp}+E^{(20)}_{\rm exch-disp}{\rm (scaled)}$ by the $E^{(20)}_{\rm disp}$-F12$+E^{(20)}_{\rm exch-disp}$-F12(scaled) value computed in the same orbital basis (aDZ or aTZ). 
To enable direct comparison with the error statistics of non-F12 SAPT levels presented in Ref.~\onlinecite{Parker:14}, we utilize exactly the same set of four benchmark databases. In fact, we did not even have to recalculate the non-F12 SAPT data ourselves, but imported the values obtained in Ref.~\onlinecite{Parker:14} from the BFDb database \cite{Burns:17} and added the corresponding F12 corrections computed by us. 
As a result, the mean absolute error (MAE) values for total SAPT interaction energies, presented in the next section, are directly comparable to the MAE presented in Ref.~\onlinecite{Parker:14}.

\section{Results}

\subsection{Convergence with the density-fitting basis set}

A crucial step in the DF-SAPT-F12 development is ensuring that the density-fitting approximation does not introduce any significant errors. Following our previous study \cite{Kodrycka:19a}, the set of five complexes: He--He, Ne--Ne, Ar--Ar, H$_2$O--H$_2$O, and CH$_4$--CH$_4$ at their van der Waals minimum geometries was the subject of our preliminary tests.
For these systems,  we carried out the frozen-core DF-$E^{(20)}_{\rm disp}$-F12(ODA) and DF-$E^{(20)}_{\rm exch-disp}$-F12(ODA) calculations using the  a$X$Z/a$X$Z-RIFIT ($X$=D,T,Q,5) basis sets for orbital/CABS in conjunction with the increasing cardinal number of the a$Y$Z-MP2FIT ($Y$=D,T,Q,5) set for density fitting, with $Y \geq X$.  In the case of the helium dimer, basis sets up to a6Z, a6Z-RIFIT, and a6Z-MP2FIT were used. 
The resulting DF and non-DF explicitly correlated $E^{(20)}_{\rm disp}$ and  $E^{(20)}_{\rm exch-disp}$ energies are presented in Tables~\ref{table:H2OH2O} and \ref{table:Ch4Ch4} for the water and methane dimers, respectively; the corresponding results for the rare gas dimers are given in the Supporting
Information, which also contains tables
displaying the convergence of the $E^{(20)}_{\rm disp}+E^{(20)}_{\rm exch-disp}$ sum. 
We have also considered an alternative choice of auxiliary basis set for DF, in which the a$Y$Z-MP2FIT set is used for the DF of all integrals except for those occurring in the Fock and exchange operators present in Eqs.~(\ref{dfCklab}) and (\ref{dfBklmn}), in which case the a$Y$Z-JKFIT basis is used instead. 
The resulting explicitly correlated SAPT corrections are presented in the Supporting Information.

\begin{table}
\scriptsize
\centering
\caption{Frozen-core non-DF and DF $E_{\rm disp}^{(20)}$-F12(ODA) and $E_{\rm exch-disp}^{(20)}$-F12(ODA) values (in kcal/mol) for the water dimer for different combinations of the orbital and MP2FIT  auxiliary basis. The benchmark  values given below the F12 data were computed as conventional DF-$E_{\rm disp}^{(20)}$/DF-$E_{\rm exch-disp}^{(20)}$
 at the (a6Z+(a6Z),a7Z+(a7Z)) extrapolated level.}
\label{table:H2OH2O}
\begin{tabular}{llllllll}
\hline
\hline    

Basis & CABS Set & DF-$E^{(20)}_{\rm disp}$ & \multicolumn{5}{c}{DF-$E^{(20)}_{\rm disp}$-F12} \\
\multicolumn{3}{r}{DF Set:}  &             aDZ-MP2FIT &            aTZ-MP2FIT &            aQZ-MP2FIT &            a5Z-MP2FIT& non-DF \\
\hline
aDZ & aDZ-RIFIT   &  -2.1504 &  -2.5652 &  -2.5656 &  -2.5658 &  -2.5657& -2.5657\\
aTZ & aTZ-RIFIT   &  -2.4601 &          &  -2.5924 &  -2.5924 &  -2.5924& -2.5924\\
aQZ & aQZ-RIFIT   &  -2.5484 &          &          &  -2.6054 &  -2.6054&  -2.6054\\
a5Z & a5Z-RIFIT   &  -2.5776 &          &          &          &  -2.6086&  -2.6086\\

\hline
\hline
\multicolumn{2}{l}{\textbf{CBS(a6Z+(a6Z),a7Z+(a7Z))}}    & \textbf{-2.6126}  & $\pm$ 0.0069       &          &          &   &  \\
\hline
\hline
Basis & CABS Set & DF-$E^{(20)}_{\rm exch-disp}$ & \multicolumn{5}{c}{DF-$E^{(20)}_{\rm exch-disp}$-F12} \\
\multicolumn{3}{r}{DF Set:}  &             aDZ-MP2FIT &            aTZ-MP2FIT &            aQZ-MP2FIT &            a5Z-MP2FIT& non-DF \\
\hline
aDZ & aDZ-RIFIT   &  0.3837 &   0.4530 &  0.4542 &  0.4543 &  0.4543& 0.4543  \\
aTZ & aTZ-RIFIT   &  0.4407 &          &  0.4737 &  0.4740 &  0.4740& 0.4740 \\
aQZ & aQZ-RIFIT   &  0.4655 &          &         &  0.4830 &  0.4831& 0.4831 \\
a5Z & a5Z-RIFIT   &  0.4753 &          &         &         &  0.4863& 0.4864\\

\hline
\hline
\multicolumn{2}{l}{\textbf{CBS(a6Z+(a6Z),a7Z+(a7Z))}}   &\textbf{0.4885}  & $\pm$ 0.0024          &          &          &   & \\
\hline
\hline
\end{tabular}
\end{table}

\begin{table}
\scriptsize
\centering
\caption{Frozen-core non-DF and DF $E_{\rm disp}^{(20)}$-F12(ODA) and $E_{\rm exch-disp}^{(20)}$-F12(ODA) values (in kcal/mol) for the methane dimer for different combinations of the orbital and MP2FIT  auxiliary basis. The benchmark  values given below the F12 data were computed as conventional DF-$E_{\rm disp}^{(20)}$/DF-$E_{\rm exch-disp}^{(20)}$
 at the (a5Z+(a5Z),a6Z+(a6Z)) extrapolated level.}
\label{table:Ch4Ch4}
\begin{tabular}{llllllll}
\hline
\hline  

Basis & CABS Set & DF-$E^{(20)}_{\rm disp}$ & \multicolumn{5}{c}{DF-$E^{(20)}_{\rm disp}$-F12} \\
\multicolumn{3}{r}{DF Set:}  &             aDZ-MP2FIT &            aTZ-MP2FIT &            aQZ-MP2FIT &            a5Z-MP2FIT& non-DF \\
\hline
aDZ & aDZ-RIFIT   &  -1.0086 &  -1.1367 &  -1.1379 &  -1.1380 &  -1.1380& -1.1380 \\
aTZ & aTZ-RIFIT   &  -1.1084 &          &  -1.1471 &  -1.1472 &  -1.1471& -1.1471 \\
aQZ & aQZ-RIFIT   &  -1.1331 &          &          &  -1.1498 &  -1.1498& -1.1498 \\
a5Z & a5Z-RIFIT   &  -1.1420 &          &          &          &  -1.1506&  \\

\hline
\hline
\multicolumn{2}{l}{\textbf{CBS(a5Z+(a5Z),a6Z+(a6Z))}}    &\textbf{-1.1513}  &  $\pm$ 0.0024        &          &          &    &\\
\hline
\hline
Basis & CABS Set & DF-$E^{(20)}_{\rm exch-disp}$ & \multicolumn{5}{c}{DF-$E^{(20)}_{\rm exch-disp}$-F12} \\
\multicolumn{3}{r}{DF Set:}  &             aDZ-MP2FIT &            aTZ-MP2FIT &            aQZ-MP2FIT &            a5Z-MP2FIT& non-DF \\
\hline
aDZ & aDZ-RIFIT   &  0.0690 &  0.0822 &  0.0829 &  0.0830 &  0.0830& 0.0830  \\
aTZ & aTZ-RIFIT   &  0.0787 &         &  0.0851 &  0.0852 &  0.0852&  0.0852\\
aQZ & aQZ-RIFIT   &  0.0831 &         &         &  0.0868 &  0.0868&  0.0869\\
a5Z & a5Z-RIFIT   &  0.0851 &         &         &         &  0.0876 \\

\hline
\hline
\multicolumn{2}{l}{\textbf{CBS(a5Z+(a5Z),a6Z+(a6Z))}}    &\textbf{0.0882}  & $\pm$ 0.0008      &          &          &    &\\
\hline
\hline
\end{tabular}
\end{table}  

The convergence of the non-DF $E^{(20)}_{\rm disp}$-F12 and $E^{(20)}_{\rm exch-disp}$-F12 calculations for the same set of complexes has been thoroughly investigated in Ref.~\onlinecite{Kodrycka:19a}. In particular, it was found that while for the water and methane dimers the results are insensitive to the choice of CABS, the small CABS sets (especially aDZ-RIFIT) are not adequate for the noble gas dimers, leading to approximate dispersion energies below the variational limit. While in this work we focus on the errors introduced by density fitting rather
than those arising from an incomplete RI space, to complement the findings of Ref.~\onlinecite{Kodrycka:19a},
we performed $E^{(20)}_{\rm disp}$-F12 and $E^{(20)}_{\rm exch-disp}$-F12 calculations for the same five
complexes using another popular CABS choice, the a$X$Z-OPTRI sets designed specifically to decribe the orthogonal
complement of the a$X$Z orbital basis with the same $X$ \cite{Yousaf:09}. The resulting DF-SAPT-F12 corrections (with the a$Y$Z-MP2FIT sets, $Y\geq X$, used for DF) are presented in Tables SXVI--SXXX in the Supporting Information. Overall, the $E^{(20)}_{\rm disp}$-F12 and $E^{(20)}_{\rm exch-disp}$-F12 accuracy afforded by the two kinds of CABS bases is quite similar: while the $E^{(20)}_{\rm disp}$-F12/a$X$Z/a$X$Z-MP2FIT results are more accurate than the $E^{(20)}_{\rm disp}$-F12/a$X$Z/a$X$Z-OPTRI ones for the neon dimer, the opposite is true for the methane dimer. Thus, it appears that the smaller a$X$Z-OPTRI basis sets are also a sensible choice for CABS in SAPT-F12 calculations, however, a more extensive assessment of the performance of both auxiliary bases is required to make definite recommendations.

Out of the multitude of a$X$Z/a$Y$Z-RIFIT orbital and CABS set combinations investigated in Ref.~\onlinecite{Kodrycka:19a}, we have selected only 
the ``diagonal'' a$X$Z/a$X$Z-RIFIT ones in Tables~\ref{table:H2OH2O}--\ref{table:Ch4Ch4}, focusing on the convergence of results with the DF basis. 
For all considered dimers, the DF-$E^{(20)}_{\rm disp}$-F12 calculations in the aDZ/aDZ-RIFIT/aDZ-MP2FIT combination of orbital, CABS, and DF bases lead to errors in the range of 0.0005-0.0013 kcal/mol compared to the non-DF value, with an exception of He--He, for which the error amounts to 0.03 kcal/mol. 
This error is consistently reduced down to 0.0001 kcal/mol (0.0012 kcal/mol for the helium dimer) when one utilizes the aDZ/aDZ-RIFIT sets combined with the aTZ-MP2FIT set for DF. 
This observation confirms that the aDZ/aDZ-RIFIT/aDZ-MP2FIT combination is somewhat inaccurate for DF-SAPT-F12 calculations, which is in agreement with our earlier study \cite{Kodrycka:19a} (albeit sometimes, an error cancellation may occur). 
For the triple-zeta and larger bases, the default a$X$Z-MP2FIT choice for the DF basis, with $X$ the same as for the orbital set, is always adequate, with the DF-$E^{(20)}_{\rm disp}$-F12 errors not exceeding 0.0002 kcal/mol (0.0008 kcal/mol for the helium dimer).

Based on the DF-$E^{(20)}_{\rm exch-disp}$-F12 energies presented in Tables~\ref{table:H2OH2O}--\ref{table:Ch4Ch4} and in the Supporting Information, we again observe that the aDZ/aDZ-RIFIT/aDZ-MP2FIT combination is the source of the largest DF error (0.0003--0.0036 kcal/mol).  
The error is reduced to just 0.0001--0.0002 kcal/mol (0.0015 kcal/mol for Ar--Ar) when the aDZ/aDZ-RIFIT/aTZ-MP2FIT basis  sets are employed. 
Analyzing the bases with higher cardinal numbers, the DF-$E^{(20)}_{\rm exch-disp}$-F12 energies are virtually converged to their non-DF counterparts when utilizing the default a$X$Z-MP2FIT DF sets, with $X$ the same as for the orbital basis, except for the argon dimer (where the aTZ/aTZ-RIFIT/aTZ-MP2FIT combination produces the largest error of 0.0011 kcal/mol).

The results in Tables~\ref{table:H2OH2O}--\ref{table:Ch4Ch4} demonstrate that the DF error is negligible and well-controlled in the DF-$E^{(20)}_{\rm disp}$-F12 and  DF-$E^{(20)}_{\rm exch-disp}$-F12 calculations.  
Moreover, the choice of the auxiliary basis set family does not affect the accuracy of energies, and both  MP2FIT and JKFIT sets (the results for the latter are presented in the Supporting Information) are adequate for fitting the Fock and exchange matrices in the DF-SAPT-F12 context. Since the MP2FIT basis set is usually smaller than JKFIT, and at the same time does not compromise the quality of results, it was selected for fitting all four-index quantities in all remaining calculations. 

\subsection{Performance of DF-$E^{(20)}_{\rm disp}$-F12 and 
DF-$E^{(20)}_{\rm exch-disp}$-F12 on the A24 database}

\begin{figure}[!h]
	\centering
	\caption{Convergence of the frozen-core non-DF and DF $E_{\rm disp}^{(20)}$-F12 results as a function of basis set for the Ar--CH$_4$ complex from the A24 database. 
The hydrogenic functions from the same a$X$Z orbital basis set or the constant (3s3p2d2f) set of functions are chosen for midbond functions. 
The reference value marked by a black dashed line was obtained at the $E_{\rm disp}^{(20)}$/(a5Z+(a5Z),a6Z+(a6Z)) extrapolated level.
 The yellow dashed lines indicate the benchmark uncertainty. }
	\label{figure:arch4_DFE20disp}
	\includegraphics[height=0.42\textheight]{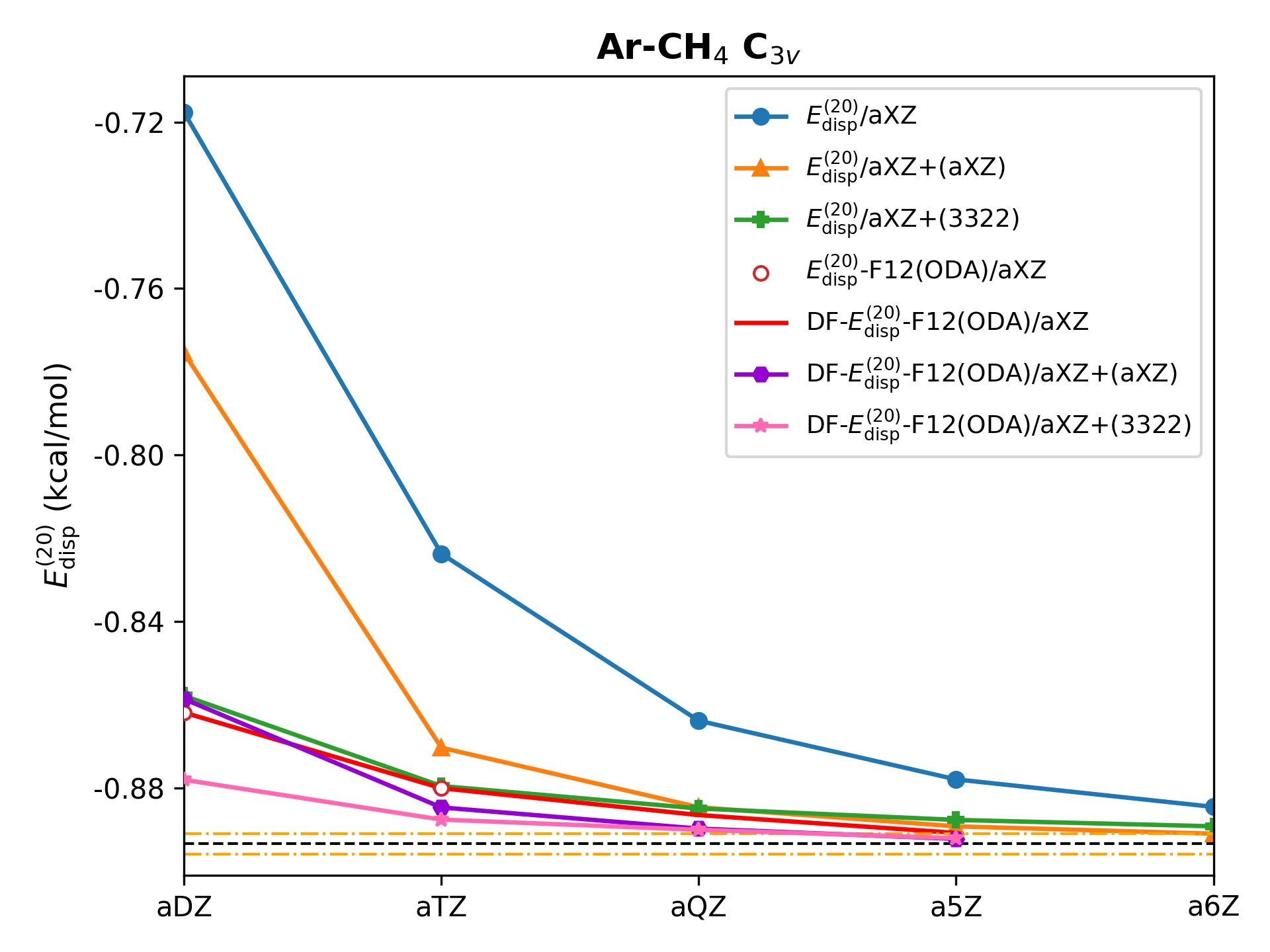}
\end{figure}

\begin{figure}[!h]
	\centering
	\caption{Convergence of the frozen-core non-DF and DF $E_{\rm exch-disp}^{(20)}$-F12 results as a function of basis set for the Ar--CH$_4$ complex from the A24 database. 
The hydrogenic functions from the same a$X$Z orbital basis set or the constant (3s3p2d2f) set of functions are chosen for midbond functions. 
The reference value marked by a black dashed line was obtained at the $E_{\rm exch-disp}^{(20)}$/(a5Z+(a5Z),a6Z+(a6Z)) extrapolated level.
 The yellow dashed lines indicate the benchmark uncertainty. }
	\label{figure:arch4_E20exchdisp}
	\includegraphics[height=0.42\textheight]{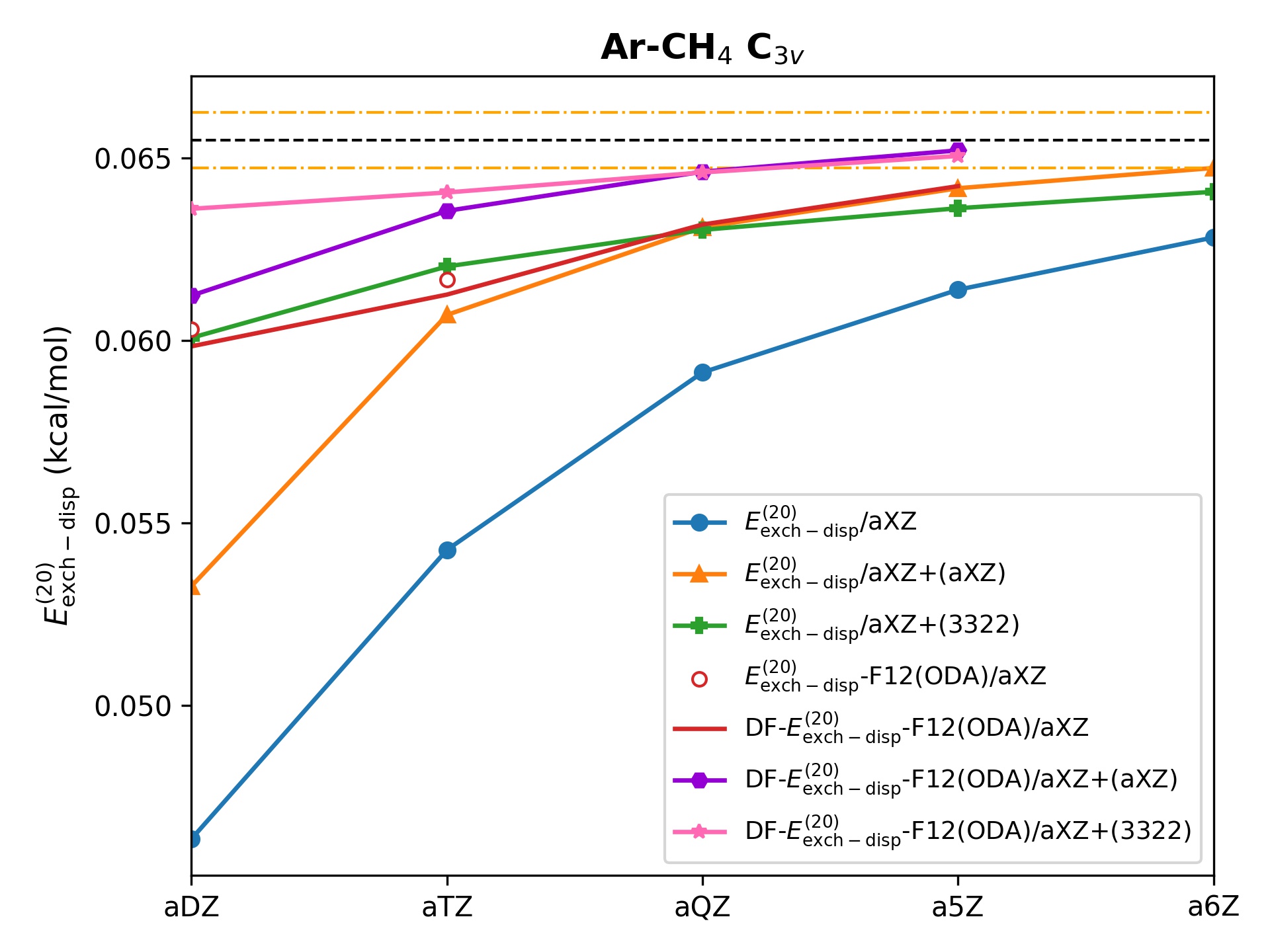}
\end{figure}

In order to investigate the performance of DF-SAPT-F12 on a broader sample of complexes, further tests were carried out on the A24 database \cite{Rezac:13a}. 
In contrast to our previous work \cite{Kodrycka:19a} where we were only able to run A24 calculations in basis sets up to aTZ,  the current DF implementation allows us to obtain all energies in bases up to a5Z with and without midbond functions.  
The basis set convergence of dispersion and exchange dispersion corrections, and of the corresponding sum $E^{(20)}_{\rm disp}+E^{(20)}_{\rm exch-disp}$,  for individual A24 systems is presented in Figs.~\ref{figure:h2odim_DFE20disp}--\ref{figure:arch4_E20exchdisp} and S1--S6 (Supporting Information).
Analyzing Figs. S1 and S2, it is clear that our newly proposed DF-SAPT-F12 method shows the same fast convergence rate as its non-DF counterpart: in fact, the non-DF and DF results are indistinguishable on the scale of these figures. We further see that the  DF-$E^{(20)}_{\rm disp}$-F12/aTZ and DF-$E^{(20)}_{\rm exch-disp}$-F12/aQZ  energies are either converged or almost converged to the reference CBS value. 
It turns out that the most challenging systems observed in our previous study \cite{Kodrycka:19a}, namely Ar--CH$_4$ (Figure~\ref{figure:arch4_DFE20disp}) and Ar--C$_2$H$_4$ (Figure S2), require a basis set with one cardinal number larger (aQZ) to reach the benchmark level for the DF-$E^{(20)}_{\rm disp}$-F12 calculations. Although the DF-$E^{(20)}_{\rm exch-disp}$-F12/a5Z energies are still not fully converged to the reference value for these two systems, we definitely observe a substantial improvement. 
Another way to improve the DF-$E^{(20)}_{\rm disp}$-F12 and DF-$E^{(20)}_{\rm exch-disp}$-F12 performance for these challenging systems is to include additional compact basis functions centered on the argon atom, such as the functions from a polarized core and valence basis sets, aug-cc-pwCV$X$Z \cite{Peterson:02}. Such sets are normally used for
calculations with all electrons correlated, but can provide additional flexibility to our frozen-core calculations as well. 
As illustrated in Figs. S7--S10 in the Supporting Information, enlarging the argon basis set from a$X$Z to aug-cc-pwCV$X$Z (and using aug-cc-pwCV$X$Z-MP2FIT as the CABS and DF set for argon), while keeping the a$X$Z basis set on all other atoms, leads to a noticeable improvement in the accuracy of DF-$E^{(20)}_{\rm disp}$-F12 and a small improvement in the accuracy of DF-$E^{(20)}_{\rm exch-disp}$-F12. Thus, it appears that the standard a$X$Z sets for argon might not contain sufficiently tight functions to maximize the benefits of the F12 approach.

It is worth emphasizing that the DF-$E^{(20)}_{\rm disp}$-F12 results are consistently  converged to the CBS limit in the aQZ basis, whereas that precision level is still not reproduced by the non-F12 approach even in a very large orbital basis such as a6Z+(a6Z).
To further illustrate the superior convergence of the DF-$E^{(20)}_{\rm disp}$-F12 and DF-$E^{(20)}_{\rm exch-disp}$-F12 values over their non-F12 counterparts (even those computed with midbonds), we selected the water dimer and compared both approaches against benchmark data of increased quality. 
For this purpose, we performed standard SAPT calculations (augmented with hydrogenic midbond functions) utilizing the aug-cc-pV7Z and aug-mcc-pV7Z bases for the oxygen and hydrogen atoms, respectively, and extrapolated the results from the a6Z+(a6Z) and a7Z+(a7Z) data. 
Figure~\ref{figure:h2odim_DFE20disp} displays the convergence of DF-$E^{(20)}_{\rm disp}$-F12 H$_2$O--H$_2$O values to this new improved CBS reference, and it demonstrates that the DF-$E^{(20)}_{\rm disp}$-F12 numbers are very well converged to the extra-precise CBS value. 
We therefore conclude that DF-$E^{(20)}_{\rm disp}$-F12 converges to the actual CBS limit, which is still not reproduced to that precision by a very large orbital basis such as a7Z+(a7Z). The corresponding behavior of the
exchange-dispersion energies relative to the H$_2$O--H$_2$O benchmark value of increased precision is illustrated in Figure~\ref{figure:h2odim_DFsum}, once again confirming superior convergence of the F12 data.

Another way of alleviating the slow basis set convergence of dispersion energy is the utilization of midbond functions.  It has been demonstrated that this technique in conjunction with explicitly correlated methods works very well for supermolecular CCSD(T)-F12 interaction energies \cite{Dutta:18}. 
Thus, we decided to investigate the influence of midbond functions on the SAPT and SAPT-F12 convergence using the A24 dataset as an example. It is worth emphasizing that the midbond and F12 approaches to dispersion are complementary, not competitive. 

The combination of DF-$E^{(20)}_{\rm disp}$-F12 with the +(a$X$Z) midbond set leads to a moderate improvement of the accuracy in the aDZ and aTZ bases. 
On the other hand, the DF-$E^{(20)}_{\rm disp}$-F12 values obtained with the +(3322) midbond set show results converged to the a6Z+(a6Z) level, or even to the reference value, already in the aDZ basis with an exception of Ar--CH$_4$ and Ar--C$_2$H$_4$. 
While the superior performance of the aDZ+(3322) results might be to some extent fortuitous, it is encouraging that the accuracy is consistent, similar to the consistently high accuracy (in particular, on the same A24 database) provided by the properly selected supermolecular CCSD(T)-F12/aDZ+(3322) variants \cite{Dutta:18}. 

The effect of hydrogenic midbond functions is more pronounced for the DF-$E^{(20)}_{\rm exch-disp}$-F12 correction. In Figs.~\ref{figure:h2odim_DFsum}, \ref{figure:arch4_E20exchdisp}, S3, and S4 we observe a significantly improved recovery of the CBS values in aDZ+(aDZ), aTZ+(aTZ), and aQZ+(aQZ) relative to the corresponding midbondless bases. 
Overall, the $E^{(20)}_{\rm exch-disp}$-F12/aTZ+(aTZ) energies are converged as well, or better, than standard $E^{(20)}_{\rm exch-disp}$ in the a5Z+(a5Z) basis. 
Considering the aQZ+(aQZ) basis set, the F12 energies are consistently converged above the conventional a6Z+(a6Z) level (or to that level in the case of Ar--CH$_4$ and Ar--C$_2$H$_4$, Figure~\ref{figure:arch4_E20exchdisp} and S4) and some further enhancement is attained in the a5Z+(a5Z) basis.

When evaluating DF-$E^{(20)}_{\rm exch-disp}$-F12 with the constant +(3322) set of midbond functions, we  observe superior convergence at the aDZ and aTZ levels compared to DF-$E^{(20)}_{\rm exch-disp}$-F12 with and without the hydrogenic midbond. 
In a few cases (e.g., for the NH$_3$--C$_2$H$_4$ complex), the addition of constant midbonds provides results nearly converged to the reference value already in the aDZ+(3322) basis. 
The $E^{(20)}_{\rm exch-disp}$-F12/aTZ+(3322) results attain the conventional $E^{(20)}_{\rm exch-disp}$/a6Z+(a6Z)-level accuracy across nearly the entire A24 database. The challenging systems containing the argon atom require basis sets with one cardinal number higher to reproduce the same precision. 

\begin{figure}[!h]
	\centering
	\caption{Relative errors on the A24 database for frozen-core $E_{\rm disp}^{(20)}$-F12 and DF-$E_{\rm disp}^{(20)}$-F12 computed with the aDZ orbital basis set, the aTZ-RIFIT CABS set, and the aTZ-MP2FIT DF set. The notation +(a$X$Z) and +(3322) signifies that the hydrogenic a$X$Z and the constant (3s3p2d2f) set of midbond functions has been added to the a$X$Z atom-centered basis, respectively.  The reference values were obtained at the $E_{\rm disp}^{(20)}$/(a5Z+(a5Z),a6Z+(a6Z)) extrapolated level.}
	\label{figure:Error_Edisp20F12_aDZ_aTZRI}
	\includegraphics[height=0.42\textheight]{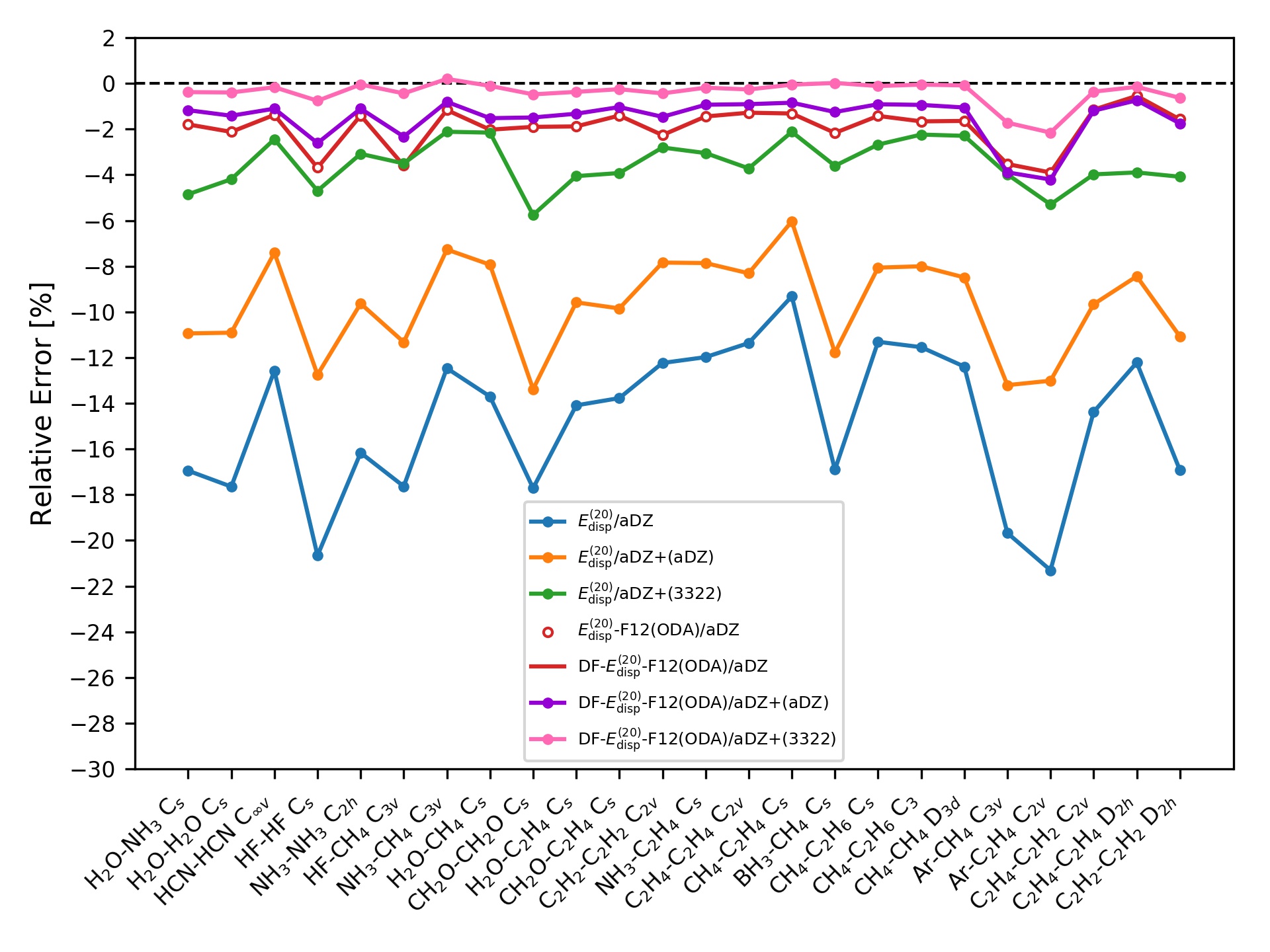}
\end{figure}

\begin{figure}[!h]
	\centering
	\caption{Relative errors on the A24 database for frozen-core $E_{\rm disp}^{(20)}$-F12 and DF-$E_{\rm disp}^{(20)}$-F12 computed with the aTZ orbital basis set, the aTZ-RIFIT CABS set, and the aTZ-MP2FIT DF set. The notation +(a$X$Z) and +(3322) signifies that the hydrogenic a$X$Z and the constant (3s3p2d2f) set of midbond functions has been added to the a$X$Z atom-centered basis, respectively. The reference values were obtained at the $E_{\rm disp}^{(20)}$/(a5Z+(a5Z),a6Z+(a6Z)) extrapolated level. The error bars indicate the benchmark uncertainty. }
	\label{figure:Error_Edisp20F12_aTZ_aTZRI}
	\includegraphics[height=0.42\textheight]{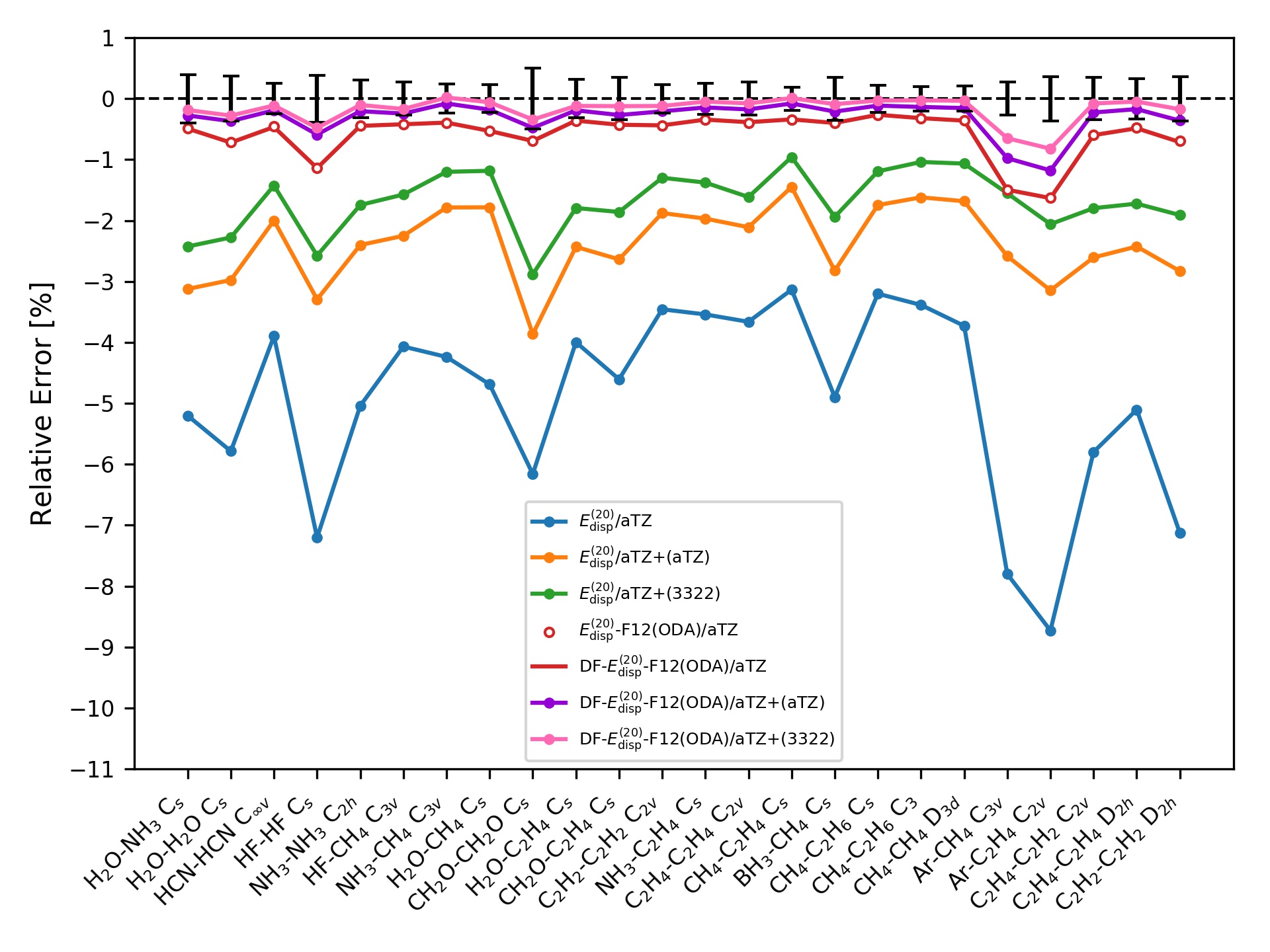}
\end{figure}

\begin{figure}[!h]
	\centering
	\caption{Relative errors on the A24 database for frozen-core $E_{\rm disp}^{(20)}$-F12 and DF-$E_{\rm disp}^{(20)}$-F12 computed with the aQZ orbital basis set, the aQZ-RIFIT CABS set, and the aQZ-MP2FIT DF set. The notation +(a$X$Z) and +(3322) signifies that the hydrogenic a$X$Z and the constant (3s3p2d2f) set of midbond functions has been added to the a$X$Z atom-centered basis, respectively. The reference values were obtained at the $E_{\rm disp}^{(20)}$/(a5Z+(a5Z),a6Z+(a6Z)) extrapolated level. The error bars indicate the benchmark uncertainty.}
	\label{figure:Error_Edisp20F12_aQZ_aQZRI}
	\includegraphics[height=0.42\textheight]{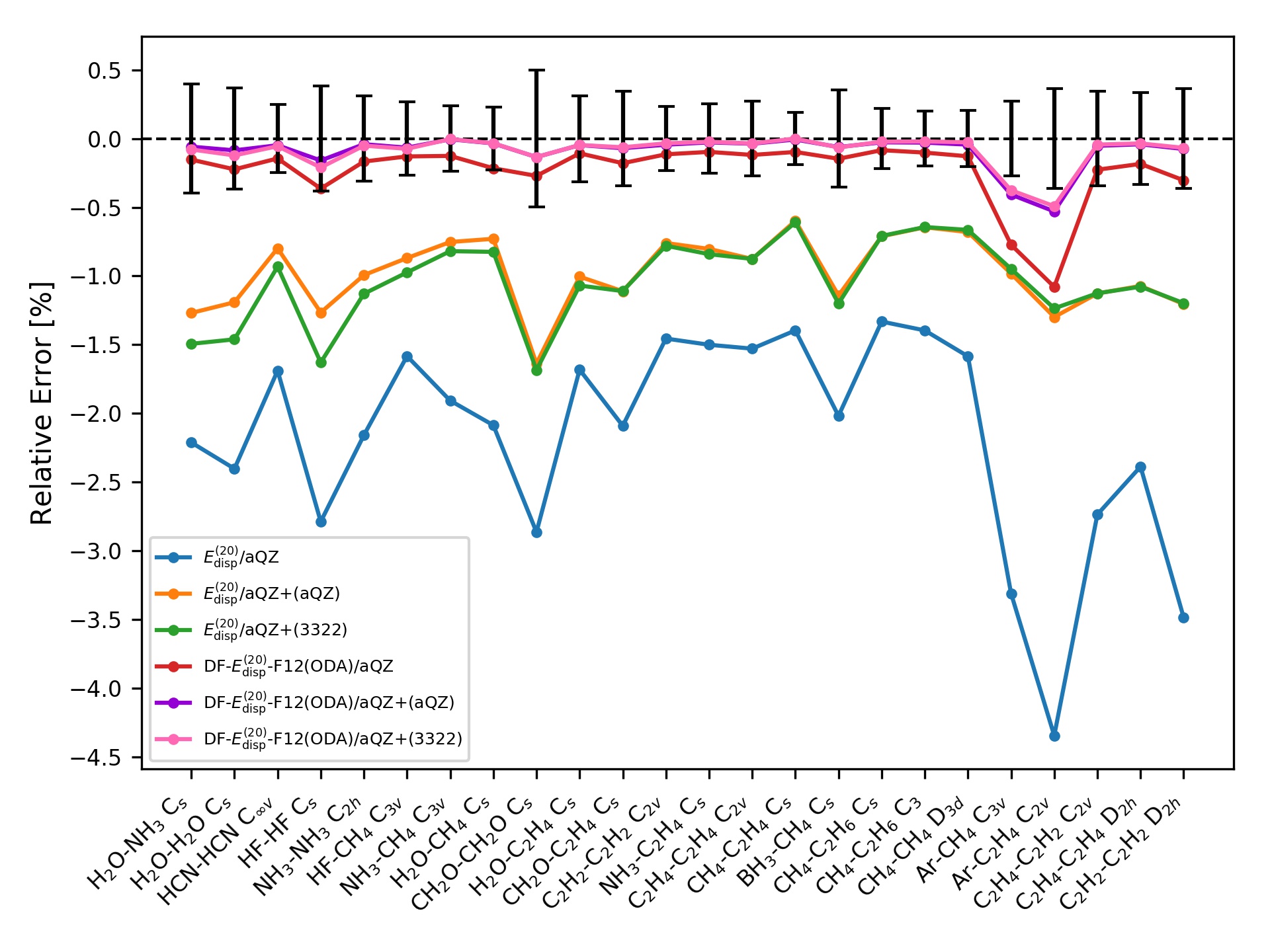}
\end{figure}

\begin{figure}[!h]
	\centering
	\caption{Relative errors on the A24 database for frozen-core $E_{\rm disp}^{(20)}$-F12 and DF-$E_{\rm disp}^{(20)}$-F12 computed with the a5Z orbital basis set, the a5Z-RIFIT CABS set, and the a5Z-MP2FIT DF set. The notation +(a$X$Z) and +(3322) signifies that the hydrogenic a$X$Z and the constant (3s3p2d2f) set of midbond functions has been added to the a$X$Z atom-centered basis, respectively. The reference values were obtained at the $E_{\rm disp}^{(20)}$/(a5Z+(a5Z),a6Z+(a6Z)) extrapolated level. The error bars indicate the benchmark uncertainty.}
	\label{figure:Error_Edisp20F12_a5Z_a5ZRI}
	\includegraphics[height=0.42\textheight]{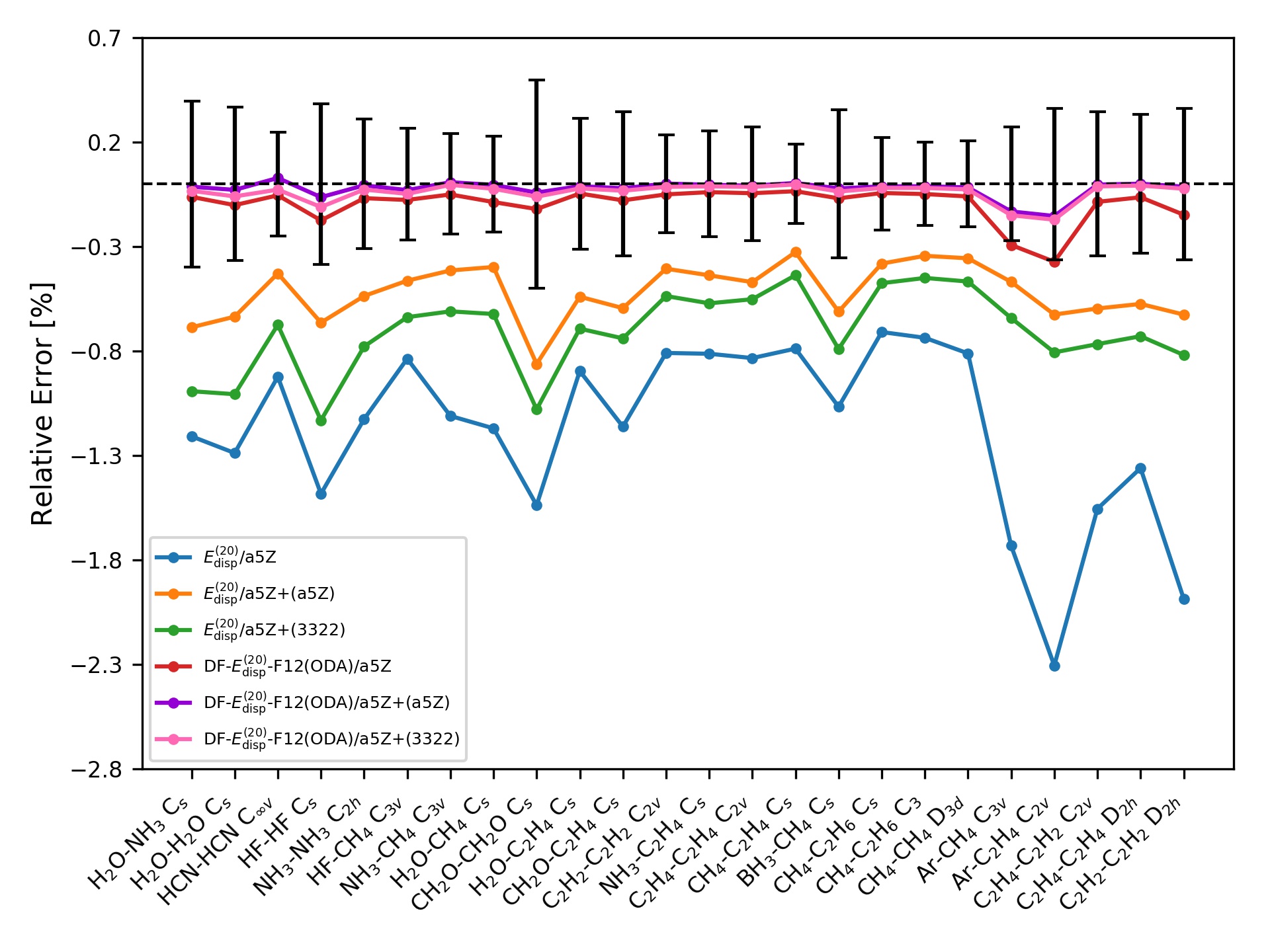}
\end{figure}

\begin{figure}[!h]
	\centering
	\caption{Relative errors on the A24 database for frozen-core $E_{\rm exch-disp}^{(20)}$-F12 and DF-$E_{\rm exch-disp}^{(20)}$-F12 computed with the aDZ orbital basis set, the aTZ-RIFIT CABS set, and the aTZ-MP2FIT DF set. The notation +(a$X$Z) and +(3322) signifies that the hydrogenic a$X$Z and the constant (3s3p2d2f) set of midbond functions has been added to the a$X$Z atom-centered basis, respectively. The reference values were obtained at the $E_{\rm exch-disp}^{(20)}$/(a5Z+(a5Z),a6Z+(a6Z)) extrapolated level.}
	\label{figure:Error_Exchdisp20F12_aDZ_aTZRI}
	\includegraphics[height=0.42\textheight]{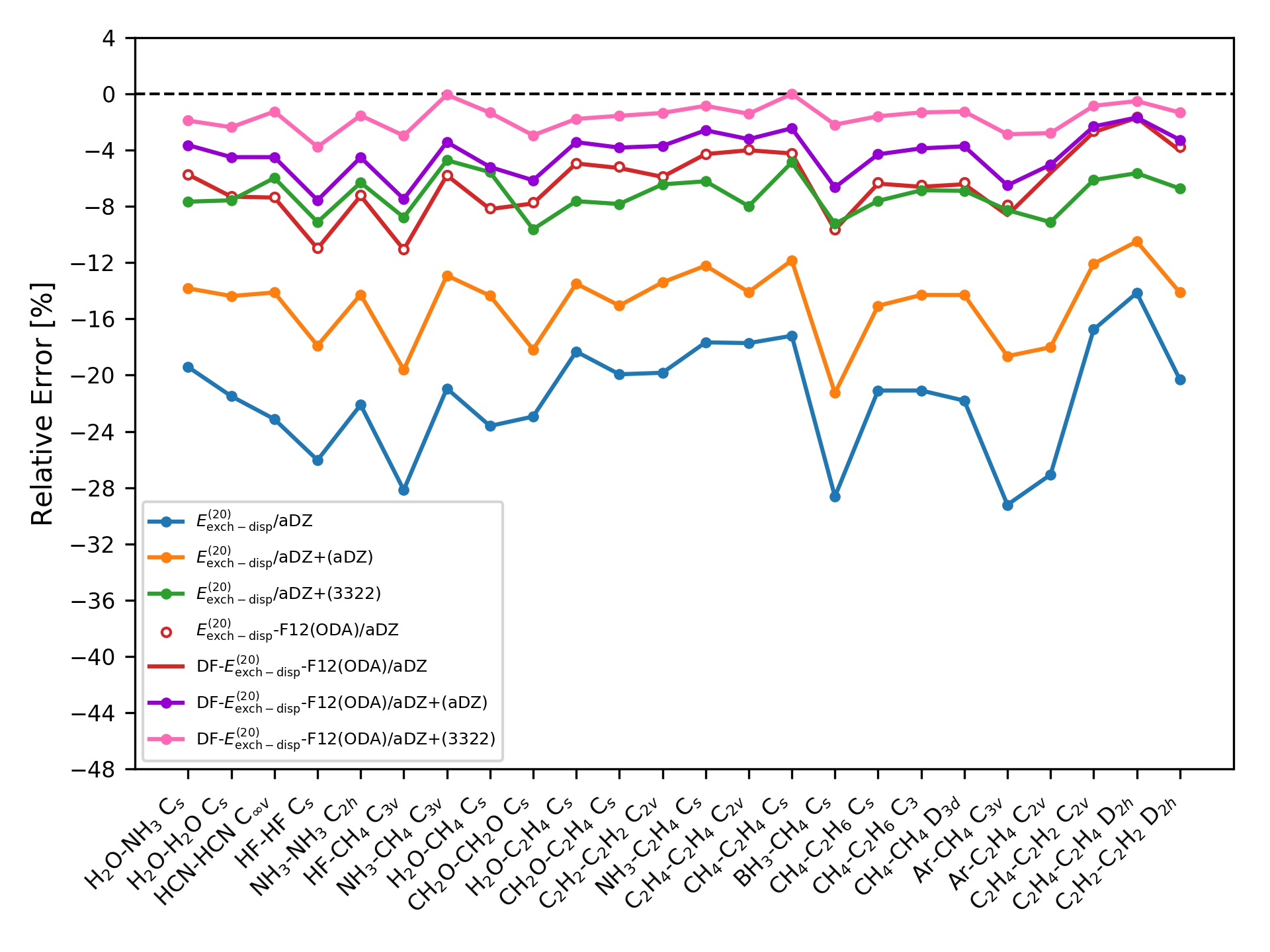}
\end{figure}

\begin{figure}[!h]
	\centering
	\caption{Relative errors on the A24 database for frozen-core $E_{\rm exch-disp}^{(20)}$-F12 and DF-$E_{\rm exch-disp}^{(20)}$-F12 computed with the aTZ orbital basis set, the aTZ-RIFIT CABS set, and the aTZ-MP2FIT DF set. The notation +(a$X$Z) and +(3322) signifies that the hydrogenic a$X$Z and the constant (3s3p2d2f) set of midbond functions has been added to the a$X$Z atom-centered basis, respectively. The reference values were obtained at the $E_{\rm exch-disp}^{(20)}$/(a5Z+(a5Z),a6Z+(a6Z)) extrapolated level.}
	\label{figure:Error_Exchdisp20F12_aTZ_aTZRI}
	\includegraphics[height=0.42\textheight]{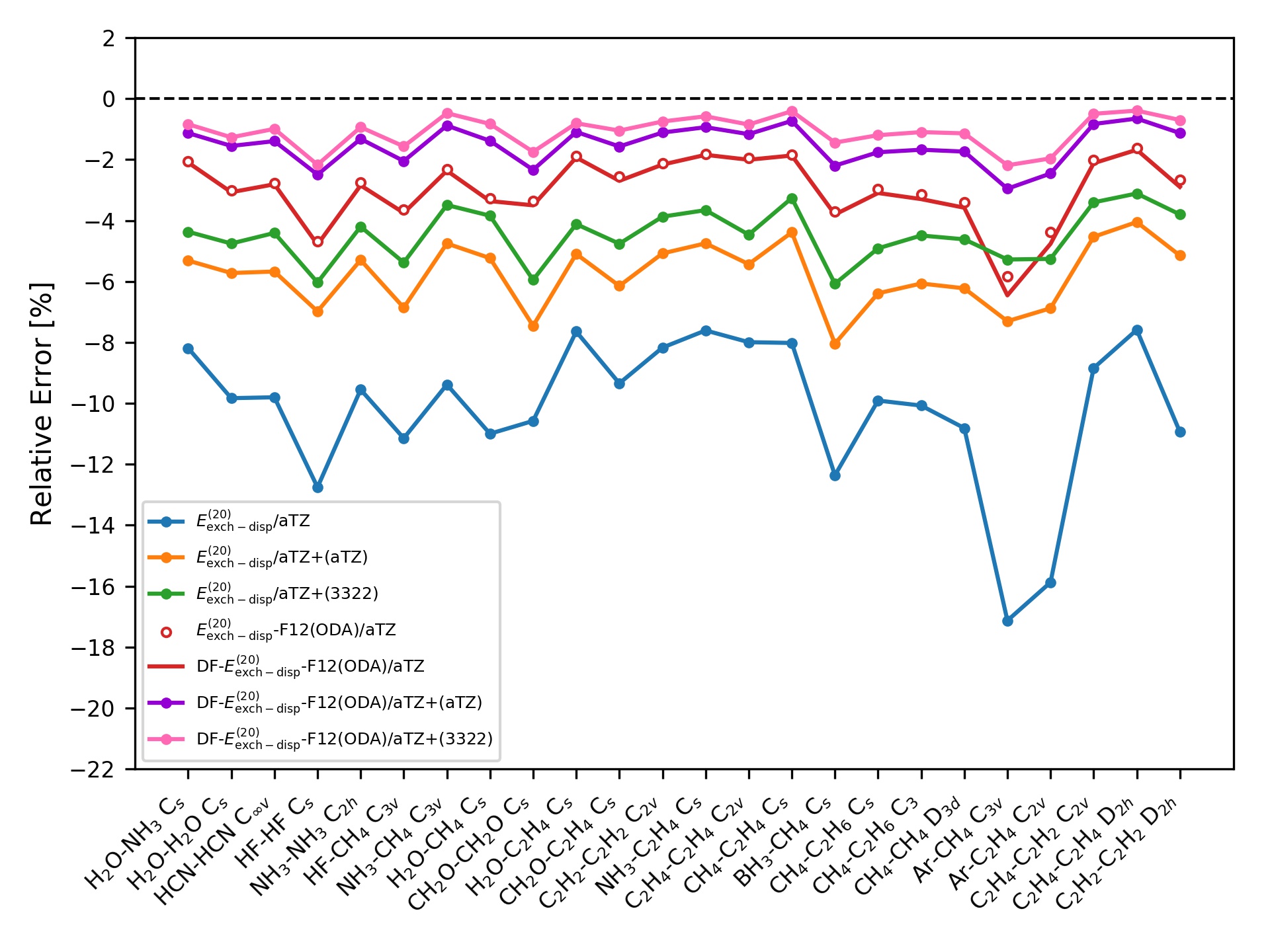}
\end{figure}

\begin{figure}[!h]
	\centering
	\caption{Relative errors on the A24 database for frozen-core $E_{\rm exch-disp}^{(20)}$-F12 and DF-$E_{\rm exch-disp}^{(20)}$-F12 computed with the aQZ orbital basis set, the aQZ-RIFIT CABS set, and the aQZ-MP2FIT DF set. The notation +(a$X$Z) and +(3322) signifies that the hydrogenic a$X$Z and the constant (3s3p2d2f) set of midbond functions has been added to the a$X$Z atom-centered basis, respectively. The reference values were obtained at the $E_{\rm exch-disp}^{(20)}$/(a5Z+(a5Z),a6Z+(a6Z)) extrapolated level. The error bars indicate the benchmark uncertainty.}
	\label{figure:Error_Exchdisp20F12_aQZ_aQZRI}
	\includegraphics[height=0.42\textheight]{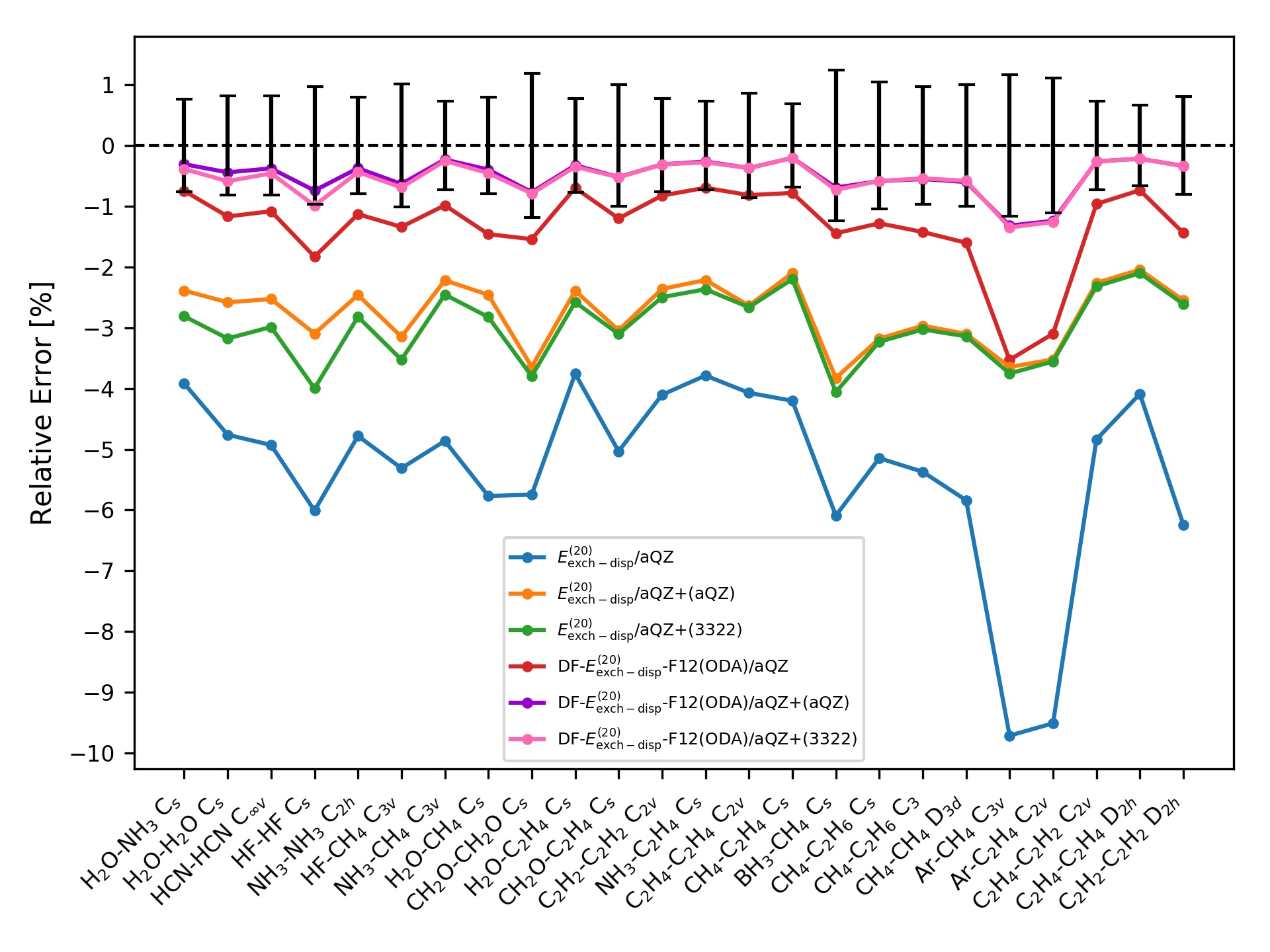}
\end{figure}

\begin{figure}[!h]
	\centering
	\caption{Relative errors on the A24 database for frozen-core $E_{\rm exch-disp}^{(20)}$-F12 and DF-$E_{\rm exch-disp}^{(20)}$-F12 computed with the a5Z orbital basis set, the a5Z-RIFIT CABS set, and the a5Z-MP2FIT DF set. The notation +(a$X$Z) and +(3322) signifies that the hydrogenic a$X$Z and the constant (3s3p2d2f) set of midbond functions has been added to the a$X$Z atom-centered basis, respectively. The reference values were obtained at the $E_{\rm exch-disp}^{(20)}$/(a5Z+(a5Z),a6Z+(a6Z)) extrapolated level. The error bars indicate the benchmark uncertainty.}
	\label{figure:Error_Exchdisp20F12_a5Z_a5ZRI}
	\includegraphics[height=0.42\textheight]{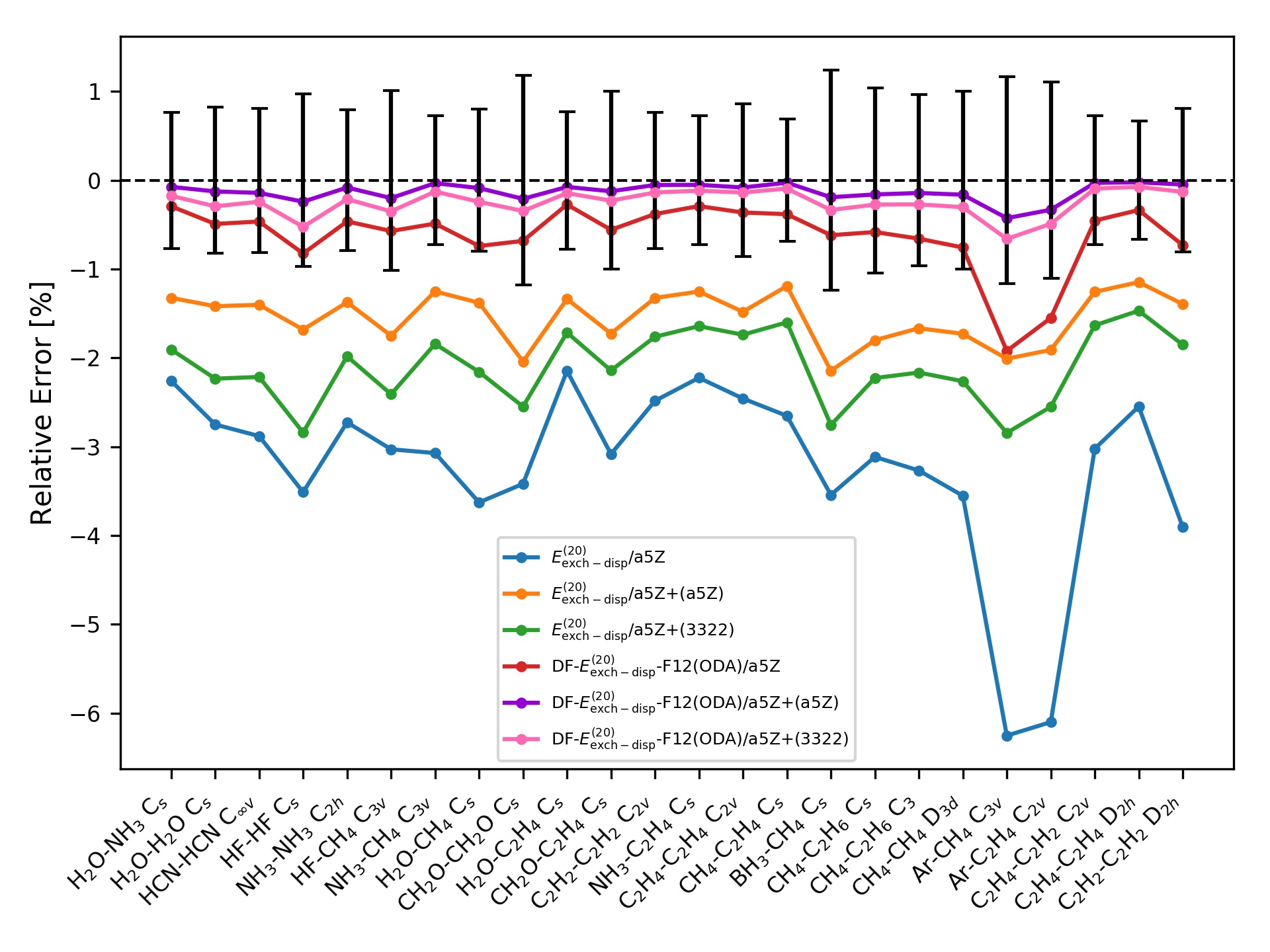}
\end{figure}

The relative errors with respect to the benchmark for the DF-$E^{(20)}_{\rm disp}$-F12 and DF-$E^{(20)}_{\rm exch-disp}$-F12 results in a range of basis sets with and without midbond functions are illustrated in Figures~\ref{figure:Error_Edisp20F12_aDZ_aTZRI}--\ref{figure:Error_Exchdisp20F12_a5Z_a5ZRI}. The corresponding relative errors for the sum $E^{(20)}_{\rm disp}+E^{(20)}_{\rm exch-disp}$ are presented in the Supporting Information (Figs. S11--S14). 
One clearly sees that the DF-$E^{(20)}_{\rm disp}$-F12 and  DF-$E^{(20)}_{\rm exch-disp}$-F12 energies computed in aDZ and aTZ  show the same performance as their non-DF counterparts.  
The only systems which exhibit a visible, though still tiny, DF error for $E^{(20)}_{\rm exch-disp}$-F12/aDZ are Ar--CH$_4$ and Ar--C$_2$H$_4$, indicating that the DF basis is not entirely converged in these cases.

Thanks to the DF algorithm, we can now investigate how well the explicitly correlated dispersion and exchange-dispersion corrections work in the aQZ and a5Z basis sets (our non-DF implementation \cite{Kodrycka:19a} could not handle some of the A24 systems in such large bases). 
Generally, the DF-$E^{(20)}_{\rm disp}$-F12/aQZ results closely reproduce the benchmark level, leading to the mean absolute percent error (MA\%E) value of 0.23\%. This error is reduced down to 0.04\% in the a5Z basis. It needs to be stressed that the systems containing the argon atom are the source of largest errors.  

Pursuing the same analysis for DF-$E^{(20)}_{\rm exch-disp}$-F12/aQZ (Fig.~\ref{figure:Error_Exchdisp20F12_aQZ_aQZRI}), we observe the enhanced  
performance compared to the aTZ level, leading to the MA\%E value of 1.33\%.  
A further improvement is obtained in the a5Z basis set (See Fig.~\ref{figure:Error_Exchdisp20F12_a5Z_a5ZRI}), showing a small MA\%E value of 0.25\%. 
Examining the sum of DF-$E^{(20)}_{\rm disp}$-F12 and DF-$E^{(20)}_{\rm exch-disp}$-F12 at the aQZ and a5Z level (Figs. S13 and S14, respectively, in the Supporting Information), we notice that the energies phenomenally reproduce the CBS limit, leading to MA\%E values of 0.08\% and 0.02\%, respectively. 

For the last test on the A24 database, we looked at the accuracy of the explicitly correlated calculations augmented with the variable (hydrogenic) and constant (3s3p2d2f) sets of midbond functions. 
Figures~\ref{figure:Error_Edisp20F12_aDZ_aTZRI} and \ref{figure:Error_Exchdisp20F12_aDZ_aTZRI} indicate that the improvement brought about by the +(3322) midbond set for the DF-SAPT-F12/aDZ calculations is nothing short of impressive. 
This level of theory faithfully reproduces the benchmark data with MA\%E of 0.41\%, 1.7\%, and 0.27\% for DF-$E^{(20)}_{\rm disp}$-F12, DF-$E^{(20)}_{\rm exch-disp}$-F12, and DF-$E^{(20)}_{\rm disp}$-F12 + DF-$E^{(20)}_{\rm exch-disp}$-F12, respectively. 
The hydrogenic midbond functions bring minimal (but still noticeable) accuracy gain for aDZ, leading to MA\%E values of 1.5\%, 4.3\%, and 1.1\% for  DF-$E^{(20)}_{\rm disp}$-F12, DF-$E^{(20)}_{\rm exch-disp}$-F12, and DF-$E^{(20)}_{\rm disp}$-F12 + DF-$E^{(20)}_{\rm exch-disp}$-F12, respectively.
These values should be contrasted with the respective MA\%E for the midbondless aDZ data, amounting to 1.9\%, 6.3\%, and 1.3\% for these three quantities.
Moving on to the aTZ basis, an additional enhancement is observed for the explicitly correlated dispersion and exchange-dispersion corrections, as well as for their sum, when either midbond type is applied. 
The  DF-$E^{(20)}_{\rm exch-disp}$-F12/aTZ+(3322) approach exhibits superior performance (a MA\%E of 1.1\% was found for this level) with respect to DF-$E^{(20)}_{\rm exch-disp}$-F12/aTZ+(aTZ) (a MA\%E of 1.5\%) and plain DF-$E^{(20)}_{\rm exch-disp}$-F12/aTZ (a MA\%E of 3.0\%). 
The  DF-$E^{(20)}_{\rm disp}$-F12/aTZ variant 
performs better with the +(3322) midbond than with the hydrogenic ones,  providing respective MA\%E values of  0.18\%. and 0.30\%. This behavior is very much expected: since the (3s3p2d2f) set is larger than the hydrogenic set for the aDZ and aTZ orbital basis sets, it results in more accurate values. 

The inclusion of midbond functions still brings improvement over the midbondless approach for the DF-$E^{(20)}_{\rm disp}$-F12 values in aQZ and a5Z, even though the energies are already well converged to the CBS limit. 
The improvement is even more pronounced for the exchange dispersion energy. For the A24 systems other than Ar--CH$_4$ and Ar--C$_2$H$_4$, converging the results to within the (very tight) uncertainty of the benchmark requires going up to either aQZ or aTZ+(midbond) for DF-$E^{(20)}_{\rm disp}$-F12 and either a5Z or aQZ+(midbond) for DF-$E^{(20)}_{\rm exch-disp}$-F12. 
For the two complexes containing argon, one needs to go all the way to a5Z+(midbond) to converge either correction to within the benchmark uncertainty.
At the aQZ level, the +(aQZ) and +(3322) midbonds perform nearly identically, which is expected given the similar composition of the two sets of bond functions. When one goes up to a5Z, the +(a5Z) midbond set becomes larger than the +(3322) one, leading to a slightly larger improvement in accuracy. 
Last but not least, Figs.~\ref{figure:Error_Edisp20F12_a5Z_a5ZRI} and \ref{figure:Error_Exchdisp20F12_a5Z_a5ZRI} show that the deviations between the (a5Z+(a5Z),a6Z+(a6Z)) extrapolated benchmarks and the largest-basis F12 results are significantly smaller than the benchmark uncertainties. This level of consistency is really gratifying, as it validates both the $X^{-3}$ extrapolation and the F12 data, and even indicates that the error estimates for the former approach are quite conservative.

\subsection{Total SAPT-F12 interaction energies}

\begin{figure}[!h]
	\centering
	\caption{Mean absolute errors (MAE) (kcal/mol) of the interaction energy averaged over the S22, HBC6, NBC10, and HSG databases for SAPT and SAPT-F12 in aDZ and aTZ basis sets.}
	\label{figure:overall_mae}
	\includegraphics[height=0.42\textheight]{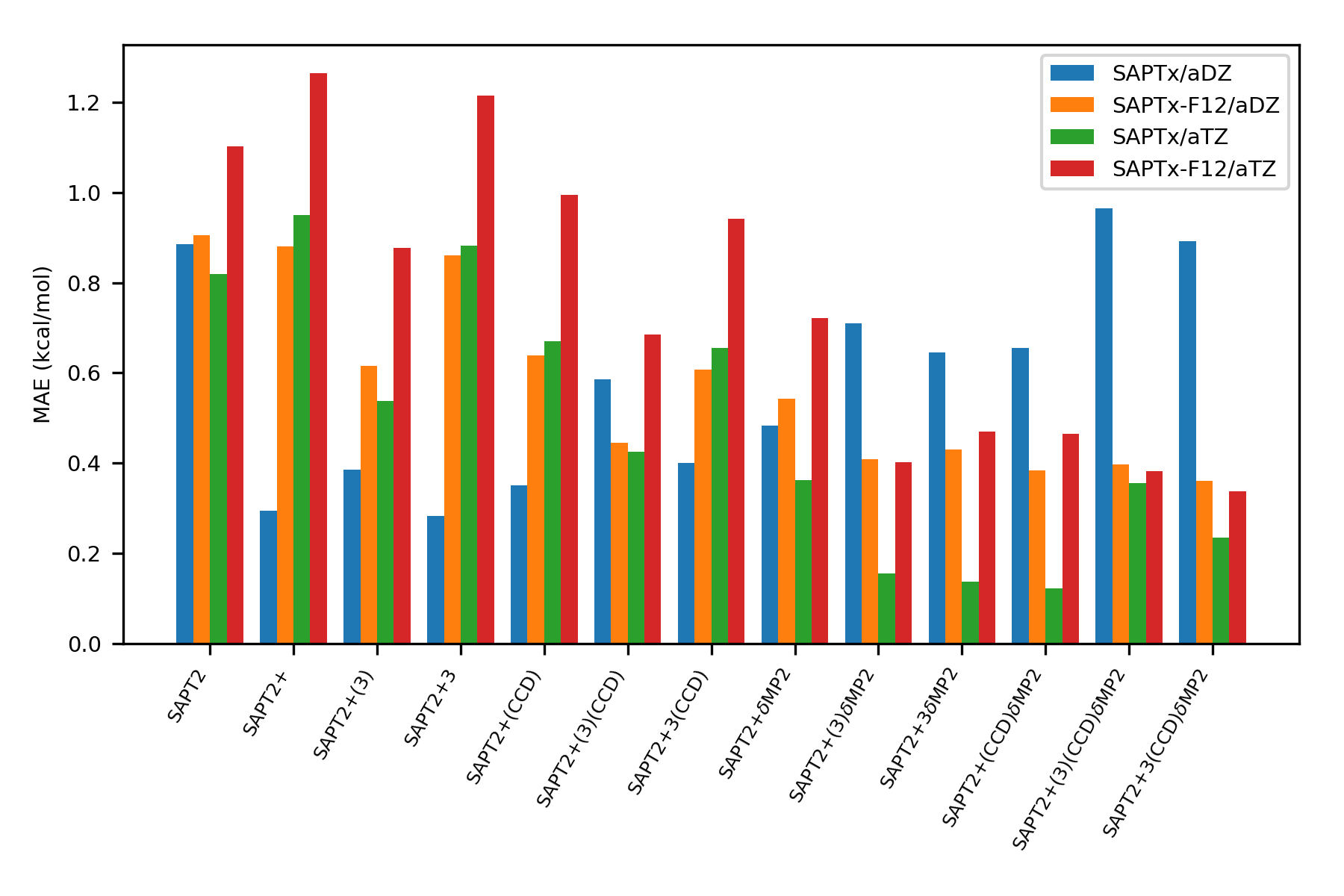}
\end{figure}

\begin{figure}[!h]
	\centering
	\caption{Mean absolute percent errors (MA\%E) of the interaction energy averaged over the S22, HBC6, NBC10, and HSG databases for SAPT and SAPT-F12 in aDZ and aTZ basis sets.}
	\label{figure:overall_mare}
	\includegraphics[height=0.42\textheight]{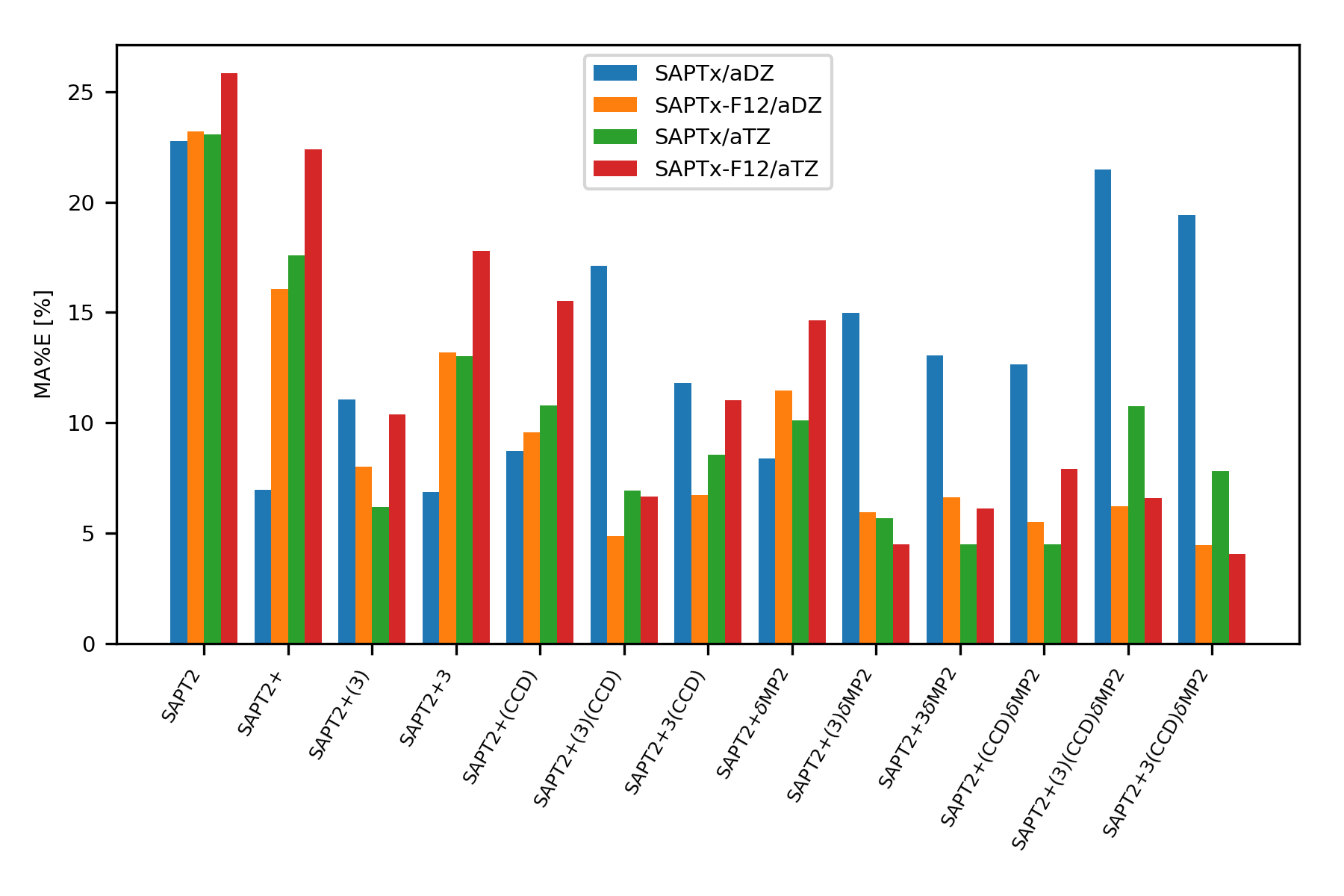}
\end{figure}

One of the significant advantages of energy decomposition methods such as SAPT is that various perturbation corrections can be studied separately in different basis sets. This allows us to reach a CBS limit of a particular correction independently from others. Specifically, the small-basis 
$E^{(20)}_{\rm disp}$ and $E^{(20)}_{\rm exch-disp}$ energies can be substituted by their  F12 counterparts in all wave function-based SAPT levels defined in Ref.~\onlinecite{Parker:14}, giving rise to new SAPT variants that will be denoted below as SAPT0-F12, SAPT2-F12, SAPT2+-F12, SAPT2+(3)-F12, and SAPT2+3-F12. 
Such a procedure is valid (no double counting occurs) and can be considered as a ``focal point" approach similar to MP2/CBS+$\delta$CCSD(T), where the lower-level estimate, converged to CBS (or at least much closer to CBS thanks to the F12 approach), is improved by a higher-level estimate in a moderate basis.
To illustrate the effect of this F12 modification on the accuracy of total SAPT interaction energies, 
Figure~\ref{figure:overall_mae} displays the mean absolute errors (MAE) of interaction energy, relative to the CCSD(T)/CBS benchmarks, averaged over the S22 \cite{Jurecka:06,Marshall:11}, HBC6 \cite{Thanthiriwatte:11}, NBC10 \cite{Sherrill:09}, and HSG \cite{Faver:11} databases for standard and explicitly correlated SAPT methods, with and without the CCD dispersion and the ``$\delta$MP2"  correction. 
All MAE values reported below refer to this database combination and are fully compatible with the errors obtained and displayed, on the very same dataset, in Ref.~\onlinecite{Parker:14}; see Sec.~\ref{sec:details} for some technical details (truncation of the HBC6 and NBC10 databases, scaling of the $E^{(20)}_{\rm exch-disp}$ and $E^{(20)}_{\rm exch-disp}$-F12 corrections) that are essential for maintaining this compatibility.
The MAE data tend to emphasize the performance on the database(s) with the largest magnitude of interaction energies (in this case, the HBC6 dataset of doubly hydrogen bonded complexes); to provide a more balanced view, Fig.~\ref{figure:overall_mare} complements Fig.~\ref{figure:overall_mae} by showing the respective 
mean absolute percent errors (MA\%E).
The corresponding statistical errors (MAE and MA\%E) for the four individual databases are presented in Figures S15--S22 in the Supporting Information. 

At first, one could think that the F12 dispersion aims to improve the SAPT0 interaction energies. However, it is just the opposite: it makes them less accurate, leading to the MAE values of 2.46 kcal/mol and 2.58 kcal/mol in aDZ and aTZ, respectively, as compared to 1.74 and 2.34 kcal/mol for conventional SAPT0.  
This behavior was foreseeable since the $E^{(20)}_{\rm disp}$ correction tends to overestimate the dispersion binding and it becomes even more negative when approaching the CBS limit. 
It has been demonstrated that the success of SAPT0 relies on error cancellation between the overestimation of dispersion by its leading $E^{(20)}_{\rm disp}$ term and the underestimation of dispersion by a small ``calendar" basis set, such as jun-cc-pVDZ \cite{Parker:14}. Therefore, the simplest SAPT0 approach is not recommended to be combined with the F12 dispersion, and it is excluded from further studies. 

The SAPT2 variant, which extends SAPT0 by including intramolecular electron correlation up to second order for electrostatic, exchange, and induction interactions, was the subject of the first tests with the F12 treatment. 
The poor performance of  SAPT2-F12  can be assigned to the lack of intramolecular correlation in the dispersion corrections. 
At this level, the F12 treatment has a negative effect on the accuracy of SAPT2, increasing the average errors in aDZ and aTZ by 0.02 kcal/mol and 0.28 kcal/mol, respectively.

When the F12 dispersion and exchange dispersion is added to SAPT2+/aDZ, the first SAPT level which includes intramolecular electron correlation for dispersion up to second order ($E^{(21)}_{\rm disp}$ and $E^{(22)}_{\rm disp}$), the errors increase quite dramatically. It is worth noting that SAPT2+/aDZ was established in Ref.~\onlinecite{Parker:14} as the silver standard of SAPT with a MAE of 0.30 kcal/mol.
Such an attractive accuracy is linked to a fortuitous error cancellation between the method error and basis set incompleteness, which no longer happens for SAPT2+-F12. Interestingly, the SAPT2+-F12/aTZ approach produces
the MAE value of 1.27 kcal/mol, which is the largest error across all considered SAPT-F12 options.

The SAPT2+(3)-F12 method exhibits better performance than SAPT2+-F12 in both aDZ and aTZ, affording errors of  0.62 kcal/mol and 0.88 kcal/mol, respectively. However, the F12 treatment still worsens the accuracy of standard SAPT2+(3) with MAEs of 0.39 kcal/mol and 0.54 kcal/mol in the same bases. 
The approach with the complete third-order correction, SAPT2+3-F12, does not outperform SAPT2+(3)-F12 either in aDZ or aTZ, leading to the MAE values of 0.86 kcal/mol and 1.22 kcal/mol, respectively.  
It follows that the higher accuracy of SAPT2+(3)-F12, just like for the non-F12 approach, is based on an error cancellation that occurs when the third-order mixed induction-dispersion effects are neglected.

It has been shown that the CCD-based construction of dispersion amplitudes in SAPT, even though not as remarkably accurate as the full CCSD treatment of dispersion \cite{Korona:08,Korona:13}, usually provides an improved description of noncovalent interactions \cite{Williams:95a,Parrish:13,Parker:14}. In this spirit, the amplitudes in the second-order dispersion energy terms ($E^{(20)}_{\rm disp}+E^{(21)}_{\rm disp}+E^{(22)}_{\rm disp}$) 
are replaced by the converged CCD amplitudes \cite{Williams:95a}, giving rise to SAPT(CCD) (this dispersion algorithm was originally denoted, more precisely, as CCD+ST(CCD) as it also includes a perturbative estimate of the singles and triples contributions to $E^{(22)}_{\rm disp}$ using converged doubles amplitudes). 
Another route to enhance the accuracy of the SAPT calculations, especially for hydrogen-bonded systems, is to include the ``$\delta$MP2" correction. This term  is computed as a difference between the counterpoise (CP) corrected MP2 interaction energy  and the SAPT2 interaction energy:
\begin{equation}
\delta {\rm MP2} = E^{\rm MP2}_{\rm int}-E^{\rm SAPT2}_{\rm int}
\end{equation}
and it accounts for third- as well as higher-order coupling between induction and dispersion \cite{Parker:14,Patkowski:20}. Both improvements, (CCD) and ``$\delta$MP2", can be employed within SAPT2+, SAPT2+(3), and SAPT2+3 together or separately. 
For all these SAPT variants, the effect of replacing standard $E^{(20)}_{\rm disp}$ and $E^{(20)}_{\rm exch-disp}$ by our nearly converged $E^{(20)}_{\rm disp}$-F12 and $E^{(20)}_{\rm exch-disp}$-F12 values was also tested.

Figure~\ref{figure:overall_mae} reveals that the more robust treatment of dispersion achieved by the CCD amplitudes reduces the overall errors at all high levels of SAPT.  We again observe that SAPT2+(CCD)-F12 and SAPT2+3(CCD)-F12 perform nearly equivalent in both basis sets, and worsen results relative to SAPT2+(3)(CCD)-F12.
The latter flavor leads to an enhancement in aDZ with respect to the standard counterpart, showing a MAE value of 0.45 kcal/mol, albeit the performance is deteriorated in aTZ, producing an average error of 0.69 kcal/mol. 

Once the $\delta$MP2 term is added to the SAPT calculations, we observe that SAPT2+-F12$\delta$MP2 still slightly underperforms SAPT2+$\delta$MP2 in aDZ, giving rise to an error of 0.54 kcal/mol, while  SAPT2+(3)-F12$\delta$MP2/aDZ  and SAPT2+3-F12$\delta$MP2/aDZ  substantially improve the accuracy of their non-F12 counterparts, yielding the MAE values of  0.41 kcal/mol and 0.43 kcal/mol, respectively. 
These values should be compared with the MAE of 0.48 kcal/mol, 0.71 kcal/mol, and 0.64 kcal/mol obtained for SAPT2+$\delta$MP2, SAPT2+(3)$\delta$MP2,  and SAPT2+3$\delta$MP2, respectively. 
The inclusion of the F12 terms spoils the excellent accuracy of SAPT2+$\delta$MP2, SAPT2+(3)$\delta$MP2,  and SAPT2+3$\delta$MP2  in aTZ, increasing the errors by 0.36 kcal/mol, 0.24 kcal/mol, and 0.33 kcal/mol, respectively. 
The F12 treatment combined with both CCD dispersion and the $\delta$MP2 correction has a highly positive influence on the accuracy in the aDZ basis, reducing the non-F12 errors from 0.66 to 0.38 kcal/mol for SAPT2+(CCD)-F12$\delta$MP2, from 0.97 to 0.40 kcal/mol for SAPT2+(3)(CCD)-F12$\delta$MP2, and  from 0.89 to 0.36 kcal/mol for SAPT2+3(CCD)-F12$\delta$MP2.
Once the basis set size is enlarged, we observe that SAPT2+(CCD)-F12$\delta$MP2/aTZ is not nearly as accurate as SAPT2+(CCD)$\delta$MP2/aTZ (the error increases from 0.12 kcal/mol to 0.47 kcal/mol). At the highest levels of theory, the F12 treatment still increases the MAE, but only by a small amount: from 0.36 kcal/mol to 0.38 kcal/mol for  SAPT2+(3)(CCD)$\delta$MP2/aTZ and from 0.24 kcal/mol to 0.34 kcal/mol for SAPT2+3(CCD)$\delta$MP2/aTZ.

A closer examination of the SAPT and SAPT-F12 errors on the individual databases (presented in the Supporting Information) reveals that the overall MAE trends follow closely the behavior observed for the dataset exhibiting the strongest binding, that is, HBC6. This happens to be the subset of the overall database for which including the F12 correction in a high-level SAPT treatment such as SAPT2+3(CCD)$\delta$MP2 performs particularly poorly (note that, at this level of SAPT, the F12 inclusion is uniformly beneficial in both aDZ and aTZ for the S22, NBC10, and HSG databases). To alleviate the dominance of the HBC6 set over the overall error statistics, 
Fig.~\ref{figure:overall_mare} presents the relative errors (MA\%E values) computed on the entire dataset.
For this statistical metric, F12 still does not provide uniform improvement of the lower theory levels of SAPT. 
However, at the highest levels, SAPT2+(3)(CCD)$\delta$MP2 and SAPT2+3(CCD)$\delta$MP2, the inclusion of the F12 term reduces the average relative errors in both aDZ and aTZ, and the lowest MA\%E value for all approaches (albeit by a very small margin) is obtained for the formally highest-level treatment, SAPT2+3(CCD)-F12$\delta$MP2/aTZ. The origins of the relatively poor performance of the highest-level SAPT-F12 variants on the strongly hydrogen bonded HBC6 set warrant further investigation, however, it is likely that a better basis set convergence of dispersion and exchange dispersion impedes a cancellation of errors coming from some non-dispersion SAPT terms. 

Overall, the performance of different SAPT variants in the aDZ basis is a net result of the basis set incompleteness errors and the intrinsic errors of a given theory level. 
Only when the latter errors are sufficiently reduced (most importantly, by including the $\delta$MP2 correction), the basis set incompleteness effects on SAPT in general, and on the dispersion and exchange-dispersion energy in particular, become the single dominant factor limiting the accuracy of the aDZ-based treatment. 
In such a case, the improvement of SAPT-F12/aDZ over the corresponding SAPT/aDZ variant, observed in Fig.~\ref{figure:overall_mae}, is quite remarkable. 
When the basis set is enlarged to aTZ, the incompleteness errors are reduced to a magnitude similar to the remaining intrinsic errors of the method. In such a case, the F12 approach, while significantly limiting one source of errors, does not provide an improvement in the accuracy of total interaction energies (it actually somewhat worsens the MAE by disturbing the error cancellation).
In particular, the accuracy of SAPT2+(3)$\delta$MP2/aTZ, which was established as the ``gold standard" of SAPT thanks to a very favorable accuracy-to-cost ratio (the accuracy is similar but the computational cost is $\sim$50\% less than that of SAPT2+(CCD)$\delta$MP2/aTZ \cite{Parker:14}), cannot be beaten by the SAPT-F12 treatment at this stage. 
However, it is likely that further enhancements, in particular, an extension of the F12 formalism to higher-order dispersion corrections including the effects of intramolecular correlation, will provide additional improvement to the SAPT-F12 accuracy by strongly reducing another source of residual errors. Such an advancement is the subject of ongoing research in our group.

\subsection{Computational cost of DF-SAPT0-F12}

\begin{figure}[!h]
	\centering
	\caption{The comparison of DF-SAPT0 and DF-SAPT0-F12 calculation times (in hours) for the benzene dimer (from the S22 database \cite{Jurecka:06}) on a single core of a 2.4 GHz Intel Xeon E5-2680 v4 machine
using the {\sc Psi4} quantum chemistry program and our {\sc Psi4NumPy} implementation. The consecutive basis sets listed for each bar pertain to the AO and DF sets for DF-SAPT0 and to the AO, CABS, and DF sets for DF-SAPT0-F12.}
	\label{fig:timings}
	\includegraphics[height=0.42\textheight]{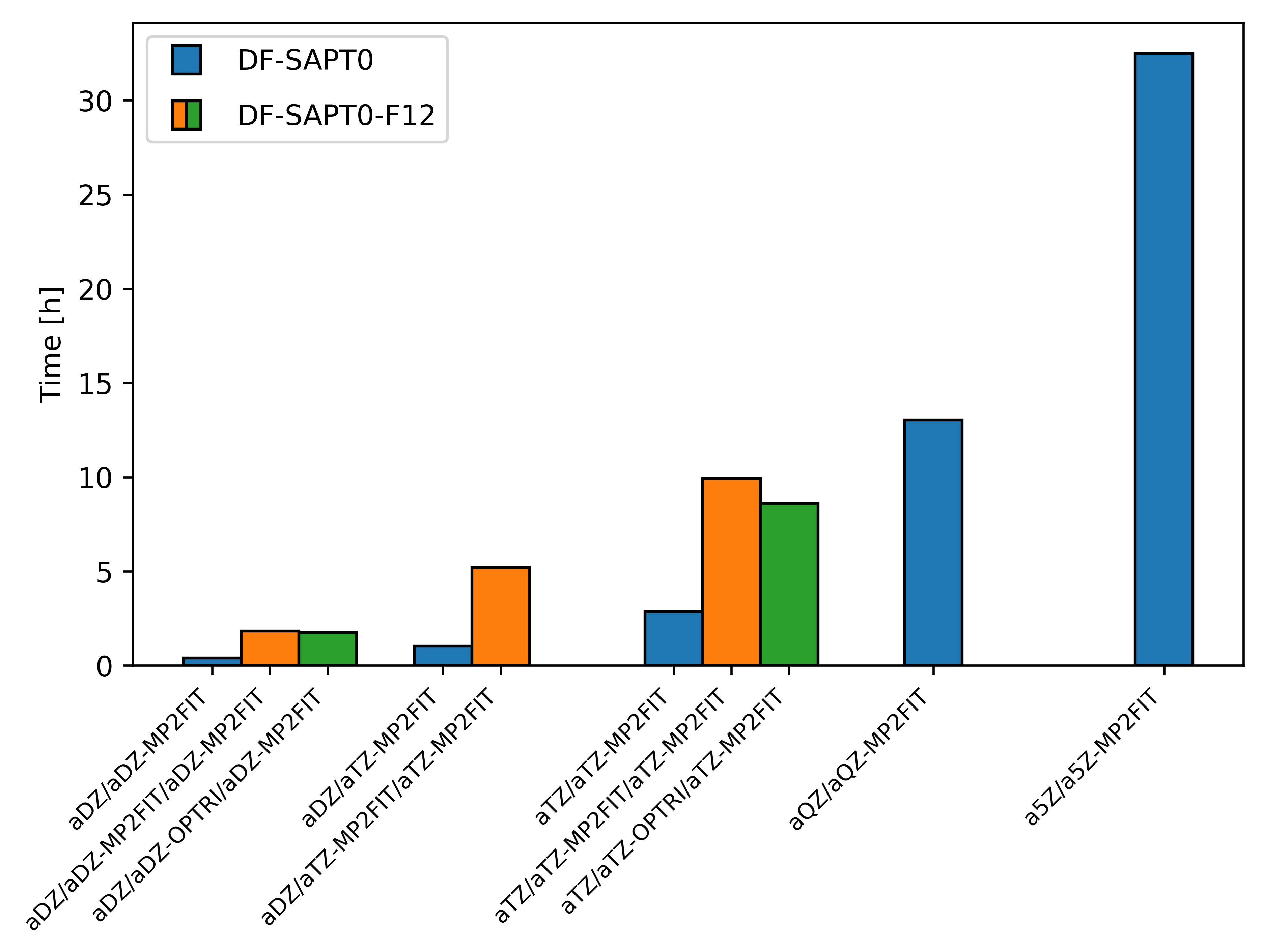}
\end{figure}

To conclude this section, we provide some evidence that, even though our current implementation is certainly not a fully optimized production code, a DF-SAPT0-F12 calculation is already more efficient than a standard DF-SAPT0 one in a larger orbital basis required to attain the same accuracy of $E^{(20)}_{\rm disp}$ and $E^{(20)}_{\rm exch-disp}$. To this end, Fig.~\ref{fig:timings} compares the wall timings of DF-SAPT0 (with the highly efficient {\sc Psi4} code \cite{Smith:20}) and DF-SAPT0-F12 (using our {\sc Psi4NumPy} \cite{Smith:18} code for the F12 part in addition to a complete {\sc Psi4} DF-SAPT0 calculation), obtained for the benzene dimer (from the S22 database \cite{Jurecka:06}) on a single core of a 2.4 GHz Intel Xeon E5-2680 v4 machine. The aDZ timings in Fig.~\ref{fig:timings} include both calculations with the recommended set of auxiliary bases (aTZ-MP2FIT for DF-SAPT0 and aTZ-RIFIT/aTZ-MP2FIT for DF-SAPT0-F12) and the ones utilizing reduced aDZ-RIFIT/aDZ-MP2FIT sets for CABS and DF. 
We see that a DF-SAPT0-F12 calculation currently takes a few times longer than a conventional DF-SAPT0 one in the same basis, with DF-SAPT0-F12/a$X$Z/a$X$Z-RIFIT slightly less expensive than DF-SAPT0/a$(X+1)$Z. 
This indicates that the F12 approach is already a viable alternative to conventional SAPT0 as far as performance is concerned. 
It should be stressed that our implementation, while completely density fitted, is quite preliminary, and the performance will still be improved when our code is merged into the new release of {\sc Psi4} that features a new F12 integral library \cite{Smith:20}. Moreover, the overhead associated with calculating $E^{(20)}_{\rm disp}$-F12 and $E^{(20)}_{\rm exch-disp}$-F12 is minor compared to the cost of computing correlated SAPT corrections in case the F12 effects are meant to augment a higher-level SAPT result such as SAPT2+3.
 
\section{Summary}

In this work, we have transformed the explicitly correlated SAPT0 corrections $E^{(20)}_{\rm disp}$-F12 and $E^{(20)}_{\rm exch-disp}$-F12, proposed recently in Ref.~\onlinecite{Kodrycka:19a} (and, in part, earlier in Ref.~\onlinecite{Przybytek:18}), into a practical closed-shell SAPT enhancement, significantly improving the basis set convergence relative to conventional $E^{(20)}_{\rm disp}$ and $E^{(20)}_{\rm exch-disp}$. While the proof-of-concept SAPT0-F12 implementation of Ref.~\onlinecite{Kodrycka:19a} employed resolution of identity to avoid three-electron integrals, it was not competitive to the conventional SAPT0 approach due to the need to explicitly consider many different two-electron integrals over a combination of orbital and complementary auxiliary basis indices. 
In the present work, all the two-electron integrals have been decomposed into three-index quantities by means of a robust density fitting, in which the error in the fitted integral decays quadratically with the errors in the fitted densities \cite{Manby:03}. Thus, our development of DF-$E^{(20)}_{\rm disp}$-F12 bears many similarities to the DF-MP2-F12 formalism, and we employ the so-called {\em approximation 3C} in the calculation of the F12 intermediates in exactly the same way as in the MP2-F12 formulation of Ref.~\onlinecite{Werner:07}. 
Out of several {\em Ans\"atze} proposed for the explicitly correlated dispersion amplitudes in Ref.~\onlinecite{Kodrycka:19a}, we employ the ODA variant which has been shown to provide an excellent approximation to the results obtained using fully optimized amplitudes, but without the steep scaling and numerical instabilities of the latter.

Numerical tests for several small complexes indicate that the DF approximation to $E^{(20)}_{\rm disp}$-F12 and $E^{(20)}_{\rm exch-disp}$-F12 works very well, with virtually no loss in accuracy as long as the frozen-core approximation is applied. For an a$X$Z orbital basis, the standard a$X$Z-MP2FIT choice for the DF basis is very appropriate, except that the aDZ/aTZ-MP2FIT combination is slightly superior to aDZ/aDZ-MP2FIT. Similar to explicitly correlated electronic structure methods such as MP2-F12 and CCSD(T)-F12, a separate DF basis (from the a$X$Z-JKFIT family) can be used for the fitting of the Fock and exchange matrices appearing in some F12 intermediates, however, the results are of exactly the same quality as when the a$X$Z-MP2FIT set is used in all DF contexts.

Further tests on the entire A24 database of small noncovalent complexes \cite{Rezac:13a} confirm that the errors introduced by the DF approximation are insignificant in all cases. However, thanks to the vastly superior computational performance of DF-SAPT0-F12 over the non-DF version, we were able to extend the aDZ and aTZ calculations of Ref.~\onlinecite{Kodrycka:19a} to basis sets as large as a5Z+(a5Z), that is, a5Z augmented by a set of midbond functions. 
The resulting $E^{(20)}_{\rm disp}$-F12 and $E^{(20)}_{\rm exch-disp}$-F12 values are so well converged to the CBS limit that better reference values than those in Ref.~\onlinecite{Kodrycka:19a} became necessary. Thus, we compared the $E^{(20)}_{\rm disp}$-F12 and $E^{(20)}_{\rm exch-disp}$-F12 data for the A24 complexes with conventional $E^{(20)}_{\rm disp}$ and $E^{(20)}_{\rm exch-disp}$ values obtained by the (a5Z+(a5Z),a6Z+(a6Z)) extrapolation, or even (a6Z+(a6Z),a7Z+(a7Z)) for the water dimer. 
However, the CBS convergence of the SAPT-F12 corrections is so fast that nearly all $E^{(20)}_{\rm disp}$-F12/aQZ values are within the tight error bars of the benchmark, and the $E^{(20)}_{\rm exch-disp}$-F12/aQZ data are not far outside the benchmark range. The only exception are the two complexes containing an argon atom for which the convergence is somewhat slower. 
The CBS convergence of the SAPT-F12 corrections can be further enhanced by the addition of midbond functions. In particular, the +(3322) set of bond functions makes even the $E^{(20)}_{\rm disp}$-F12/aDZ values highly accurate, very often (perhaps fortuitously) within the error bars of the benchmark.

Having established that the F12 approach significantly speeds up the basis set convergence of the individual $E^{(20)}_{\rm disp}$ and $E^{(20)}_{\rm exch-disp}$ corrections, we turned our attention to the total SAPT interaction energies. In this case, a replacement of the conventional $E^{(20)}_{\rm disp}+E^{(20)}_{\rm exch-disp}$ value by its F12 counterpart
corresponds to combining nearly converged $E^{(20)}_{\rm disp}$ and $E^{(20)}_{\rm exch-disp}$ data with the remaining SAPT corrections computed in a moderate basis, in the spirit of the focal-point MP2/CBS+$\delta$CCSD(T) approach. In this way, we constructed F12-enhanced variants of all standard SAPT levels from SAPT0 to SAPT2+3(CCD)$\delta$MP2 \cite{Parker:14} and examined the SAPT accuracy, relative to the CCSD(T)/CBS benchmark data, on the same set of complexes as in Ref.~\onlinecite{Parker:14}.
We found that low levels of SAPT do not exhibit improvement when $E^{(20)}_{\rm disp}$-F12 and $E^{(20)}_{\rm exch-disp}$-F12 are used in place of $E^{(20)}_{\rm disp}$ and $E^{(20)}_{\rm exch-disp}$: in particular, the performance of SAPT0 deteriorates due to a breakdown of the error cancellation between the theory level and basis set incompleteness.
Only when the method errors are minimized (which happens at high levels such as SAPT2+(3) or SAPT2+3 including the $\delta$MP2 correction), the basis set incompleteness effects determine the overall accuracy. Under those circumstances, SAPT-F12/aDZ is much more accurate than the corresponding SAPT/aDZ approach.
In the aTZ basis, incompleteness errors no longer dominate the picture and the F12 and non-F12 variants of SAPT exhibit similar average deviations from the benchmark CCSD(T)/CBS data.

\section*{Associated Content}

{\bf Supporting Information.} Extended versions of Tables \ref{table:H2OH2O}--\ref{table:Ch4Ch4} including data for noble gas dimers, results with midbond functions, values obtained with the a$X$Z-JKFIT auxiliary basis in place of a$X$Z-MP2FIT in the construction of Fock and exchange matrices, values computed with the a$X$Z-OPTRI CABS sets, and the $E^{(20)}_{\rm disp}+E^{(20)}_{\rm exch-disp}$ sums. Figures displaying the basis set convergence of $E^{(20)}_{\rm disp}$-F12, $E^{(20)}_{\rm exch-disp}$-F12, and the sum of the two for all individual A24 complexes. Mean absolute errors and mean absolute percent errors of different SAPT and SAPT-F12 variants for the individual databases (S22, HBC6, NBC10, and HSG). 
This material 
is available free of charge
via the Internet at http://pubs.acs.org/.

\section*{Acknowledgments}

This work was supported by the U.S. National Science Foundation (NSF) awards CHE-1351978 and CHE-1955328.
M. K. acknowledges a fellowship from the Molecular Sciences Software Institute (MolSSI) under NSF Grant
ACI-1547580. We thank Dr. Wim Klopper and Dr. Christof Holzer for helpful discussions.


\providecommand{\latin}[1]{#1}
\makeatletter
\providecommand{\doi}
  {\begingroup\let\do\@makeother\dospecials
  \catcode`\{=1 \catcode`\}=2 \doi@aux}
\providecommand{\doi@aux}[1]{\endgroup\texttt{#1}}
\makeatother
\providecommand*\mcitethebibliography{\thebibliography}
\csname @ifundefined\endcsname{endmcitethebibliography}
  {\let\endmcitethebibliography\endthebibliography}{}

\newpage

\begin{figure}[h]
   \centering
   \includegraphics[width = \textwidth]{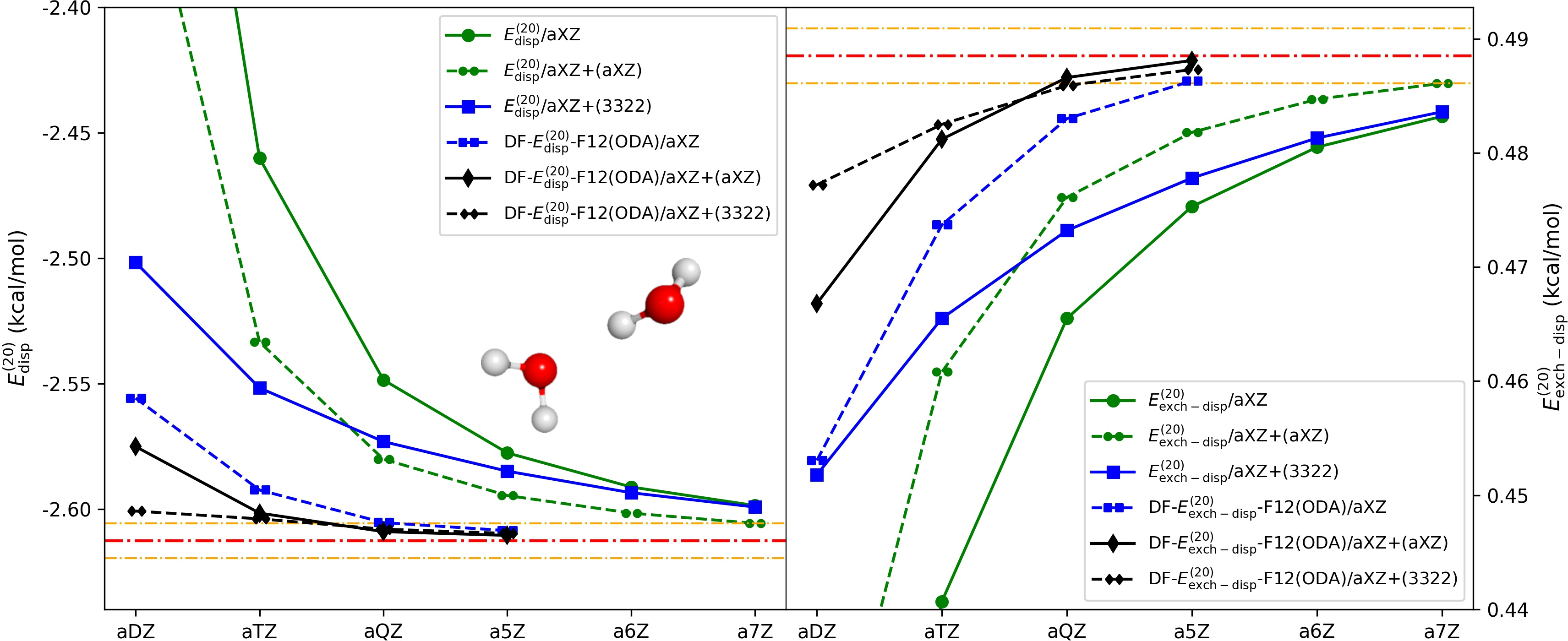}
   \caption{TOC Graphic}
\label{fig:toc}
\end{figure}


\begin{mcitethebibliography}{86}
\providecommand*\natexlab[1]{#1}
\providecommand*\mciteSetBstSublistMode[1]{}
\providecommand*\mciteSetBstMaxWidthForm[2]{}
\providecommand*\mciteBstWouldAddEndPuncttrue
  {\def\EndOfBibitem{\unskip.}}
\providecommand*\mciteBstWouldAddEndPunctfalse
  {\let\EndOfBibitem\relax}
\providecommand*\mciteSetBstMidEndSepPunct[3]{}
\providecommand*\mciteSetBstSublistLabelBeginEnd[3]{}
\providecommand*\EndOfBibitem{}
\mciteSetBstSublistMode{f}
\mciteSetBstMaxWidthForm{subitem}{(\alph{mcitesubitemcount})}
\mciteSetBstSublistLabelBeginEnd
  {\mcitemaxwidthsubitemform\space}
  {\relax}
  {\relax}

\bibitem[Klein \latin{et~al.}(2017)Klein, Shagam, Skomorowski, \.Zuchowski,
  Pawlak, Janssen, Moiseyev, {van de Meerakker}, {van der Avoird}, Koch, and
  Narevicius]{Klein:17}
Klein,~A.; Shagam,~Y.; Skomorowski,~W.; \.Zuchowski,~P.~S.; Pawlak,~M.;
  Janssen,~L. M.~C.; Moiseyev,~N.; {van de Meerakker},~S. Y.~T.; {van der
  Avoird},~A.; Koch,~C.~P.; Narevicius,~E. Directly probing anisotropy in
  atom-molecule collisions through quantum scattering resonances. \emph{Nature
  Phys.} \textbf{2017}, \emph{13}, 35--38\relax
\mciteBstWouldAddEndPuncttrue
\mciteSetBstMidEndSepPunct{\mcitedefaultmidpunct}
{\mcitedefaultendpunct}{\mcitedefaultseppunct}\relax
\EndOfBibitem
\bibitem[Beran(2016)]{Beran:16}
Beran,~G. J.~O. Modeling Polymorphic Molecular Crystals with Electronic
  Structure Theory. \emph{Chem. Rev.} \textbf{2016}, \emph{116},
  5567--5613\relax
\mciteBstWouldAddEndPuncttrue
\mciteSetBstMidEndSepPunct{\mcitedefaultmidpunct}
{\mcitedefaultendpunct}{\mcitedefaultseppunct}\relax
\EndOfBibitem
\bibitem[Kraka \latin{et~al.}(2013)Kraka, Freindorf, and Cremer]{Kraka:13}
Kraka,~E.; Freindorf,~M.; Cremer,~D. Chiral Discrimination by Vibrational
  Spectroscopy Utilizing Local Modes. \emph{Chirality} \textbf{2013},
  \emph{25}, 185--196\relax
\mciteBstWouldAddEndPuncttrue
\mciteSetBstMidEndSepPunct{\mcitedefaultmidpunct}
{\mcitedefaultendpunct}{\mcitedefaultseppunct}\relax
\EndOfBibitem
\bibitem[Hobza \latin{et~al.}(1996)Hobza, Selzle, and Schlag]{Hobza:96}
Hobza,~P.; Selzle,~H.~L.; Schlag,~E.~W. Potential Energy Surface for the
  Benzene Dimer. Results of ab Initio {CCSD(T)} Calculations Show Two Nearly
  Isoenergetic Structures: {T}-Shaped and Parallel-Displaced. \emph{J. Phys.
  Chem.} \textbf{1996}, \emph{100}, 18790--18794\relax
\mciteBstWouldAddEndPuncttrue
\mciteSetBstMidEndSepPunct{\mcitedefaultmidpunct}
{\mcitedefaultendpunct}{\mcitedefaultseppunct}\relax
\EndOfBibitem
\bibitem[Kodrycka and Patkowski(2019)Kodrycka, and Patkowski]{Kodrycka:19}
Kodrycka,~M.; Patkowski,~K. Platinum, gold, and silver standards of
  intermolecular interaction energy calculations. \emph{J. Chem. Phys.}
  \textbf{2019}, \emph{151}, 070901\relax
\mciteBstWouldAddEndPuncttrue
\mciteSetBstMidEndSepPunct{\mcitedefaultmidpunct}
{\mcitedefaultendpunct}{\mcitedefaultseppunct}\relax
\EndOfBibitem
\bibitem[Raghavachari \latin{et~al.}(1989)Raghavachari, Trucks, Pople, and
  Head-Gordon]{Raghavachari:89}
Raghavachari,~K.; Trucks,~G.~W.; Pople,~J.~A.; Head-Gordon,~M. A 5th-Order
  Perturbation Comparison of Electron Correlation Theories. \emph{Chem. Phys.
  Lett.} \textbf{1989}, \emph{157}, 479--483\relax
\mciteBstWouldAddEndPuncttrue
\mciteSetBstMidEndSepPunct{\mcitedefaultmidpunct}
{\mcitedefaultendpunct}{\mcitedefaultseppunct}\relax
\EndOfBibitem
\bibitem[Halkier \latin{et~al.}(1998)Halkier, Helgaker, {J\o}rgensen, Klopper,
  Koch, Olsen, and Wilson]{Halkier:98}
Halkier,~A.; Helgaker,~T.; {J\o}rgensen,~P.; Klopper,~W.; Koch,~H.; Olsen,~J.;
  Wilson,~A.~K. Basis-Set Convergence in Correlated Calculations on {Ne, N$_2$,
  and H$_2$O}. \emph{Chem. Phys. Lett.} \textbf{1998}, \emph{286},
  243--252\relax
\mciteBstWouldAddEndPuncttrue
\mciteSetBstMidEndSepPunct{\mcitedefaultmidpunct}
{\mcitedefaultendpunct}{\mcitedefaultseppunct}\relax
\EndOfBibitem
\bibitem[Kutzelnigg and Klopper(1991)Kutzelnigg, and Klopper]{Kutzelnigg:91}
Kutzelnigg,~W.; Klopper,~W. Wave functions with terms linear in the
  interelectronic coordinates to take care of the correlation cusp. I. General
  theory. \emph{J. Chem. Phys.} \textbf{1991}, \emph{94}, 1985--2001\relax
\mciteBstWouldAddEndPuncttrue
\mciteSetBstMidEndSepPunct{\mcitedefaultmidpunct}
{\mcitedefaultendpunct}{\mcitedefaultseppunct}\relax
\EndOfBibitem
\bibitem[Tew and Klopper(2005)Tew, and Klopper]{Tew:05}
Tew,~D.~P.; Klopper,~W. New correlation factors for explicitly correlated
  electronic wave functions. \emph{J. Chem. Phys.} \textbf{2005}, \emph{123},
  074101\relax
\mciteBstWouldAddEndPuncttrue
\mciteSetBstMidEndSepPunct{\mcitedefaultmidpunct}
{\mcitedefaultendpunct}{\mcitedefaultseppunct}\relax
\EndOfBibitem
\bibitem[Werner \latin{et~al.}(2007)Werner, Adler, and Manby]{Werner:07}
Werner,~H.-J.; Adler,~T.~B.; Manby,~F.~R. General orbital invariant MP2-F12
  theory. \emph{J. Chem. Phys.} \textbf{2007}, \emph{126}, 164102\relax
\mciteBstWouldAddEndPuncttrue
\mciteSetBstMidEndSepPunct{\mcitedefaultmidpunct}
{\mcitedefaultendpunct}{\mcitedefaultseppunct}\relax
\EndOfBibitem
\bibitem[Shiozaki \latin{et~al.}(2008)Shiozaki, Kamiya, Hirata, and
  Valeev]{Shiozaki:08}
Shiozaki,~T.; Kamiya,~M.; Hirata,~S.; Valeev,~E.~F. Explicitly correlated
  coupled-cluster singles and doubles method based on complete diagrammatic
  equations. \emph{J. Chem. Phys.} \textbf{2008}, \emph{129}, 071101\relax
\mciteBstWouldAddEndPuncttrue
\mciteSetBstMidEndSepPunct{\mcitedefaultmidpunct}
{\mcitedefaultendpunct}{\mcitedefaultseppunct}\relax
\EndOfBibitem
\bibitem[{K\"{o}hn}(2009)]{Kohn:09}
{K\"{o}hn},~A. Explicitly correlated connected triple excitations in
  coupled-cluster theory. \emph{J. Chem. Phys.} \textbf{2009}, \emph{130},
  131101\relax
\mciteBstWouldAddEndPuncttrue
\mciteSetBstMidEndSepPunct{\mcitedefaultmidpunct}
{\mcitedefaultendpunct}{\mcitedefaultseppunct}\relax
\EndOfBibitem
\bibitem[Shiozaki and Werner(2010)Shiozaki, and Werner]{Shiozaki:10}
Shiozaki,~T.; Werner,~H.-J. Second-order multireference perturbation theory
  with explicit correlation: CASPT2-F12. \emph{J. Chem. Phys.} \textbf{2010},
  \emph{133}, 141103\relax
\mciteBstWouldAddEndPuncttrue
\mciteSetBstMidEndSepPunct{\mcitedefaultmidpunct}
{\mcitedefaultendpunct}{\mcitedefaultseppunct}\relax
\EndOfBibitem
\bibitem[Shiozaki \latin{et~al.}(2011)Shiozaki, Knizia, and
  Werner]{Shiozaki:11}
Shiozaki,~T.; Knizia,~G.; Werner,~H.-J. Explicitly correlated multireference
  configuration interaction: MRCI-F12. \emph{J. Chem. Phys.} \textbf{2011},
  \emph{134}, 034113\relax
\mciteBstWouldAddEndPuncttrue
\mciteSetBstMidEndSepPunct{\mcitedefaultmidpunct}
{\mcitedefaultendpunct}{\mcitedefaultseppunct}\relax
\EndOfBibitem
\bibitem[Tao(2001)]{Tao:01}
Tao,~F.-M. Bond functions, basis set superposition errors and other practical
  issues with ab initio calculations of intermolecular potentials. \emph{Int.
  Rev. Phys. Chem.} \textbf{2001}, \emph{20}, 617--643\relax
\mciteBstWouldAddEndPuncttrue
\mciteSetBstMidEndSepPunct{\mcitedefaultmidpunct}
{\mcitedefaultendpunct}{\mcitedefaultseppunct}\relax
\EndOfBibitem
\bibitem[Williams \latin{et~al.}(1995)Williams, Mas, Szalewicz, and
  Jeziorski]{Williams:95}
Williams,~H.~L.; Mas,~E.~M.; Szalewicz,~K.; Jeziorski,~B. On the effectiveness
  of monomer-, dimer-, and bond-centered basis functions in calculations of
  intermolecular interaction energies. \emph{J. Chem. Phys.} \textbf{1995},
  \emph{103}, 7374--7391\relax
\mciteBstWouldAddEndPuncttrue
\mciteSetBstMidEndSepPunct{\mcitedefaultmidpunct}
{\mcitedefaultendpunct}{\mcitedefaultseppunct}\relax
\EndOfBibitem
\bibitem[Jeziorska \latin{et~al.}(2008)Jeziorska, Cencek, Patkowski, Jeziorski,
  and Szalewicz]{Jeziorska:08}
Jeziorska,~M.; Cencek,~W.; Patkowski,~K.; Jeziorski,~B.; Szalewicz,~K. Complete
  Basis Set Extrapolations of Dispersion, Exchange, and Coupled-Clusters
  Contributions to the Interaction Energy: A Helium Dimer Study. \emph{Int. J.
  Quantum Chem.} \textbf{2008}, \emph{108}, 2053--2075\relax
\mciteBstWouldAddEndPuncttrue
\mciteSetBstMidEndSepPunct{\mcitedefaultmidpunct}
{\mcitedefaultendpunct}{\mcitedefaultseppunct}\relax
\EndOfBibitem
\bibitem[Dutta and Patkowski(2018)Dutta, and Patkowski]{Dutta:18}
Dutta,~N.~N.; Patkowski,~K. Improving `Silver-Standard' Benchmark Interaction
  Energies with Bond Functions. \emph{J. Chem. Theory Comput.} \textbf{2018},
  \emph{14}, 3053--3070\relax
\mciteBstWouldAddEndPuncttrue
\mciteSetBstMidEndSepPunct{\mcitedefaultmidpunct}
{\mcitedefaultendpunct}{\mcitedefaultseppunct}\relax
\EndOfBibitem
\bibitem[Stone(2013)]{Stone:13}
Stone,~A.~J. \emph{The Theory of Intermolecular Forces, Second Edition}; Oxford
  University Press, 2013\relax
\mciteBstWouldAddEndPuncttrue
\mciteSetBstMidEndSepPunct{\mcitedefaultmidpunct}
{\mcitedefaultendpunct}{\mcitedefaultseppunct}\relax
\EndOfBibitem
\bibitem[Jeziorski \latin{et~al.}(1994)Jeziorski, Moszy\'nski, and
  Szalewicz]{Jeziorski:94}
Jeziorski,~B.; Moszy\'nski,~R.; Szalewicz,~K. Perturbation Theory Approach to
  Intermolecular Potential Energy Surfaces of van der Waals Complexes.
  \emph{Chem. Rev.} \textbf{1994}, \emph{94}, 1887--1930\relax
\mciteBstWouldAddEndPuncttrue
\mciteSetBstMidEndSepPunct{\mcitedefaultmidpunct}
{\mcitedefaultendpunct}{\mcitedefaultseppunct}\relax
\EndOfBibitem
\bibitem[Patkowski(2020)]{Patkowski:20}
Patkowski,~K. Recent developments in symmetry-adapted perturbation theory.
  \emph{WIREs Comput. Mol. Sci.} \textbf{2020}, \emph{10}, e1452\relax
\mciteBstWouldAddEndPuncttrue
\mciteSetBstMidEndSepPunct{\mcitedefaultmidpunct}
{\mcitedefaultendpunct}{\mcitedefaultseppunct}\relax
\EndOfBibitem
\bibitem[Korona \latin{et~al.}(1997)Korona, Moszy{\'{n}}ski, and
  Jeziorski]{Korona:97a}
Korona,~T.; Moszy{\'{n}}ski,~R.; Jeziorski,~B. Convergence of symmetry-adapted
  perturbation theory for the interaction between helium atoms and between a
  hydrogen molecule and a helium atom. \emph{Adv. Quantum Chem.} \textbf{1997},
  \emph{28}, 171--188\relax
\mciteBstWouldAddEndPuncttrue
\mciteSetBstMidEndSepPunct{\mcitedefaultmidpunct}
{\mcitedefaultendpunct}{\mcitedefaultseppunct}\relax
\EndOfBibitem
\bibitem[Patkowski \latin{et~al.}(2001)Patkowski, Korona, and
  Jeziorski]{Patkowski:01}
Patkowski,~K.; Korona,~T.; Jeziorski,~B. Convergence behavior of the
  symmetry-adapted perturbation theory for states submerged in {P}auli
  forbidden continuum. \emph{J. Chem. Phys.} \textbf{2001}, \emph{115},
  1137--1152\relax
\mciteBstWouldAddEndPuncttrue
\mciteSetBstMidEndSepPunct{\mcitedefaultmidpunct}
{\mcitedefaultendpunct}{\mcitedefaultseppunct}\relax
\EndOfBibitem
\bibitem[Cha{\l}asi{\'n}ski and Szcz\c{e}\'sniak(1988)Cha{\l}asi{\'n}ski, and
  Szcz\c{e}\'sniak]{Chalasinski:88}
Cha{\l}asi{\'n}ski,~G.; Szcz\c{e}\'sniak,~M.~M. On the connection between the
  supermolecular {M{\o}ller-Plesset} treatment of the interaction energy and
  the perturbation theory of intermolecular forces. \emph{Mol. Phys.}
  \textbf{1988}, \emph{63}, 205--224\relax
\mciteBstWouldAddEndPuncttrue
\mciteSetBstMidEndSepPunct{\mcitedefaultmidpunct}
{\mcitedefaultendpunct}{\mcitedefaultseppunct}\relax
\EndOfBibitem
\bibitem[Rybak \latin{et~al.}(1991)Rybak, Jeziorski, and Szalewicz]{Rybak:91}
Rybak,~S.; Jeziorski,~B.; Szalewicz,~K. Many-body symmetry-adapted perturbation
  theory of intermolecular interactions - {H$_2$O and HF} dimers. \emph{J.
  Chem. Phys.} \textbf{1991}, \emph{95}, 6579--6601\relax
\mciteBstWouldAddEndPuncttrue
\mciteSetBstMidEndSepPunct{\mcitedefaultmidpunct}
{\mcitedefaultendpunct}{\mcitedefaultseppunct}\relax
\EndOfBibitem
\bibitem[Williams \latin{et~al.}(1995)Williams, Szalewicz, Moszy\'nski, and
  Jeziorski]{Williams:95a}
Williams,~H.~L.; Szalewicz,~K.; Moszy\'nski,~R.; Jeziorski,~B. Dispersion
  Energy in the Coupled Pair Approximation with Noniterative Inclusion of
  Single and Triple Excitations. \emph{J. Chem. Phys.} \textbf{1995},
  \emph{103}, 4586--4599\relax
\mciteBstWouldAddEndPuncttrue
\mciteSetBstMidEndSepPunct{\mcitedefaultmidpunct}
{\mcitedefaultendpunct}{\mcitedefaultseppunct}\relax
\EndOfBibitem
\bibitem[Korona and Jeziorski(2008)Korona, and Jeziorski]{Korona:08}
Korona,~T.; Jeziorski,~B. Dispersion energy from density-fitted density
  susceptibilities of singles and doubles coupled cluster theory. \emph{J.
  Chem. Phys.} \textbf{2008}, \emph{128}, 144107\relax
\mciteBstWouldAddEndPuncttrue
\mciteSetBstMidEndSepPunct{\mcitedefaultmidpunct}
{\mcitedefaultendpunct}{\mcitedefaultseppunct}\relax
\EndOfBibitem
\bibitem[Hesselmann \latin{et~al.}(2005)Hesselmann, Jansen, and
  Sch{\"u}tz]{Hesselmann:05}
Hesselmann,~A.; Jansen,~G.; Sch{\"u}tz,~M. Density-functional
  theory-symmetry-adapted intermolecular perturbation theory with density
  fitting: {A} new efficient method to study intermolecular interaction
  energies. \emph{J. Chem. Phys.} \textbf{2005}, \emph{122}, 014103\relax
\mciteBstWouldAddEndPuncttrue
\mciteSetBstMidEndSepPunct{\mcitedefaultmidpunct}
{\mcitedefaultendpunct}{\mcitedefaultseppunct}\relax
\EndOfBibitem
\bibitem[Misquitta \latin{et~al.}(2005)Misquitta, Podeszwa, Jeziorski, and
  Szalewicz]{Misquitta:05a}
Misquitta,~A.~J.; Podeszwa,~R.; Jeziorski,~B.; Szalewicz,~K. Intermolecular
  potentials based on symmetry-adapted perturbation theory including dispersion
  energies from time-dependent density functional calculations. \emph{J. Chem.
  Phys.} \textbf{2005}, \emph{123}, 214103\relax
\mciteBstWouldAddEndPuncttrue
\mciteSetBstMidEndSepPunct{\mcitedefaultmidpunct}
{\mcitedefaultendpunct}{\mcitedefaultseppunct}\relax
\EndOfBibitem
\bibitem[Holzer and Klopper(2017)Holzer, and Klopper]{Holzer:17a}
Holzer,~C.; Klopper,~W. Communication: Symmetry-adapted perturbation theory
  with intermolecular induction and dispersion energies from the Bethe-Salpeter
  equation. \emph{J. Chem. Phys.} \textbf{2017}, \emph{147}, 181101\relax
\mciteBstWouldAddEndPuncttrue
\mciteSetBstMidEndSepPunct{\mcitedefaultmidpunct}
{\mcitedefaultendpunct}{\mcitedefaultseppunct}\relax
\EndOfBibitem
\bibitem[Hapka \latin{et~al.}(2019)Hapka, Przybytek, and Pernal]{Hapka:19}
Hapka,~M.; Przybytek,~M.; Pernal,~K. Second-Order Dispersion Energy Based on
  Multireference Description of Monomers. \emph{J. Chem. Theory Comput.}
  \textbf{2019}, \emph{15}, 1016--1027\relax
\mciteBstWouldAddEndPuncttrue
\mciteSetBstMidEndSepPunct{\mcitedefaultmidpunct}
{\mcitedefaultendpunct}{\mcitedefaultseppunct}\relax
\EndOfBibitem
\bibitem[Korona(2009)]{Korona:09a}
Korona,~T. Exchange-Dispersion Energy: A Formulation in Terms of Monomer
  Properties and Coupled Cluster Treatment of Intramonomer Correlation.
  \emph{J. Chem. Theory Comput.} \textbf{2009}, \emph{5}, 2663--2678\relax
\mciteBstWouldAddEndPuncttrue
\mciteSetBstMidEndSepPunct{\mcitedefaultmidpunct}
{\mcitedefaultendpunct}{\mcitedefaultseppunct}\relax
\EndOfBibitem
\bibitem[{Sch\"{a}ffer} and Jansen(2013){Sch\"{a}ffer}, and
  Jansen]{Schaffer:13}
{Sch\"{a}ffer},~R.; Jansen,~G. Single-determinant-based symmetry-adapted
  perturbation theory without single-exchange approximation. \emph{Mol. Phys.}
  \textbf{2013}, \emph{111}, 2570--2584\relax
\mciteBstWouldAddEndPuncttrue
\mciteSetBstMidEndSepPunct{\mcitedefaultmidpunct}
{\mcitedefaultendpunct}{\mcitedefaultseppunct}\relax
\EndOfBibitem
\bibitem[Marchetti and Werner(2008)Marchetti, and Werner]{Marchetti:08}
Marchetti,~O.; Werner,~H.-J. Accurate calculations of intermolecular
  interaction energies using explicitly correlated wave functions. \emph{Phys.
  Chem. Chem. Phys.} \textbf{2008}, \emph{10}, 3400--3409\relax
\mciteBstWouldAddEndPuncttrue
\mciteSetBstMidEndSepPunct{\mcitedefaultmidpunct}
{\mcitedefaultendpunct}{\mcitedefaultseppunct}\relax
\EndOfBibitem
\bibitem[Marchetti and Werner(2009)Marchetti, and Werner]{Marchetti:09}
Marchetti,~O.; Werner,~H.-J. Accurate Calculations of Intermolecular
  Interaction Energies Using Explicitly Correlated Coupled Cluster Wave
  Functions and a Dispersion-Weighted MP2 Method. \emph{J. Phys. Chem. A}
  \textbf{2009}, \emph{113}, 11580--11585\relax
\mciteBstWouldAddEndPuncttrue
\mciteSetBstMidEndSepPunct{\mcitedefaultmidpunct}
{\mcitedefaultendpunct}{\mcitedefaultseppunct}\relax
\EndOfBibitem
\bibitem[Patkowski(2013)]{Patkowski:13}
Patkowski,~K. Basis Set Converged Weak Interaction Energies from Conventional
  and Explicitly Correlated Coupled-Cluster Approach. \emph{J. Chem. Phys.}
  \textbf{2013}, \emph{138}, 154101\relax
\mciteBstWouldAddEndPuncttrue
\mciteSetBstMidEndSepPunct{\mcitedefaultmidpunct}
{\mcitedefaultendpunct}{\mcitedefaultseppunct}\relax
\EndOfBibitem
\bibitem[Sirianni \latin{et~al.}(2017)Sirianni, Burns, and
  Sherrill]{Sirianni:17}
Sirianni,~D.~A.; Burns,~L.~A.; Sherrill,~C.~D. Comparison of Explicitly
  Correlated Methods for Computing High-Accuracy Benchmark Energies for
  Noncovalent Interactions. \emph{J. Chem. Theory Comput.} \textbf{2017},
  \emph{13}, 86--99\relax
\mciteBstWouldAddEndPuncttrue
\mciteSetBstMidEndSepPunct{\mcitedefaultmidpunct}
{\mcitedefaultendpunct}{\mcitedefaultseppunct}\relax
\EndOfBibitem
\bibitem[Szalewicz and Jeziorski(1979)Szalewicz, and Jeziorski]{Szalewicz:79}
Szalewicz,~K.; Jeziorski,~B. Symmetry-adapted double-perturbation analysis of
  intramolecular correlation effects in weak intermolecular interactions: the
  He-He interaction. \emph{Mol. Phys.} \textbf{1979}, \emph{38}, 191--208\relax
\mciteBstWouldAddEndPuncttrue
\mciteSetBstMidEndSepPunct{\mcitedefaultmidpunct}
{\mcitedefaultendpunct}{\mcitedefaultseppunct}\relax
\EndOfBibitem
\bibitem[Korona \latin{et~al.}(1997)Korona, Williams, Bukowski, Jeziorski, and
  Szalewicz]{Korona:97}
Korona,~T.; Williams,~H.~L.; Bukowski,~R.; Jeziorski,~B.; Szalewicz,~K. Helium
  dimer potential from symmetry-adapted perturbation theory calculations using
  large Gaussian geminal and orbital basis sets. \emph{J. Chem. Phys.}
  \textbf{1997}, \emph{106}, 5109--5122\relax
\mciteBstWouldAddEndPuncttrue
\mciteSetBstMidEndSepPunct{\mcitedefaultmidpunct}
{\mcitedefaultendpunct}{\mcitedefaultseppunct}\relax
\EndOfBibitem
\bibitem[Jeziorska \latin{et~al.}(2007)Jeziorska, Cencek, Patkowski, Jeziorski,
  and Szalewicz]{Jeziorska:07}
Jeziorska,~M.; Cencek,~W.; Patkowski,~K.; Jeziorski,~B.; Szalewicz,~K. Pair
  potential for helium from symmetry-adapted perturbation theory calculations
  and from supermolecular data. \emph{J. Chem. Phys.} \textbf{2007},
  \emph{127}, 124303\relax
\mciteBstWouldAddEndPuncttrue
\mciteSetBstMidEndSepPunct{\mcitedefaultmidpunct}
{\mcitedefaultendpunct}{\mcitedefaultseppunct}\relax
\EndOfBibitem
\bibitem[Mitroy \latin{et~al.}(2013)Mitroy, Bubin, Horiuchi, Suzuki, Adamowicz,
  Cencek, Szalewicz, Komasa, Blume, and Varga]{Mitroy:13}
Mitroy,~J.; Bubin,~S.; Horiuchi,~W.; Suzuki,~Y.; Adamowicz,~L.; Cencek,~W.;
  Szalewicz,~K.; Komasa,~J.; Blume,~D.; Varga,~K. Theory and application of
  explicitly correlated Gaussians. \emph{Rev. Mod. Phys.} \textbf{2013},
  \emph{85}, 693--749\relax
\mciteBstWouldAddEndPuncttrue
\mciteSetBstMidEndSepPunct{\mcitedefaultmidpunct}
{\mcitedefaultendpunct}{\mcitedefaultseppunct}\relax
\EndOfBibitem
\bibitem[Przybytek(2018)]{Przybytek:18}
Przybytek,~M. Dispersion Energy of Symmetry-Adapted Perturbation Theory from
  the Explicitly Correlated F12 Approach. \emph{J. Chem. Theory Comput.}
  \textbf{2018}, \emph{14}, 5105--5117\relax
\mciteBstWouldAddEndPuncttrue
\mciteSetBstMidEndSepPunct{\mcitedefaultmidpunct}
{\mcitedefaultendpunct}{\mcitedefaultseppunct}\relax
\EndOfBibitem
\bibitem[Kendall \latin{et~al.}(1992)Kendall, {Dunning Jr.}, and
  Harrison]{Kendall:92}
Kendall,~R.~A.; {Dunning Jr.},~T.~H.; Harrison,~R.~J. Electron Affinities of
  the 1st-Row Atoms Revisited - Systematic Basis Sets and Wave Functions.
  \emph{J. Chem. Phys.} \textbf{1992}, \emph{96}, 6796--6806\relax
\mciteBstWouldAddEndPuncttrue
\mciteSetBstMidEndSepPunct{\mcitedefaultmidpunct}
{\mcitedefaultendpunct}{\mcitedefaultseppunct}\relax
\EndOfBibitem
\bibitem[Kodrycka \latin{et~al.}(2019)Kodrycka, Holzer, Klopper, and
  Patkowski]{Kodrycka:19a}
Kodrycka,~M.; Holzer,~C.; Klopper,~W.; Patkowski,~K. Explicitly correlated
  dispersion and exchange dispersion energies in symmetry-adapted perturbation
  theory. \emph{J. Chem. Theory Comput.} \textbf{2019}, \emph{15},
  5965--5986\relax
\mciteBstWouldAddEndPuncttrue
\mciteSetBstMidEndSepPunct{\mcitedefaultmidpunct}
{\mcitedefaultendpunct}{\mcitedefaultseppunct}\relax
\EndOfBibitem
\bibitem[Valeev(2004)]{Valeev:04}
Valeev,~E.~F. Improving on the resolution of the identity in linear R12 ab
  initio theories. \emph{Chem. Phys. Lett.} \textbf{2004}, \emph{395},
  190--195\relax
\mciteBstWouldAddEndPuncttrue
\mciteSetBstMidEndSepPunct{\mcitedefaultmidpunct}
{\mcitedefaultendpunct}{\mcitedefaultseppunct}\relax
\EndOfBibitem
\bibitem[Whitten(1973)]{Whitten:73}
Whitten,~J.~L. Coulombic potential energy integrals and approximations.
  \emph{J. Chem. Phys.} \textbf{1973}, \emph{58}, 4496--4501\relax
\mciteBstWouldAddEndPuncttrue
\mciteSetBstMidEndSepPunct{\mcitedefaultmidpunct}
{\mcitedefaultendpunct}{\mcitedefaultseppunct}\relax
\EndOfBibitem
\bibitem[Dunlap \latin{et~al.}(1979)Dunlap, Connolly, and Sabin]{Dunlap:79}
Dunlap,~B.~I.; Connolly,~J. W.~D.; Sabin,~J.~R. On first-row diatomic molecules
  and local density models. \emph{J. Chem. Phys.} \textbf{1979}, \emph{71},
  4993--4999\relax
\mciteBstWouldAddEndPuncttrue
\mciteSetBstMidEndSepPunct{\mcitedefaultmidpunct}
{\mcitedefaultendpunct}{\mcitedefaultseppunct}\relax
\EndOfBibitem
\bibitem[Manby(2003)]{Manby:03}
Manby,~F.~R. Density fitting in second-order linear-R12 M{\o}ller-Plesset
  perturbation theory. \emph{J. Chem. Phys.} \textbf{2003}, \emph{119},
  4607--4613\relax
\mciteBstWouldAddEndPuncttrue
\mciteSetBstMidEndSepPunct{\mcitedefaultmidpunct}
{\mcitedefaultendpunct}{\mcitedefaultseppunct}\relax
\EndOfBibitem
\bibitem[May and Manby(2004)May, and Manby]{May:04a}
May,~A.~J.; Manby,~F.~R. An explicitly correlated second order
  M{\o}ller-Plesset theory using a frozen Gaussian geminal. \emph{J. Chem.
  Phys.} \textbf{2004}, \emph{121}, 4479--4485\relax
\mciteBstWouldAddEndPuncttrue
\mciteSetBstMidEndSepPunct{\mcitedefaultmidpunct}
{\mcitedefaultendpunct}{\mcitedefaultseppunct}\relax
\EndOfBibitem
\bibitem[Adler \latin{et~al.}(2007)Adler, Knizia, and Werner]{Adler:07}
Adler,~T.~B.; Knizia,~G.; Werner,~H.-J. A Simple and Efficient {CCSD(T)-F12}
  Approximation. \emph{J. Chem. Phys.} \textbf{2007}, \emph{127}, 221106\relax
\mciteBstWouldAddEndPuncttrue
\mciteSetBstMidEndSepPunct{\mcitedefaultmidpunct}
{\mcitedefaultendpunct}{\mcitedefaultseppunct}\relax
\EndOfBibitem
\bibitem[Knizia \latin{et~al.}(2009)Knizia, Adler, and Werner]{Knizia:09}
Knizia,~G.; Adler,~T.~B.; Werner,~H.-J. Simplified {CCSD(T)-F12} Methods:
  Theory and Benchmarks. \emph{J. Chem. Phys.} \textbf{2009}, \emph{130},
  054104\relax
\mciteBstWouldAddEndPuncttrue
\mciteSetBstMidEndSepPunct{\mcitedefaultmidpunct}
{\mcitedefaultendpunct}{\mcitedefaultseppunct}\relax
\EndOfBibitem
\bibitem[{H\"{a}ttig} \latin{et~al.}(2010){H\"{a}ttig}, Tew, and
  {K\"{o}hn}]{Hattig:10}
{H\"{a}ttig},~C.; Tew,~D.~P.; {K\"{o}hn},~A. Accurate and efficient
  approximations to explicitly correlated coupled-cluster singles and doubles,
  {CCSD-F12}. \emph{J. Chem. Phys.} \textbf{2010}, \emph{132}, 231102\relax
\mciteBstWouldAddEndPuncttrue
\mciteSetBstMidEndSepPunct{\mcitedefaultmidpunct}
{\mcitedefaultendpunct}{\mcitedefaultseppunct}\relax
\EndOfBibitem
\bibitem[Papajak \latin{et~al.}(2011)Papajak, Zheng, Xu, Leverentz, and
  Truhlar]{Papajak:11}
Papajak,~E.; Zheng,~J.; Xu,~X.; Leverentz,~H.~R.; Truhlar,~D.~G. Perspectives
  on Basis Sets Beautiful: Seasonal Plantings of Diffuse Basis Functions.
  \emph{J. Chem. Theory Comput.} \textbf{2011}, \emph{7}, 3027--3034\relax
\mciteBstWouldAddEndPuncttrue
\mciteSetBstMidEndSepPunct{\mcitedefaultmidpunct}
{\mcitedefaultendpunct}{\mcitedefaultseppunct}\relax
\EndOfBibitem
\bibitem[Parker \latin{et~al.}(2014)Parker, Burns, Parrish, Ryno, and
  Sherrill]{Parker:14}
Parker,~T.~M.; Burns,~L.~A.; Parrish,~R.~M.; Ryno,~A.~G.; Sherrill,~C.~D.
  Levels of Symmetry Adapted Perturbation Theory (SAPT). I. Efficiency and
  Performance for Interaction Energies. \emph{J. Chem. Phys.} \textbf{2014},
  \emph{140}, 094106\relax
\mciteBstWouldAddEndPuncttrue
\mciteSetBstMidEndSepPunct{\mcitedefaultmidpunct}
{\mcitedefaultendpunct}{\mcitedefaultseppunct}\relax
\EndOfBibitem
\bibitem[Riley \latin{et~al.}(2010)Riley, {Pito\v{n}\'ak}, {Jure\v{c}ka}, and
  Hobza]{Riley:10}
Riley,~K.~E.; {Pito\v{n}\'ak},~M.; {Jure\v{c}ka},~P.; Hobza,~P. Stabilization
  and Structure Calculations for Noncovalent Interactions in Extended Molecular
  Systems Based on Wave Function and Density Functional Theories. \emph{Chem.
  Rev.} \textbf{2010}, \emph{110}, 5023--5063\relax
\mciteBstWouldAddEndPuncttrue
\mciteSetBstMidEndSepPunct{\mcitedefaultmidpunct}
{\mcitedefaultendpunct}{\mcitedefaultseppunct}\relax
\EndOfBibitem
\bibitem[Marshall \latin{et~al.}(2011)Marshall, Burns, and
  Sherrill]{Marshall:11}
Marshall,~M.~S.; Burns,~L.~A.; Sherrill,~C.~D. Basis Set Convergence of the
  Coupled-Cluster Correction, {$\delta_{\rm MP2}^{\rm CCSD(T)}$}: Best
  Practices for Benchmarking Non-Covalent Interactions and the Attendant
  Revision of the {S22, NBC10, HBC6, and HSG} Databases. \emph{J. Chem. Phys.}
  \textbf{2011}, \emph{135}, 194102\relax
\mciteBstWouldAddEndPuncttrue
\mciteSetBstMidEndSepPunct{\mcitedefaultmidpunct}
{\mcitedefaultendpunct}{\mcitedefaultseppunct}\relax
\EndOfBibitem
\bibitem[Tajti \latin{et~al.}(2004)Tajti, Szalay, {Cs\'asz\'ar}, {K\'allay},
  Gauss, Valeev, Flowers, {V\'azquez}, and Stanton]{Tajti:04}
Tajti,~A.; Szalay,~P.~G.; {Cs\'asz\'ar},~A.~G.; {K\'allay},~M.; Gauss,~J.;
  Valeev,~E.~F.; Flowers,~B.~A.; {V\'azquez},~J.; Stanton,~J.~F. {HEAT}: High
  accuracy extrapolated ab initio thermochemistry. \emph{J. Chem. Phys.}
  \textbf{2004}, \emph{121}, 11599--11613\relax
\mciteBstWouldAddEndPuncttrue
\mciteSetBstMidEndSepPunct{\mcitedefaultmidpunct}
{\mcitedefaultendpunct}{\mcitedefaultseppunct}\relax
\EndOfBibitem
\bibitem[Karton \latin{et~al.}(2006)Karton, Rabinovich, Martin, and
  Ruscic]{Karton:06}
Karton,~A.; Rabinovich,~E.; Martin,~J. M.~L.; Ruscic,~B. W4 theory for
  computational thermochemistry: In pursuit of confident sub-kJ/mol
  predictions. \emph{J. Chem. Phys.} \textbf{2006}, \emph{125}, 144108\relax
\mciteBstWouldAddEndPuncttrue
\mciteSetBstMidEndSepPunct{\mcitedefaultmidpunct}
{\mcitedefaultendpunct}{\mcitedefaultseppunct}\relax
\EndOfBibitem
\bibitem[Smith \latin{et~al.}(2018)Smith, Burns, Sirianni, Nascimento, Kumar,
  James, Schriber, Zhang, Zhang, Abbott, Berquist, Lechner, Cunha, Heide,
  Waldrop, Takeshita, Alenaizan, Neuhauser, King, Simmonett, Turney, Schaefer,
  Evangelista, {DePrince III}, Crawford, Patkowski, and Sherrill]{Smith:18}
Smith,~D. G.~A. \latin{et~al.}  {\sc Psi4NumPy}: An Interactive Quantum
  Chemistry Programming Environment for Reference Implementations and Rapid
  Development. \emph{J. Chem. Theory Comput.} \textbf{2018}, \emph{14},
  3504--3511\relax
\mciteBstWouldAddEndPuncttrue
\mciteSetBstMidEndSepPunct{\mcitedefaultmidpunct}
{\mcitedefaultendpunct}{\mcitedefaultseppunct}\relax
\EndOfBibitem
\bibitem[Parrish \latin{et~al.}(2017)Parrish, Burns, Smith, Simmonett,
  {DePrince, III}, Hohenstein, Bozkaya, Sokolov, {Di Remigio}, Richard,
  Gonthier, James, {McAlexander}, Kumar, Saitow, Wang, Pritchard, Verma,
  {Schaefer, III}, Patkowski, King, Valeev, Evangelista, Turney, Crawford, and
  Sherrill]{Parrish:17}
Parrish,~R.~M. \latin{et~al.}  {\sc Psi}4 1.1: An Open-Source Electronic
  Structure Program Emphasizing Automation, Advanced Libraries, and
  Interoperability. \emph{J. Chem. Theory Comput.} \textbf{2017}, \emph{13},
  3185--3197\relax
\mciteBstWouldAddEndPuncttrue
\mciteSetBstMidEndSepPunct{\mcitedefaultmidpunct}
{\mcitedefaultendpunct}{\mcitedefaultseppunct}\relax
\EndOfBibitem
\bibitem[Smith \latin{et~al.}(2020)Smith, Burns, Simmonett, Parrish, Schieber,
  Galvelis, Kraus, Kruse, {Di Remigio}, Alenaizan, James, Lehtola, Misiewicz,
  Scheurer, Shaw, Schriber, Xie, Glick, Sirianni, {O'Brien}, Waldrop, Kumar,
  Hohenstein, Pritchard, Brooks, {Schaefer III}, Sokolov, Patkowski,
  {DePrince}, Bozkaya, King, Evangelista, Turney, Crawford, and
  Sherrill]{Smith:20}
Smith,~D. G.~A. \latin{et~al.}  {\sc Psi4} 1.4: Open-source software for
  high-throughput quantum chemistry. \emph{J. Chem. Phys.} \textbf{2020},
  \emph{152}, 184108\relax
\mciteBstWouldAddEndPuncttrue
\mciteSetBstMidEndSepPunct{\mcitedefaultmidpunct}
{\mcitedefaultendpunct}{\mcitedefaultseppunct}\relax
\EndOfBibitem
\bibitem[{\v{R}ez\'a\v{c}} and Hobza(2013){\v{R}ez\'a\v{c}}, and
  Hobza]{Rezac:13a}
{\v{R}ez\'a\v{c}},~J.; Hobza,~P. Describing Noncovalent Interactions beyond the
  Common Approximations: How Accurate Is the 'Gold Standard,' {CCSD(T)} at the
  Complete Basis Set Limit? \emph{J. Chem. Theory Comput.} \textbf{2013},
  \emph{9}, 2151--2155\relax
\mciteBstWouldAddEndPuncttrue
\mciteSetBstMidEndSepPunct{\mcitedefaultmidpunct}
{\mcitedefaultendpunct}{\mcitedefaultseppunct}\relax
\EndOfBibitem
\bibitem[{H\"attig} \latin{et~al.}(2012){H\"attig}, Klopper, {K\"ohn}, and
  Tew]{Hattig:12}
{H\"attig},~C.; Klopper,~W.; {K\"ohn},~A.; Tew,~D.~P. Explicitly Correlated
  Electrons in Molecules. \emph{Chem. Rev.} \textbf{2012}, \emph{112},
  4--74\relax
\mciteBstWouldAddEndPuncttrue
\mciteSetBstMidEndSepPunct{\mcitedefaultmidpunct}
{\mcitedefaultendpunct}{\mcitedefaultseppunct}\relax
\EndOfBibitem
\bibitem[Tew and Klopper(2006)Tew, and Klopper]{Tew:06}
Tew,~D.~P.; Klopper,~W. A comparison of linear and nonlinear correlation
  factors for basis set limit M{\o}ller-Plesset second order binding energies
  and structures of He$_2$, Be$_2$, and Ne$_2$. \emph{J. Chem. Phys.}
  \textbf{2006}, \emph{125}, 094302\relax
\mciteBstWouldAddEndPuncttrue
\mciteSetBstMidEndSepPunct{\mcitedefaultmidpunct}
{\mcitedefaultendpunct}{\mcitedefaultseppunct}\relax
\EndOfBibitem
\bibitem[{Ten-no}(2004)]{Tenno:04}
{Ten-no},~S. Initiation of explicitly correlated Slater-type geminal theory.
  \emph{Chem. Phys. Lett.} \textbf{2004}, \emph{398}, 56--61\relax
\mciteBstWouldAddEndPuncttrue
\mciteSetBstMidEndSepPunct{\mcitedefaultmidpunct}
{\mcitedefaultendpunct}{\mcitedefaultseppunct}\relax
\EndOfBibitem
\bibitem[{H\"{o}fener} \latin{et~al.}(2008){H\"{o}fener}, Bischoff,
  {Gl\"{o}\ss}, and Klopper]{Hoefener:08}
{H\"{o}fener},~S.; Bischoff,~F.~A.; {Gl\"{o}\ss},~A.; Klopper,~W. Slater-type
  geminals in explicitly-correlated perturbation theory: application to
  $n$-alkanols and analysis of errors and basis-set requirements. \emph{Phys.
  Chem. Chem. Phys.} \textbf{2008}, \emph{10}, 3390--3399\relax
\mciteBstWouldAddEndPuncttrue
\mciteSetBstMidEndSepPunct{\mcitedefaultmidpunct}
{\mcitedefaultendpunct}{\mcitedefaultseppunct}\relax
\EndOfBibitem
\bibitem[{Dunning Jr.}(1989)]{Dunning:89}
{Dunning Jr.},~T.~H. Gaussian-Basis Sets for Use in Correlated Molecular
  Calculations. 1. The Atoms Boron through Neon and Hydrogen. \emph{J. Chem.
  Phys.} \textbf{1989}, \emph{90}, 1007--1023\relax
\mciteBstWouldAddEndPuncttrue
\mciteSetBstMidEndSepPunct{\mcitedefaultmidpunct}
{\mcitedefaultendpunct}{\mcitedefaultseppunct}\relax
\EndOfBibitem
\bibitem[Weigend \latin{et~al.}(2002)Weigend, {K\"{o}hn}, and
  {H\"{a}ttig}]{Weigend:02}
Weigend,~F.; {K\"{o}hn},~A.; {H\"{a}ttig},~C. Efficient use of the correlation
  consistent basis sets in resolution of the identity {MP2} calculations.
  \emph{J. Chem. Phys.} \textbf{2002}, \emph{116}, 3175--3183\relax
\mciteBstWouldAddEndPuncttrue
\mciteSetBstMidEndSepPunct{\mcitedefaultmidpunct}
{\mcitedefaultendpunct}{\mcitedefaultseppunct}\relax
\EndOfBibitem
\bibitem[{H\"{a}ttig}(2005)]{Hattig:05}
{H\"{a}ttig},~C. Optimization of auxiliary basis sets for {RI-MP2 and RI-CC2}
  calculations: Core-valence and quintuple-zeta basis sets for {H to Ar and
  QZVPP} basis sets for {Li to Kr}. \emph{Phys. Chem. Chem. Phys.}
  \textbf{2005}, \emph{7}, 59--66\relax
\mciteBstWouldAddEndPuncttrue
\mciteSetBstMidEndSepPunct{\mcitedefaultmidpunct}
{\mcitedefaultendpunct}{\mcitedefaultseppunct}\relax
\EndOfBibitem
\bibitem[Weigend(2002)]{Weigend:02b}
Weigend,~F. A fully direct {RI-HF} algorithm: Implementation, optimised
  auxiliary basis sets, demonstration of accuracy and efficiency. \emph{Phys.
  Chem. Chem. Phys.} \textbf{2002}, \emph{4}, 4285--4291\relax
\mciteBstWouldAddEndPuncttrue
\mciteSetBstMidEndSepPunct{\mcitedefaultmidpunct}
{\mcitedefaultendpunct}{\mcitedefaultseppunct}\relax
\EndOfBibitem
\bibitem[Hohenstein and Sherrill(2010)Hohenstein, and Sherrill]{Hohenstein:10}
Hohenstein,~E.~G.; Sherrill,~C.~D. Density fitting and Cholesky decomposition
  approximations in symmetry-adapted perturbation theory: Implementation and
  application to probe the nature of pi-pi interactions in linear acenes.
  \emph{J. Chem. Phys.} \textbf{2010}, \emph{132}, 184111\relax
\mciteBstWouldAddEndPuncttrue
\mciteSetBstMidEndSepPunct{\mcitedefaultmidpunct}
{\mcitedefaultendpunct}{\mcitedefaultseppunct}\relax
\EndOfBibitem
\bibitem[Helgaker \latin{et~al.}(1997)Helgaker, Klopper, Koch, and
  Noga]{Helgaker:97}
Helgaker,~T.; Klopper,~W.; Koch,~H.; Noga,~J. Basis-set convergence of
  correlated calculations on water. \emph{J. Chem. Phys.} \textbf{1997},
  \emph{106}, 9639--9646\relax
\mciteBstWouldAddEndPuncttrue
\mciteSetBstMidEndSepPunct{\mcitedefaultmidpunct}
{\mcitedefaultendpunct}{\mcitedefaultseppunct}\relax
\EndOfBibitem
\bibitem[Mielke \latin{et~al.}(2002)Mielke, Garrett, and Peterson]{Mielke:02}
Mielke,~S.~L.; Garrett,~B.~C.; Peterson,~K.~A. A hierarchical family of global
  analytic Born-Oppenheimer potential energy surfaces for the H+H$_2$ reaction
  ranging in quality from double-zeta to the complete basis set limit. \emph{J.
  Chem. Phys.} \textbf{2002}, \emph{116}, 4142--4161\relax
\mciteBstWouldAddEndPuncttrue
\mciteSetBstMidEndSepPunct{\mcitedefaultmidpunct}
{\mcitedefaultendpunct}{\mcitedefaultseppunct}\relax
\EndOfBibitem
\bibitem[Podeszwa \latin{et~al.}(2006)Podeszwa, Bukowski, and
  Szalewicz]{Podeszwa:06a}
Podeszwa,~R.; Bukowski,~R.; Szalewicz,~K. Potential energy surface for the
  benzene dimer and perturbational analysis of $\pi-\pi$ interactions. \emph{J.
  Phys. Chem. A} \textbf{2006}, \emph{110}, 10345--10354\relax
\mciteBstWouldAddEndPuncttrue
\mciteSetBstMidEndSepPunct{\mcitedefaultmidpunct}
{\mcitedefaultendpunct}{\mcitedefaultseppunct}\relax
\EndOfBibitem
\bibitem[Akin-Ojo \latin{et~al.}(2003)Akin-Ojo, Bukowski, and
  Szalewicz]{Akin-Ojo:03}
Akin-Ojo,~O.; Bukowski,~R.; Szalewicz,~K. Ab Initio Studies of {He--HCCCN}
  Interaction. \emph{J. Chem. Phys.} \textbf{2003}, \emph{119},
  8379--8396\relax
\mciteBstWouldAddEndPuncttrue
\mciteSetBstMidEndSepPunct{\mcitedefaultmidpunct}
{\mcitedefaultendpunct}{\mcitedefaultseppunct}\relax
\EndOfBibitem
\bibitem[Patkowski(2017)]{Patkowski:17}
Patkowski,~K. Benchmark Databases of Intermolecular Interaction Energies:
  Design, Construction, and Significance. In \emph{Annual Reports in
  Computational Chemistry}; Dixon,~D.~A., Ed.; Elsevier, Amsterdam, 2017;
  Vol.~13; pp 3--91\relax
\mciteBstWouldAddEndPuncttrue
\mciteSetBstMidEndSepPunct{\mcitedefaultmidpunct}
{\mcitedefaultendpunct}{\mcitedefaultseppunct}\relax
\EndOfBibitem
\bibitem[Parrish \latin{et~al.}(2013)Parrish, Hohenstein, and
  Sherrill]{Parrish:13}
Parrish,~R.~M.; Hohenstein,~E.~G.; Sherrill,~C.~D. Tractability gains in
  symmetry-adapted perturbation theory including coupled double excitations:
  CCD+ST(CCD) dispersion with natural orbital truncations. \emph{J. Chem.
  Phys.} \textbf{2013}, \emph{139}, 174102\relax
\mciteBstWouldAddEndPuncttrue
\mciteSetBstMidEndSepPunct{\mcitedefaultmidpunct}
{\mcitedefaultendpunct}{\mcitedefaultseppunct}\relax
\EndOfBibitem
\bibitem[{Jure\v{c}ka} \latin{et~al.}(2006){Jure\v{c}ka}, {\v{S}poner},
  {\v{C}ern\'y}, and Hobza]{Jurecka:06}
{Jure\v{c}ka},~P.; {\v{S}poner},~J.; {\v{C}ern\'y},~J.; Hobza,~P. Benchmark
  database of accurate ({MP2} and {CCSD(T)} complete basis set limit)
  interaction energies of small model complexes, {DNA} base pairs, and amino
  acid pairs. \emph{Phys. Chem. Chem. Phys.} \textbf{2006}, \emph{8},
  1985--1993\relax
\mciteBstWouldAddEndPuncttrue
\mciteSetBstMidEndSepPunct{\mcitedefaultmidpunct}
{\mcitedefaultendpunct}{\mcitedefaultseppunct}\relax
\EndOfBibitem
\bibitem[Thanthiriwatte \latin{et~al.}(2011)Thanthiriwatte, Hohenstein, Burns,
  and Sherrill]{Thanthiriwatte:11}
Thanthiriwatte,~K.~S.; Hohenstein,~E.~G.; Burns,~L.~A.; Sherrill,~C.~D.
  Assessment of the Performance of DFT and DFT-D Methods for Describing
  Distance Dependence of Hydrogen-Bonded Interactions. \emph{J. Chem. Theory
  Comput.} \textbf{2011}, \emph{7}, 88--96\relax
\mciteBstWouldAddEndPuncttrue
\mciteSetBstMidEndSepPunct{\mcitedefaultmidpunct}
{\mcitedefaultendpunct}{\mcitedefaultseppunct}\relax
\EndOfBibitem
\bibitem[Sherrill \latin{et~al.}(2009)Sherrill, Takatani, and
  Hohenstein]{Sherrill:09}
Sherrill,~C.~D.; Takatani,~T.; Hohenstein,~E.~G. An Assessment of Theoretical
  Methods for Nonbonded Interactions: Comparison to Complete Basis Set Limit
  Coupled-Cluster Potential Energy Curves for the Benzene Dimer, the Methane
  Dimer, Benzene-Methane, and Benzene-H$_2$S. \emph{J. Phys. Chem. A}
  \textbf{2009}, \emph{113}, 10146--10159\relax
\mciteBstWouldAddEndPuncttrue
\mciteSetBstMidEndSepPunct{\mcitedefaultmidpunct}
{\mcitedefaultendpunct}{\mcitedefaultseppunct}\relax
\EndOfBibitem
\bibitem[Faver \latin{et~al.}(2011)Faver, Benson, He, Roberts, Wang, Marshall,
  Kennedy, Sherrill, and {Merz, Jr.}]{Faver:11}
Faver,~J.~C.; Benson,~M.~L.; He,~X.; Roberts,~B.~P.; Wang,~B.; Marshall,~M.~S.;
  Kennedy,~M.~R.; Sherrill,~C.~D.; {Merz, Jr.},~K.~M. Formal Estimation of
  Errors in Computed Absolute Interaction Energies of Protein-Ligand Complexes.
  \emph{J. Chem. Theory Comput.} \textbf{2011}, \emph{7}, 790--797\relax
\mciteBstWouldAddEndPuncttrue
\mciteSetBstMidEndSepPunct{\mcitedefaultmidpunct}
{\mcitedefaultendpunct}{\mcitedefaultseppunct}\relax
\EndOfBibitem
\bibitem[Burns \latin{et~al.}(2017)Burns, Faver, Zheng, Marshall, Smith,
  Vanommeslaeghe, {MacKerell, Jr.}, {Merz, Jr.}, and Sherrill]{Burns:17}
Burns,~L.~A.; Faver,~J.~C.; Zheng,~Z.; Marshall,~M.~S.; Smith,~D. G.~A.;
  Vanommeslaeghe,~K.; {MacKerell, Jr.},~A.~D.; {Merz, Jr.},~K.~M.;
  Sherrill,~C.~D. The BioFragment Database {(BFDb)}: An open-data platform for
  computational chemistry analysis of noncovalent interactions. \emph{J. Chem.
  Phys.} \textbf{2017}, \emph{147}, 161727\relax
\mciteBstWouldAddEndPuncttrue
\mciteSetBstMidEndSepPunct{\mcitedefaultmidpunct}
{\mcitedefaultendpunct}{\mcitedefaultseppunct}\relax
\EndOfBibitem
\bibitem[Yousaf and Peterson(2009)Yousaf, and Peterson]{Yousaf:09}
Yousaf,~K.~E.; Peterson,~K.~A. Optimized complementary auxiliary basis sets for
  explicitly correlated methods: aug-cc-pVnZ orbital basis sets. \emph{Chem.
  Phys. Lett.} \textbf{2009}, \emph{476}, 303--307\relax
\mciteBstWouldAddEndPuncttrue
\mciteSetBstMidEndSepPunct{\mcitedefaultmidpunct}
{\mcitedefaultendpunct}{\mcitedefaultseppunct}\relax
\EndOfBibitem
\bibitem[Peterson and {Dunning, Jr.}(2002)Peterson, and {Dunning,
  Jr.}]{Peterson:02}
Peterson,~K.~A.; {Dunning, Jr.},~T.~H. Accurate correlation consistent basis
  sets for molecular core-valence correlation effects: The second row atoms
  Al-Ar, and the first row atoms B-Ne revisited. \emph{J. Chem. Phys.}
  \textbf{2002}, \emph{117}, 10548--10560\relax
\mciteBstWouldAddEndPuncttrue
\mciteSetBstMidEndSepPunct{\mcitedefaultmidpunct}
{\mcitedefaultendpunct}{\mcitedefaultseppunct}\relax
\EndOfBibitem
\bibitem[Korona(2013)]{Korona:13}
Korona,~T. A coupled cluster treatment of intramonomer electron correlation
  within symmetry-adapted perturbation theory: benchmark calculations and a
  comparison with a density-functional theory description. \emph{Mol. Phys.}
  \textbf{2013}, \emph{111}, 3705--3715\relax
\mciteBstWouldAddEndPuncttrue
\mciteSetBstMidEndSepPunct{\mcitedefaultmidpunct}
{\mcitedefaultendpunct}{\mcitedefaultseppunct}\relax
\EndOfBibitem
\end{mcitethebibliography}
\end{document}